%% file: paper.tex
\documentclass[11pt]{article}
\usepackage{amsfonts,amsmath,amsthm}
\usepackage{algorithm2e,framed,algorithmic}
\usepackage{enumerate,bm}
\usepackage{wrapfig}
\usepackage[pdftex]{graphicx}
\usepackage{hyperref}

\input{newcmds}

\input{macros_short.tex}

\newlength{\defbaselineskip}
\def\Pione{\Pi_1}
\def\roneone{r_1}
\def\mynorm#1{\lVert #1 \rVert}
\DeclareMathOperator{\expon}{exp}
\DeclareMathOperator{\argmin}{argmin}
\DeclareMathOperator{\median}{median}
\DeclareMathOperator{\colsp}{colsp}
\newcommand{\eps}{\varepsilon}

\DeclareMathOperator{\poly}{poly}
\newcommand\remove[1]{}
\newcommand{\tH}{\tilde{H}}
\DeclareMathSymbol{\R}{\mathbin}{AMSb}{"52}
\def\RR{{R^{-1}}} 

\begin{document}

\input{title_abst}

\input{intro}

\input{background}

\input{mainalg}

\input{resultsL1}

\input{resultsLp}

\input{numerical-long}

\input{conc}


\appendix
\input{appdx-techlemmas.tex}

\input{appdx-pfFCTlcm.tex}

\input{appdx-pfFCTfjl.tex}

\input{appdx-pfL1basis.tex}
\input{appdx-pfL1fast.tex}

\input{appdx-pfL1faster.tex}
\input{appdx-pfEllRnd.tex}

\input{appdx-pfLpCond.tex}
\input{appdx-pfSWfixed.tex}
\input{appdx-pfSubspace.tex}

\end{document}

%% file: newcmds.tex
\newcommand{\FNormS}[1]{\mbox{}\left\|#1\right\|_F^2}

\newcommand{\TNorm }[1]{\mbox{}\left\|#1\right\|_2  }

\newcommand{\VTTNorm }[1]{\mbox{}\left\|#1\right\|_2  }

\newcommand{\setlinespacing}[1]%
           {\setlength{\baselineskip}{#1 \defbaselineskip}}

\newcommand{\rank}[1]{{\bf rank}{\left(#1\right)}}
\newcommand{\abs }[1]{\left|#1\right|}

\newtheorem{definition}{Definition}
\newtheorem{proposition}{Proposition}
\newtheorem{lemma}{Lemma}
\newtheorem{theorem}{Theorem}
\newtheorem{corollary}{Corollary}

\long\def\killtext#1{}

\def\Prob{\hbox{\bf{Pr}}}

\def\Exp{\hbox{\bf{E}}}

\def\var{\hbox{\bf{Var}}}



%% file: macros_short.tex
\def\math#1{$#1$}

\def\mand#1{$$#1$$}

\def\frac#1#2{{#1\over #2}}

\def\mld#1{\begin{equation}
#1
\end{equation}}
\def\eqar#1{\begin{eqnarray}
#1
\end{eqnarray}}
\def\eqan#1{\begin{eqnarray*}
#1
\end{eqnarray*}}

\def\cl#1{{\cal #1}}

\def\norm#1{{\|#1\|}}
\def\Norm#1{{\left\|#1\right\|}}

\def\r#1{{(\ref{#1})}}

\def\dotfil{\leaders\hbox to 1.5mm{.}\hfill}
\newcounter{rmnum}
\def\RN#1{\setcounter{rmnum}{#1}\uppercase\expandafter{\romannumeral\value{rmnum}}}
\def\rn#1{\setcounter{rmnum}{#1}\expandafter{\romannumeral\value{rmnum}}}

%% file: title_abst.tex
\title{The Fast Cauchy Transform and Faster Robust Linear Regression%
\footnote{A conference version of this paper appears under the same title in the Proceedings of the 2013 ACM-SIAM Symposium on Discrete Algorithms.}
}

\author{
Kenneth L. Clarkson
\thanks{
IBM Almaden Research Center,
650 Harry Road,
San Jose, CA 95120.
Email: klclarks@us.ibm.com
}
\and
Petros Drineas
\thanks{
Dept.~of Computer Science,
Rensselaer Polytechnic Institute,
Troy, NY 12180.
Email: drinep@cs.rpi.edu
}
\and
Malik Magdon-Ismail
\thanks{
Dept.~of Computer Science,
Rensselaer Polytechnic Institute,
Troy, NY 12180.
Email: magdon@cs.rpi.edu
}
\and
Michael W. Mahoney
\thanks{
Dept.~of Mathematics,
Stanford University,
Stanford, CA 94305.
Email: mmahoney@cs.stanford.edu
}
\and
Xiangrui Meng
\thanks{
ICME,
Stanford University,
Stanford, CA 94305.
Email: mengxr@stanford.edu
}
\and
David P. Woodruff
\thanks{
IBM Almaden Research Center,
650 Harry Road,
San Jose, CA 95120.
Email: dpwoodru@us.ibm.com
}
}

\date{}
\maketitle


\begin{abstract}%
\noindent
We provide fast algorithms for overconstrained $\ell_p$ regression and 
related problems: for an $n\times d$ input matrix $A$ and vector $b\in\R^n$, 
in $O(nd\log n)$ time we reduce the problem $\min_{x\in\R^d} \norm{Ax-b}_p$ 
to the same problem with input matrix $\tilde A$ of dimension $s \times d$ 
and corresponding \math{\tilde b} of dimension \math{s\times 1}.
Here,  \math{\tilde A} and \math{\tilde b} are a \emph{coreset} for the 
problem, consisting of sampled and rescaled rows of \math{A} and \math{b}; 
and $s$ is independent of $n$ and polynomial in $d$.
Our results improve on the best previous algorithms when $n\gg d$, for all 
$p\in [1,\infty)$ except $p=2$; in particular, they improve the 
\math{O(nd^{1.376+})} running time of Sohler and Woodruff (STOC, 2011) for 
$p=1$, that uses asymptotically fast matrix multiplication, and the 
$O(nd^5\log n)$ time of Dasgupta \emph{et al.} (SICOMP, 2009) for general 
$p$, that uses ellipsoidal rounding.
We also provide a suite of improved results for finding well-conditioned 
bases via ellipsoidal rounding, illustrating tradeoffs between running time 
and conditioning quality, including a one-pass conditioning algorithm for 
general $\ell_p$ problems.

To complement this theory, we provide a detailed empirical evaluation of 
implementations of our algorithms for $p=1$, comparing them with several 
related algorithms.
Among other things, our empirical results clearly show that, in the 
asymptotic regime, the theory is a very good guide to the practical 
performance of these algorithms.
Our algorithms use our faster constructions of well-conditioned bases for 
$\ell_p$ spaces and, for $p=1$, a fast subspace embedding of independent 
interest that we call the Fast Cauchy Transform:
a distribution over matrices $\Pi: \R^n\mapsto \R^{O(d\log d)}$, found obliviously to $A$, that 
approximately preserves the $\ell_1$ norms:
that is, with large probability, simultaneously for all $x$, $\norm{Ax}_1 \approx \norm{\Pi Ax}_1$, with distortion 
$O(d^{2+\eta} )$, for an arbitrarily small constant $\eta>0$;
and, moreover, $\Pi A$ can be computed in 
\math{O(nd\log d)} time. 
The techniques underlying our Fast Cauchy Transform include fast 
Johnson-Lindenstrauss transforms, low-coherence matrices, and rescaling by 
Cauchy random~variables.
\end{abstract}


%% file: intro.tex
\section{Introduction}
\label{sxn:intro}

Random sampling, random projection, and other embedding methods have 
proven to be very useful in recent years in the development of improved 
worst-case algorithms for a range of linear algebra problems.
For example, Gaussian random projections provide low-distortion subspace 
embeddings in the \math{\ell_2} norm, mapping an arbitrary 
\math{d}-dimensional subspace in \math{\R^n} into a \math{d}-dimensional 
subspace in \math{\R^r}, with $r = O(d)$, and distorting the \math{\ell_2} 
norm of each vector in the subspace by at most a constant factor. 
Importantly for many applications, the embedding is oblivious in the sense 
that it is implemented by a linear mapping chosen from a distribution on 
mappings that is independent of the input subspace. 
Such low-distortion embeddings can be used to speed up various geometric 
algorithms, if they can be computed sufficiently quickly.
As an example, the \emph{Fast Johnson Lindenstrauss transform} (FJLT) is 
one such embedding;
the FJLT is computable in $O(n\log d)$ time, using a variant of the fast 
Hadamard transform~\cite{AC06-JRNL09}.
Among other things, use of the FJLT leads to faster algorithms for 
constructing orthonormal bases, \math{\ell_2} regression, and 
\math{\ell_2} subspace approximation, which in turn lead to faster 
algorithms for a range of related problems including low-rank matrix 
approximation~\cite{DMMS07_FastL2_NM10,Mah-mat-rev_BOOK,DMMW12_ICML}.

In this paper, we use $\ell_1$ and $\ell_p$ extensions of these methods to 
provide faster algorithms for the classical $\ell_p$ regression problem and
several other related problems.
Recall the overconstrained $\ell_p$ regression~problem.
\begin{definition}
Given a matrix $A \in \mathbb{R}^{n \times d}$, with $n > d$, a 
vector $b \in \mathbb{R}^n$, and a norm $\|\cdot\|_p$, the 
\emph{$\ell_p$ regression problem} is to find an optimal solution to:
\begin{equation}
\label{eq:lp_reg}
   \min_{x \in \mathbb{R}^d} \|A x - b\|_p.
\end{equation}
\end{definition}

\noindent
In this paper, we are most interested in the case $p=1$, although many of 
our results hold more generally, and so we state several of our results for 
general $p$.
The $\ell_1$ regression problem, also known as the Least Absolute Deviations 
or Least Absolute Errors problem, is especially of interest as a more robust 
alternative to the $\ell_2$ regression or Least Squares 
Approximation problem.

It is well-known that for $p \ge 1$, the $\ell_p$ regression problem is a 
convex optimization problem; and for $p=1$ and $p=\infty$, it is an instance 
of linear programming.
Recent work has focused on using sampling, projection, and other embedding 
methods to solve these problems more quickly than with general 
convex programming or linear programming methods. 
Most relevant for our work is the work of Clarkson~\cite{Cla05} on solving
the $\ell_1$ regression problem with subgradient and sampling methods; the 
work of Dasgupta \emph{et al.}~\cite{DDHKM09_lp_SICOMP} on using 
well-conditioned bases and subspace-preserving sampling algorithms to solve 
general $\ell_p$ regression problems; and the work of Sohler and 
Woodruff~\cite{SW11} on using the Cauchy Transform to obtain improved 
$\ell_1$ embeddings, thereby leading to improved algorithms for the $\ell_1$ 
regression problem.
The Cauchy Transform of~\cite{SW11} provides low-distortion embeddings for 
the \math{\ell_1} norm, and thus it is an $\ell_1$ analog of the Gaussian 
projection for \math{\ell_2}.
It consists of a dense matrix of Cauchy random variables, and so it is 
``slow'' to apply to an arbitrary matrix $A$; but since it provides the 
first analog of the Johnson-Lindenstrauss embedding for the $\ell_1$ norm, 
it can be used to speed up randomized algorithms for problems such as 
$\ell_1$ regression and $\ell_1$ subspace approximation \cite{SW11}.

In this paper, we provide fast algorithms for overconstrained $\ell_p$ 
regression and several related problems.
Our algorithms use our faster constructions of well-conditioned bases for 
$\ell_p$ spaces; and, for $p=1$, our algorithms use a fast subspace 
embedding of independent interest that we call the Fast Cauchy Transform 
(FCT).
We also provide a detailed empirical evaluation of the FCT and its use at 
computing $\ell_1$ well-conditioned bases and solving $\ell_1$ regression 
problems.

The FCT is our main technical result, and it is essentially an $\ell_1$ 
analog of the FJLT.
The FCT can be represented by a distribution over matrices $\Pi: \R^n\mapsto \R^{O(d\log d)}$, 
found obliviously to $A$ (in the sense that its construction does not depend 
on any information in $A$), that approximately preserves the $\ell_1$ norms 
of all vectors in $\{Ax\mid x\in\R^d\}$.
That is, with large probability, simultaneously for all $x$, $\norm{Ax}_1 \approx \norm{\Pi Ax}_1$, with distortion 
$O(d^{2+\eta} \log d)$, for an arbitrarily small constant $\eta>0$;
(see Theorem~\ref{theorem:FCT}); and, moreover, 
$\Pi A$ can be computed in \math{O(nd\log d)} time. 
We actually provide two related constructions of the FCT (see 
Theorems~\ref{theorem:FCT2} and~\ref{theorem:FCT}).
The techniques underlying our FCTs include FJLTs, low-coherence matrices, 
and rescaling by Cauchy random~variables.

Our main application of the FCT embedding is to constructing the current 
fastest algorithm for computing a \emph{well-conditioned basis} for 
\math{\ell_1}  (see Theorem~\ref{thm:basisL1}).
Such a basis is an analog for the $\ell_1$ norm of what an orthonormal basis 
is for the $\ell_2$ norm, and our result improves the result in~\cite{SW11}. 
We also provide a generalization of this result to constructing 
\math{\ell_p} well-conditioned bases (see Theorem~\ref{thm:basisLp}).
The main application for well-conditioned bases is to regression:
if the rows of $A$ are sampled according to probabilities derived from the 
norms of the rows of such a basis, the resulting sample of rows (and 
corresponding entries of $b$) are with high probability a \emph{coreset} for 
the regression problem; see, \emph{e.g.},~\cite{DDHKM09_lp_SICOMP}.
That is, for an $n\times d$ input matrix $A$ and vector $b\in\R^n$, we can 
reduce an $\ell_p$ regression problem to another $\ell_p$ regression problem 
with input matrix $\tilde A$ of dimension $s \times d$ and corresponding 
\math{\tilde b} of dimension \math{s\times 1}.
Here, \math{\tilde A} and \math{\tilde b} consist of sampled and rescaled 
rows of \math{A} and \math{b}; and $s$ is independent of $n$ and polynomial 
in $d$.
We point out that our construction uses as a black box an
FJLT, which means that any improvement in the
running time of the FJLT (for example exploiting the sparsity of \math{A})
results in a corresponding improvement to the running times of our
\math{\ell_p} regression.

Based on our constructions of well-conditioned bases, we give the fastest 
known construction of coresets for $\ell_p$ regression, for all 
$p\in [1,\infty)$, except $p=2$.
In particular, for $\ell_1$ regression, we construct a coreset of size 
\math{\frac{1}{\eps^2}\poly(d,\log\frac{1}{\eps})} that achieves a 
\math{(1+\eps)}-approximation guarantee (see 
Theorem~\ref{theorem:FCRsimp}).  
Our construction runs in \math{O(nd\log n)} time, improving the previous 
best algorithm of Sohler and Woodruff~\cite{SW11}, which has an
\math{O(nd^{1.376+})} running time.
Our extension to finding an \math{\ell_p} well-conditioned basis also leads 
to an \math{O(nd\log n)} time algorithm for a 
\math{(1+\eps)}-approximation to the \math{\ell_p} regression problem
(see Theorem~\ref{thm:lprunning}), improving the \math{O(nd^5\log n)} 
algorithm of Dasgupta \emph{et al.}~\cite{DDHKM09_lp_SICOMP}.
For $p=1$, extensions of our basic methods yield improved algorithms for 
several related problems.
For example, we actually further optimize the running time for $p=1$ to 
\math{O(nd\log(\eps^{-1}d\log n))} (see Theorem~\ref{theorem:FCRsimpFaster}).
In addition, we generalize our $\ell_1$ result to solving the multiple 
regression problem (see Theorem~\ref{theorem:FCRmult}); and we use this to give 
the current fastest algorithm for computing a 
\math{(1+\eps)}-approximation for the \math{\ell_1}
subspace approximation problem (see Theorem~\ref{theorem:L1Sub}).

In addition to our construction of \math{\ell_p} well-conditioned bases 
(see Theorem~\ref{thm:basisLp}) and their use in providing a 
\math{(1+\eps)}-approximation to the \math{\ell_p} regression problem
(see Theorem~\ref{thm:lprunning}), we also provide a suite of improved 
results for finding well-conditioned bases via ellipsoidal rounding for 
general $\ell_p$ problems, illustrating tradeoffs between running time 
and conditioning quality.
These methods complement the FCT-based methods in the sense that the FCT 
may be viewed as a tool to compute a good basis in an oblivious manner, and 
the ellipsoid-based methods provide an alternate way to compute a good basis
in a data-dependent manner.
In particular, we prove that we can obtain an ellipsoidal rounding matrix 
in at most $O(n d^3 \log n)$ time that provides a $ 2 d$-rounding
(see Theorem~\ref{thm:lp_cond_2d}).
This is much faster than the algorithm of Lov{\'a}sz~\cite{lovasz} that 
computes a $(d (d+1))^{1/2}$-rounding in $O(n d^5 \log n)$ time.
We also present an optimized algorithm that uses an FJLT to compute a 
well-conditioned basis of $A$ in $O(n d \log n)$ time 
(see Theorem~\ref{thm:basisLp}).
When $p=1$, these $\ell_p$ rounding algorithms are competitive with or 
better than previous algorithms that were developed for $\ell_1$.

Finally, we also provide the first empirical evaluation for this class of 
randomized algorithms.  
In particular, we provide a detailed evaluation of a numerical 
implementation of both FCT constructions, and we compare the results
with an implementation of the (slow) Cauchy Transform, as well as a Gaussian 
Transform and an FJLT.
These latter two are $\ell_2$-based projections.
We evaluate the quality of the $\ell_1$ well-conditioned basis, the core 
component in all our geometric algorithms, on a suite of matrices designed 
to test the limits of these randomized algorithms, and we also evaluate how 
the method performs in the context of $\ell_1$ regression.
This latter evaluation includes an implementation on a nearly terabyte-scale 
problem, where we achieve a \math{10^{-3}} relative-error approximation to 
the optimal solution, a task that was infeasible prior to our work.
Among other things, our empirical results clearly show that, in the 
asymptotic regime, the theory is a very good guide to the practical 
performance of these algorithms.

Since this paper is long and detailed, we provide here a brief outline.
We start in Section~\ref{sxn:prelim} with some preliminaries, including 
several technical results that we will use in our analysis and that are of
independent interest.
Then, in Section~\ref{sxn:main}, we will present our main technical results
for the Fast Cauchy Transform; and 
in Section~\ref{sxn:resultsL1}, we will describe applications of it to 
$\ell_1$ well-conditioned basis construction and $\ell_1$ leverage score 
approximation, to solving the $\ell_1$ regression problem, and to solving 
the $\ell_1$ norm subspace approximation problem.
Then, in Section~\ref{sxn:extensionlp}, we describe extensions of these 
ideas to general $\ell_p$ problems.
Section~\ref{sec:numer} will contain a detailed empirical evaluation of our
algorithms for $\ell_1$-based problems, including the construction of 
$\ell_1$ well-conditioned bases and both small-scale and large-scale
$\ell_1$ regression problems.
Section~\ref{sxn:conc} will then contain a brief conclusion.
For simplicity of presentation, the proofs of our main results have been 
moved to Appendices~\ref{sec:proofs-technical} 
through~\ref{sxn:app-pf-thm-subspace}.

%% file: background.tex
\section{Preliminaries}
\label{sxn:prelim}

Let \math{A\in\R^{n\times d}} be an $n \times d$ input matrix, where we 
assume $n \gg d$ and $A$ has full column rank.
The task of \emph{linear regression} is to find a vector $x^*\in \R^d$ that 
minimizes \math{\norm{Ax-b}} with respect to \math{x}, for a given 
\math{b\in\R^n} and norm \math{\norm{\cdot}}. 
In this paper, our focus is mostly on the \math{\ell_1} norm, although we 
also discuss extensions to \math{\ell_p}, for any $p \geq 1$. 
Recall that, for \math{p\in[1,\infty]}, the \math{\ell_p} norm of a vector 
\math{x} is 
\math{
\norm{x}_p=\left(\sum_i|x_i|^p\right)^{1/p}
}, defined to be \math{\max_i |x_i|} for \math{p=\infty}.
Let $[n]$ denote the set $\{1,2,\ldots,n\}$; and let \math{A_{(i)}} and 
\math{A^{(j)}} be the \math{i}th row vector and \math{j}th column vector 
of \math{A}, respectively. 
For matrices, we use the Frobenius norm 
$\FNormS{A} = \sum_{i=1}^n \sum_{j=1}^d A_{ij}^2$,
the \math{\ell_2}-operator (or spectral) norm
$\TNorm{A} = \sup_{\ \VTTNorm{x}=1} \VTTNorm{Ax}$, 
and the entrywise \math{\ell_p} norm  
\math{\norm{X}_{p}= (\sum_{i,j}|X_{ij}|^p)^{1/p}}.
(The exception to this is \math{p=2}, 
where this notation is used 
for the spectral norm and 
the entrywise 2-norm is the 
Frobenius norm.)
Finally, the standard inner product between vectors $x,y$ is 
$\langle x,y \rangle = x^Ty$;
$e_i$ are standard basis vectors of the relevant dimension;
$I_n$ denotes the $n \times n$ identity matrix; 
and
\math{c} refers to a generic constant whose specific value may
vary throughout the paper.

\paragraph{Two Useful Tail Inequalities.}
The following two Bernstein-type tail inequalities are useful because they 
give tail bounds without reference to the number of i.i.d. trials.
The first bound is due to Maurer~\cite{Maurer}, and the second is an 
immediate application of the~first.

\begin{lemma}[\cite{Maurer}]
\label{lem:bernstein}
Let \math{X_i\ge 0} be independent random variables with 
\math{\sum_{i}\Exp[X_i^2]<\infty}, and define
\math{X=\sum_{i}X_i}. 
Then, for any~\math{t>0},
\mand{
\Prob[X\le \Exp[X]-t]\le\expon\left(\frac{-t^2}{2\sum_i\Exp[X_i^2]}\right).
}
\end{lemma}

\begin{lemma}
\label{lem:bernoulli}
Let \math{x_i} be i.i.d. Bernoulli random variables with probability 
\math{p}, and let \math{X=\sum_{i\in[n]}\xi_ix_i}, where 
\math{\xi_i\ge 0}, with \math{\sum_{i\in[n]}\xi_i=\xi} and 
\math{\sum_{i\in[n]}\xi_i^2\le\xi^2/\beta^2}. 
Then, for any~\math{t>0},
\mand{\Prob[X\ge \xi(p+t)]\le \expon\left(-\frac{\beta^2t^2}{2(1-p)}\right).
}
\end{lemma}
\begin{proof}
The proof is a straightforward application of Lemma~\ref{lem:bernstein} to
\math{Z=\sum_{i\in[n]}\xi_i(1-x_i)}.
\end{proof}

\paragraph{Sums of Cauchy Random Variables.} 
The Cauchy distribution, having density 
\math{p(x)=\frac{1}{\pi}\frac{1}{1+x^2}}, 
is the unique $1$-stable distribution.
If \math{C_1,\ldots,C_M} are independent Cauchys, then 
\math{\sum_{i\in[M]}\gamma_iC_i} is distributed as a Cauchy scaled by
\math{\gamma=\sum_{i\in[M]}|\gamma_i|}.
The Cauchy distribution will factor heavily in our discussion, and bounds 
for sums of Cauchy random variables will be used throughout. We note that the
Cauchy distribution has undefined expectation and infinite variance.

The following upper and lower tail inequalities for sums of Cauchy random 
variables are proved in Appendix~\ref{sec:proofs-technical}. The proof of
Lemma \ref{lem:tail} is similar to an argument of Indyk \cite{Indyk06}, though
in that paper the Cauchy random variables are independent. As in that paper,
our argument follows by a Markov bound after conditioning 
on the magnitudes of the Cauchy
random variable summands not being too large, so that their conditional expectations are defined. 
However, in this paper, the Cauchy random
variables are dependent, and so after conditioning on a global event, the expectations
of the magnitudes need not be the same as after this conditioning in the independent case. 

Lemma \ref{lem:lower} is a simple application of Lemma \ref{lem:bernstein}, while
Lemma \ref{lem:L1Chernoff} was shown in \cite{DDHKM09_lp_SICOMP}; we include the proofs
for completeness.  

\begin{lemma}[Cauchy Upper Tail Inequality]
\label{lem:tail}
For \math{i\in[m]}, let \math{C_i} be \math{m}
(not necessarily independent)
 Cauchy random variables, 
and \math{\gamma_i>0} with \math{\gamma=\sum_{i\in[m]}\gamma_i}.
Let \math{X=\sum_{i\in[m]}\gamma_i|C_i|}.
Then, for any~\math{t \ge 1},\
\begin{eqnarray*}
\Prob\left[X>\gamma t\right]
&\le& \frac{1}{\pi t} \left(\frac{\log (1+(2mt)^2)}{1-1/(\pi t)}+1\right)   
   =  \frac{\log (m t)}{t}\left(1+{\textstyle o(1)}\right).
\end{eqnarray*}
\end{lemma}


\noindent\textbf{Remark.} 
The bound has only logarithmic dependence on the number of Cauchy random 
variables and does \emph{not} rely on any independence assumption among the 
random variables. 
Even if the Cauchys are independent, one cannot substantially improve on 
this bound due to the nature of the Cauchy distribution. 
This is because, for independent Cauchys, 
\math{\sum_i\gamma_i|C_i|\ge|\sum_i\gamma_iC_i|}, and the latter sum is 
itself distributed as a Cauchy scaled by \math{\gamma}.
Hence for independent Cauchys,
\math{\Prob[X\ge\gamma t] = \frac{2}{\pi}\tan^{-1}t=\Omega(\frac{1}{t})}.

\begin{lemma}[Cauchy Lower Tail Inequality]
\label{lem:lower}
For \math{i\in[r]}, let \math{C_i} be independent Cauchy random variables, 
and \math{\gamma_i\ge0} with \math{\gamma=\sum_{i\in[r]}\gamma_i} and
\math{\sum_{i\in[r]}{\gamma_i^2}\le \gamma^2/\beta^2}.
Let \math{X=\sum_{i\in[r]}\gamma_i|C_i|}.
Then, for any~\math{t\ge0},
\mand{
\Prob\left[X\le \gamma(1-t) \right]
\le
\expon\left(-\frac{\beta^2t^2}{3}\right)  .
}
\end{lemma}

\paragraph{An \math{\ell_1} Sampling Lemma.}
We will also need an ``\math{\ell_1}-sampling lemma,'' which is an 
application of Bernstein's inequality. 
This lemma bounds how \math{\ell_1} norms get distorted under sampling 
according to \math{\ell_1} probabilities.
The proof of this lemma is also given in Appendix~\ref{sec:proofs-technical}.

\begin{lemma}[\math{\ell_1} Sampling Lemma]
\label{lem:L1Chernoff}
Let \math{Z\in\R^{n\times k}}
and suppose that for 
\math{i\in[n]}, 
\math{t_i\ge a\norm{Z_{(i)}}_1/\norm{Z}_1}.
For \math{s>0}, define 
\math{\hat p_i=\min\{1,s\cdot t_i\}},
and let
$D\in\R^{n\times n}$ be a random diagonal matrix with $D_{ii} = 1/\hat p_i$ 
with probability $\hat p_i$, and $0$ otherwise.
Then, for any (fixed) 
\math{x \in \mathbb{R}^k}, 
with probability at least \math{1-\delta},
$$
(1-\eps)\norm{Zx}_1
\le
\norm{DZx}_1
\le
(1+\eps)\norm{Zx}_1  ,
$$
where  \math{\displaystyle\delta\le
2\expon\left(\frac{-a s\eps^2\norm{Zx}_1}{(2+\frac{2}{3}\eps)\norm{Z}_1\norm{x}_\infty}
\right).}
\end{lemma}

%% file: mainalg.tex
\section{Main Technical Result: the Fast Cauchy Transform}
\label{sxn:main}

In this section, we present the Fast Cauchy Transform (FCT), which is an
$\ell_1$-based analog of the fast Johnson-Lindenstrauss transform (FJLT).
We will actually present two related constructions, one based on using a
quickly-constructable low-coherence matrix, and one based on using a version 
of the FJLT. 
In both cases, these matrices will be rescaled by Cauchy random variables 
(hence the name \emph{Fast Cauchy Transform}).
We will also state our main results, Theorems~\ref{theorem:FCT2} 
and~\ref{theorem:FCT}, which provides running time and 
quality-of-approximation guarantees for these two FCT embeddings.

\subsection{FCT1 Construction: via a Low-coherence Matrix}
\label{sec:FCT1}

This FCT construction first preprocesses by a deterministic low-coherence 
``spreading matrix,'' then rescales by Cauchy random variables, and finally 
samples linear combinations of the rows. 
Let \math{\delta \in(0,1]} be a parameter governing the failure probability  
of our algorithm.
Then, we construct $\Pione$~as
\[
\Pione\equiv 4B C \tH ,
\]
where:
\begin{description}
\item{$B\in\R^{\roneone \times 2n}$} has each column chosen independently
and uniformly from the $\roneone$
standard basis vectors for $\R^{\roneone}$;
we will set the parameter 
\math{r_1=\alpha d\log \frac{d}{\delta}}, where
 \math{\delta} controls the probability 
that our algorithms fail and \math{\alpha} is a suitably large constant;

\item{$C\in\R^{2n\times 2n}$} is a diagonal
matrix with diagonal entries chosen independently from a Cauchy distribution;
and 
\item{$\tH \in\R^{2n\times n}$} 
is a block-diagonal matrix comprised of \math{n/s} blocks along the diagonal.
Each block is the \math{2s\times s} matrix 
\math{G_s\equiv
\left[\begin{smallmatrix} H_{s}\\ I_s\end{smallmatrix}\right]}, where 
\math{I_s} is the $s\times s$ identity matrix, and
\math{H_s} is the normalized Hadamard matrix.
We will set 
$s= r_1^6$.
(Here, for simplicity,
 we assume $s$ is a power of two and $n/s$ is an integer.) %
\[
\tH \equiv \left[
\begin{matrix}
G_s & &\\
& G_s &&\\
&& \ddots & \\
&&& G_s \\
\end{matrix}
\right]
\]
\end{description}

\noindent
(For completeness, we remind the reader that the (non-normalized) 
$n \times n$ matrix of the Hadamard transform $H_n$ may be defined 
recursively as~follows:
\begin{equation*}
H_n = \left[
\begin{array}{cc}
  H_{n/2} & H_{n/2} \\
  H_{n/2} & -H_{n/2}
\end{array}\right]   ,
\quad \mbox{with}  \quad
H_2 = \left[\begin{array}{cc}  +1 & +1 \\  +1 & -1\end{array}\right].
\end{equation*}
The $n \times n$ normalized matrix of the Hadamard transform is then equal 
to $\frac{1}{\sqrt{n}}H_n$; hereafter, we will denote this normalized matrix 
by $H_n$.)
Heuristically, the effect of $\tH$ in the above FCT construction is to 
spread the weight of a vector, so that $\tH y$ has many entries that are not 
too small.
(This is discussed in Lemma~\ref{lem:uncertainty} in the proof of 
Theorem~\ref{theorem:FCT2} below.)
This means that the vector $C\tH y$ comprises Cauchy random variables with 
scale factors that are not too small; and finally these variables are summed 
up by $B$, yielding a vector $BC\tH y$, whose $\ell_1$ norm won't be too 
small relative to $\mynorm{y}_1$.
For this version of the FCT, we have the following theorem.
The proof of this theorem may be found in Appendix~\ref{sxn:FCT2}.

\begin{theorem}[Fast Cauchy Transform (FCT1)]
\label{theorem:FCT2}
There is a distribution (given by the above construction) over matrices
\math{\Pione\in\R^{r_1\times n}}, with \math{r_1=O(d\log d+d\log\frac1\delta)},
such that for an arbitrary (but fixed) \math{A\in\R^{n\times d}}, and for
all \math{x\in\R^{d}}, the inequalities
\begin{equation}
\label{eqn:fct1-small-dist}
\norm{Ax}_1\le\norm{\Pione Ax}_1\le \kappa\norm{Ax}_1
\end{equation}
hold with probability \math{1-\delta}, where 
\begin{equation*}
\kappa=O\left(\frac{d\sqrt{s}}{\delta}\log (r_1d)\right).
\end{equation*}
Further, for any \math{y\in\R^n}, the product \math{\Pione y} can be 
computed in \math{O(n\log r_1)} time.
\end{theorem}

\noindent
Setting \math{\delta} to a small constant, since \math{\sqrt{s}=r_1^3} and 
\math{r_1=O(d\log d)}, it follows that \math{\kappa=O(d^4\log^4d)} in the 
above theorem. 

\paragraph{Remark.} 
The existence of such a $\Pione$ satisfying bounds of the form 
(\ref{eqn:fct1-small-dist}) was established by Sohler and 
Woodruff~\cite{SW11}.
Here, our contribution is to show that $\Pione$ can be factored into 
structured matrices so that the product $\Pione A$ can be computed in 
$O(nd\log d)$ time.
We also remark that, in additional theoretical bounds provided by 
the FJLT, high-quality numerical implementations of variants of the 
Hadamard transform exist, which is an additional plus for our 
empirical evaluations of Theorem~\ref{theorem:FCT2} and 
Theorem~\ref{theorem:FCT}.

\paragraph{Remark.} 
Our proof of this theorem uses a tail bound
for \math{\norm{Bg}_2^2} in terms of \math{\norm{g}_2} and \math{\norm{g}_1},
where \math{g} is any positive vector in \math{\R^n}, and 
\math{B} is the matrix used in our FCT construction.
\math{\norm{Bg}_2^2=\sum_j{\gamma_j^2}} where 
\math{\gamma_j=\sum_{i}B_{ji}g_i} are \emph{anti-correlated} random variables.
To get concentration,
we independently bounded \math{\gamma_j^2} in our proof which required 
\math{s=r_1^6} to obtain the high probability result; this resulted in 
the bound
\math{\kappa=O(d^4\log^4 d)}. 

\subsection{FCT2 Construction: via a Fast Johnson-Lindenstrauss Transform}
\label{sec:FCT2}

This FCT construction first preprocesses by a FJLT and then rescales by 
Cauchy random variables.
Recall that \math{\delta \in(0,1]} is a parameter governing the failure 
probability  of our algorithm; and let \math{\eta>0} be a generic 
arbitrarily small positive constant (whose value may change from one formula 
to another).
Let $\roneone = c\cdot d \log \frac{d}{\delta}$, 
\math{s=c'\cdot (d+\log \frac{n}{\delta})},
and 
\math{t=s^{2+\eta}},
where the parameters \math{c,c'>0} are
appropriately large constants.
Then, we construct $\Pione\in\R^{r_1\times n}$ as
\[
\Pione\equiv  \frac{8}{r_1} \sqrt{\frac{\pi t}{2s}}\cdot C \tH ,
\]
where:
\begin{description}
\item{$C\in\R^{\roneone \times ns/t}$} is a matrix of independent Cauchy
random variables; and 
\item{$\tH \in\R^{ns/t\times n}$} 
is a block-diagonal matrix comprising \math{n/t} blocks along the diagonal.
Each block is the \math{s\times t} Fast Johnson-Lindenstrauss matrix $G$.
(The important property that $G$ must satisfy is given by 
Lemmas~\ref{thm:al} and \ref{thm:subspace} in the proofs below.)
Here, for simplicity, we assume that $n/t$ is an integer. %
\[
\tH \equiv \left[
\begin{matrix}
G & &\\
& G &&\\
&& \ddots & \\
&&& G \\
\end{matrix}
\right]   .
\]
\end{description}

\noindent
Informally, the matrix \math{\tH} reduces the dimensionality of the input
space by a very small amount such that the ``slow'' Cauchy Transform $C$ 
of \cite{SW11} can be applied in the allotted time.
Then, since we are ultimately multiplying by $C$, the results of \cite{SW11} 
still hold; but since the dimensionality is slightly reduced, the running 
time is improved.
For this version of the FCT, we have the following theorem.
The proof of this theorem may be found in Appendix~\ref{sxn:FCTFJLT}.

\begin{theorem}[Fast Cauchy Transform (FCT2)]
\label{theorem:FCT}
There is a distribution (given by the above construction) over matrices
\math{\Pione\in\R^{r_1\times n}}, with \math{r_1=O(d\log \frac{d}{\delta})},
such that for arbitrary (but fixed) \math{A\in\R^{n\times d}}, and for
all \math{x\in\R^{d}}, the inequalities
\mand{\norm{Ax}_1\le\norm{\Pione Ax}_1\le \kappa \norm{Ax}_1}
hold with probability $1-\delta$, where
\math{\kappa=O(\frac{d}{\delta}(d+\log\frac{n}\delta)^{1+\eta}\log d)}.
Further, for any \math{y\in\R^n}, the product \math{\Pione y} can be 
computed in \math{O(n\log\frac{d}{\delta})} time.
\end{theorem}

\noindent
Setting \math{\delta} to be a small constant and for 
\math{\log n<d}, \math{r_1=O(d\log d)},  
\math{\kappa=O(d^{2+\eta}\log d)} and \math{\Pione A} can be computed in 
\math{O(nd\log d)} time. Thus, we have a fast linear oblivious 
mapping from 
from \math{\ell_1^{n}\mapsto\ell_1^{O(d\log d)}} 
that has distortion
\math{O(d^{2+\eta}\log d)} on any (fixed) \math{d}-dimensional subspace 
of~\math{\R^n}.

\noindent\textbf{Remark.} 
For \math{\log n<d},
FCT2 gives a better dependence of the distortion on \math{d}, but more
generally 
FCT2 has a dependence on \math{\log n}. This dependence arises because 
the random FJLT matrix does not give a deterministic guarantee for
spreading out a vector whereas the low coherence matrix used in  
FCT1 does give a deterministic guarantee. This means that in using the
union bound, we need to overcome a factor of \math{n}.

\noindent\textbf{Remark.} 
The requirement \math{t\ge s^{2+\eta}} is set by the restriction in 
Lemma~\ref{thm:al} in the proof of Theorem~\ref{theorem:FCT}. 
In the bound of Theorem~\ref{theorem:FCT}, \math{\kappa=\kappa'\sqrt{t}}, 
where \math{\kappa'=O(d\log(r_1 d))} arises from Theorem~\ref{thm:sw}, which
originally appeared in \cite{SW11}. 
If a stronger version of Lemma~\ref{thm:al} can be proved that relaxes the
restriction \math{t\ge s^{2+\eta}}, then correspondingly the bound 
of Theorem~\ref{theorem:FCT}
will~improve.

\noindent\textbf{Remark.}
This second construction has the benefit of being easily extended to 
constructing well-conditioned bases of \math{\ell_p}, for \math{p>1}; see
Section~\ref{sxn:extensionlp}.

%% file: resultsL1.tex
\section{Algorithmic Applications in $\ell_1$ of the FCT}
\label{sxn:resultsL1}

In this section, we describe three related applications of the FCT to 
$\ell_1$-based problems.
The first is to the fast construction of an $\ell_1$ well-conditioned basis 
and the fast approximation of $\ell_1$ leverage scores;
the second is a fast algorithm for the least absolute deviations or 
$\ell_1$ regression problem; and 
the third is to a fast algorithm for the $\ell_1$ norm subspace 
approximation problem.

\subsection{Fast Construction of an $\ell_1$ Well-conditioned Basis and $\ell_1$ Leverage Scores}
\label{sxn:resultsL1-basis}

We start with the following definition, adapted from~\cite{DDHKM09_lp_SICOMP},
of a basis that is ``good'' for the $\ell_1$ norm in a manner that is 
analogous to how an orthogonal matrix is ``good'' for the $\ell_2$~norm.

\begin{definition}[$\ell_1$ Well-conditioned Basis (adapted from \cite{DDHKM09_lp_SICOMP})]
\label{def:l1basis}
A basis \math{U} for the range of \math{A} 
is \emph{\math{(\alpha,\beta)}-conditioned} if 
\math{\norm{U}_{1}\le \alpha} and
for all \math{x\in\R^d},
\math{\norm{x}_\infty\le\beta\norm{Ux}_1}.
We will say that \math{U} is \emph{well-conditioned} if \math{\alpha} and 
\math{\beta} are low-degree polynomials in \math{d}, independent of~\math{n}.
\end{definition}

\noindent\textbf{Remark.} 
An Auerbach basis for \math{A} is \math{(d,1)}-conditioned, and thus we know 
that there exist well-conditioned bases for \math{\ell_1}.  
More generally, well-conditioned bases can be defined in any 
\math{\ell_p} norm, using the notion of a dual norm \math{\ell_p^*}, and 
these have proven important for solving $\ell_p$ regression 
problems~\cite{DDHKM09_lp_SICOMP}. 
Our focus in this section is the \math{\ell_1} norm, for which the dual 
norm is the \math{\ell_\infty} norm, but in Section~\ref{sxn:extensionlp} we 
will return to a discussion of extensions to the \math{\ell_p} norm.

Our main algorithm for constructing an $\ell_1$ well-conditioned basis,
\textsf{FastL1Basis}, is summarized in Figure~\ref{alg:basisL1}.
This algorithm was originally presented in~\cite{SW11}, and our main 
contribution here is to improve its running time. We note that in step 3, 
we do not explicitly compute the product of $A$ and $R^{-1}$, but rather just
return $A$ and $R^{-1}$ with the promise that $AR^{-1}$ is well-conditioned. The
leading order term in our running time to compute $R^{-1}$ is $O(n d \log d)$,
while in ~\cite{SW11} it is $O(nd^2)$, or with fast matrix multiplication, 
$O(nd^{1.376})$. 

Given an $n \times d$ matrix $A$, let \math{\Pione\in\R^{r_1\times n}} be 
any projection matrix such that for any \math{x\in\R^d},
\mld{
\norm{A x}_1\le\norm{\Pione A x}_1\le \kappa \norm{A x}_1.
\label{eq:main}
}
For example, it could be constructed with either of the FCT constructions 
described in Section~\ref{sxn:main}, or with the ``slow'' Cauchy Transform 
of~\cite{SW11}, or via some other means.
After computing the matrix \math{\Pione}, the \textsf{FastL1Basis} algorithm 
of Figure~\ref{alg:basisL1} consists of the following steps:
construct \math{\Pi_1 A} and an \math{R} such that \math{\Pi_1 A=QR}, 
where \math{Q} has orthonormal columns (for example using a QR-factorization 
of \math{\Pi_1 A}); and then return \math{U=AR^{-1}=A(Q^T\Pione A)^{-1}}.

\input{alg-L1Basis}

The next theorem and its corollary are our main results for the 
\textsf{FastL1Basis} algorithm; 
and this theorem follows by combining 
our Theorem~\ref{theorem:FCT} 
with Theorems 9 and 10 of~\cite{SW11}.
The proof of this theorem may be found in Appendix~\ref{sxn:pfL1basis}.

\begin{theorem}[Fast $\ell_1$ Well-conditioned Basis]
\label{thm:basisL1}
For any \math{A\in\R^{n\times d}}, the basis \math{U=A\RR} constructed 
by \textsf{FastL1Basis}$(A)$ of Figure~\ref{alg:basisL1} 
using any \math{\Pione} satisfying \r{eq:main} is a 
\math{(d\sqrt{r_1},\kappa)}-conditioned basis for the range of \math{A}. 
\end{theorem}

\begin{corollary}
\label{thm:basisL1-FCT2}
If \math{\Pione} is obtained from 
the FCT2 construction of
Theorem~\ref{theorem:FCT},
then the resulting \math{U} 
is an \math{(\alpha,\beta)}-conditioned basis for \math{A}, 
with \math{\alpha=O(d^{3/2}\log^{1/2}d)} and \math{\beta=O(d^{2+\eta}\log d)}, 
with probability $1-\delta$. 
The time to compute the change of basis matrix \math{\RR} is 
\math{O(nd\log d+d^3\log d)}, assuming $\log n = O(d)$ and $\delta > 0$ is a fixed constant. 
\end{corollary}

\noindent\textbf{Remark.}
Our constructions that result in \math{\Pione} satisfying \r{eq:main} do 
not require that \math{A\in\R^{n\times d}}; they only require that \math{A} 
have rank \math{d}, and so can be applied to any \math{A\in\R^{n\times m}} 
having rank \math{d}. 
In this case, a small modification is needed in the construction of 
\math{U}, because \math{R\in\R^{d\times m}}, and so we need to use 
\math{R^{\dagger}} instead of \math{R^{-1}}. 
The running time will involve terms with \math{m}. 
This can be improved by processing \math{A} quickly into a smaller matrix by 
sampling columns so that the range is preserved (as in \cite{SW11}), which 
we do not discuss further.

The notion of a well-conditioned basis plays an important role in our 
subsequent algorithms.
Basically, the reason is that these algorithms compute approximate answers 
to the problems of interest (either the $\ell_1$ regression problem or the 
$\ell_1$ subspace approximation problem) by using information in that 
basis to construct a nonuniform importance sampling distribution with which
to randomly sample.
This motivates the following definition.

\begin{definition}[$\ell_1$ Leverage Scores]
\label{def:l1leverage}
Given a well-conditioned basis \math{U} for the range of \math{A}, 
let the $n$-dimensional vector $\tilde{\lambda}$, with elements defined as
$
\tilde{\lambda}_i=||U_{(i)}||_1 ,
$
be the \emph{$\ell_1$ leverage scores} of $A$.
\end{definition}

\noindent\textbf{Remark.} 
The name \emph{$\ell_1$ leverage score} is by analogy with the 
\emph{$\ell_2$ leverage scores}, which are important in random sampling 
algorithms for $\ell_2$ regression and low-rank matrix 
approximation~\cite{CUR_PNAS,Mah-mat-rev_BOOK,DMMW12_ICML}.
As with $\ell_2$ regression and low-rank matrix approximation, our result 
for \math{\ell_1} regression and $\ell_1$ subspace approximation will 
ultimately follow from the ability to approximate these scores quickly.
Note, though, that these $\ell_1$-based scores are not well-defined for a 
given matrix $A$, in the sense that the $\ell_1$ norm is not rotationally 
invariant, and thus depending on the basis that is chosen, these scores
can differ by factors that depend on low-degree polynomials in $d$.
This contrasts with $\ell_2$, since for $\ell_2$ any orthogonal matrix 
spanning a given subspace leads to the same $\ell_2$ leverage scores.  
We will tolerate this ambiguity since these $\ell_1$ leverage scores will be 
used to construct an importance sampling distribution, and thus up to 
low-degree polynomial factors in $d$, which our analysis will take into 
account, it will not matter.

\subsection{Fast $\ell_1$ Regression}
\label{sxn:resultsL1-l1reg}

Here, we consider the $\ell_1$ regression problem, also known as the
\emph{least absolute deviations} problem, the goal of which is to minimize 
the $\ell_1$ norm of the residual vector $Ax-b$.
That is, given as input a design matrix $A \in \mathbb{R}^{n \times d}$, 
with $n > d$, and a response or target vector $b \in \mathbb{R}^n$, compute
\begin{equation}
\label{eqn:original_prob}
  \min_{x \in \mathbb{R}^d} || Ax - b ||_1,
\end{equation}
and an \math{x^*} achieving this minimum.
We start with our main algorithm and theorem for this problem; and we then 
describe how a somewhat more sophisticated version of the algorithm yields 
improved running time~bounds.

\subsubsection{Main Algorithm for Fast $\ell_1$ Regression}
\label{sxn:resultsL1-l1reg-main}

Prior work has shown that there is a diagonal sampling matrix $D$ with a 
small number of nonzero entries
so that $\hat x = \argmin\nolimits_{x\in \R^d} \mynorm{D(Ax-b)}_1$ satisfies
$$\mynorm{A\hat x-b}_1\le(1+\eps)\mynorm{Ax^*-b}_1,$$ where
\math{x^*} is an optimal solution for the minimization 
in~(\ref{eqn:original_prob}); 
see \cite{DDHKM09_lp_SICOMP,SW11}.
The matrix $D$ can be found by sampling its diagonal entries
independently according to a set of probabilities \math{p_i} that are 
proportional to the \math{\ell_1} leverage scores of $A$. 
Here, we give a fast algorithm to compute estimates \math{\hat p_i} of 
these probabilities. 
This permits us to develop an improved algorithm for \math{\ell_1} 
regression and to construct efficiently a small coreset for an arbitrary 
\math{\ell_1} regression problem.

In more detail, Figure~\ref{l1reg-mainSimp} presents the 
\textsf{FastCauchyRegression} algorithm, which we summarize here.
Let \math{X=\left[\begin{matrix}A&-b\end{matrix}\right]}.
First, a matrix \math{\Pione} satisfying \r{eq:main} is used to reduce the 
dimensionality of \math{X} to \math{\Pione X} and to obtain the 
orthogonalizer \math{R^{-1}}.
Let \math{U=X\RR} be the resulting well-conditioned basis for the range of 
\math{X}. 
The probabilities we use to sample rows are essentially the row-norms of 
\math{U}. 
However, to compute \math{XR^{-1}} explicitly takes \math{O(nd^2)} time, 
which is already too costly, and so we need to estimate 
\math{\norm{U_{(i)}}_1} without explicitly computing \math{U}.
To construct these probabilities quickly, we use a second random projection 
\math{\Pi_2}---on the \emph{right}. 
This second projection allows us to estimate the norms of the rows of
\math{XR^{-1}} efficiently to within relative error (which is all we need)
using the
median of $r_2$ independent Cauchy's, each scaled by $||U_{(i)}||_1$.
(Note that this is similar to what was done in~\cite{DMMW12_ICML} to 
approximate the $\ell_2$ leverage scores of an input matrix.)
These probabilities are then used to construct a carefully down-sampled (and 
rescaled) problem, the solution to which will give us our \math{(1+\eps)} 
approximation to the original problem.

\input{alg-L1Reg}

The next theorem summarizes our main quality-of-approximation results for
the \textsf{FastCauchyRegression} algorithm of Figure~\ref{l1reg-mainSimp}.
It improves the $O(nd^{2} + \poly(d\eps^{-1}\log n))$ algorithm 
of~\cite{SW11}, which in turn improved the result 
in~\cite{DDHKM09_lp_SICOMP}.
(Technically, the running time of~\cite{SW11} is 
\math{O(nd^{\omega^+-1}+\poly(d\eps^{-1}\log n))}, 
where \math{\omega^+} is any constant larger than the exponent for matrix 
multiplication; for practical purposes, we can set \math{\omega^+=3}.)
Our improved running time comes from using the FCT and a simple row-norm 
estimator for the row-norms of a well-conditioned basis.
The proof of this theorem may be found in Appendix~\ref{sxn:FCRsimp}.

\begin{theorem}[Fast Cauchy \math{\ell_1} Regression]
\label{theorem:FCRsimp}
Given are \math{\eps\in(0,1)}, \math{\rho>0}, 
\math{A\in\R^{n\times d}} and \math{b\in\R^{n}}.
\textsf{FastCauchyRegression}\math{(A,b)} 
constructs a coreset specified by the diagonal sampling matrix \math{D} and 
a solution vector
\math{\hat x\in\R^d} that minimizes the weighted regression objective
$\norm{D(Ax-b)}_1$.  The solution \math{\hat x} satisfies, with probability
at least \math{1-\frac{1}{d^\rho}} (\math{\rho>0} is a constant),
\mand{
\norm{A\hat x-b}_1\le\left(\frac{1+\eps}{1-\eps}\right)\norm{Ax-b}_1,
\qquad\forall x\in\R^d.} 
Further, with probability
\math{1-o(1)}, the entire algorithm to construct \math{\hat x} runs in time
\mand{
O\left(nd\log n +\phi(s,d)\right)=
O\left(nd\log n+{\textstyle\frac{1}{\eps^2}\poly(d,\log\frac d\eps)}
\right),
}
where \math{\phi(s,d)} is the time to solve an \math{\ell_1}-regression problem on
\math{s} vectors in \math{d} dimensions, and if 
FCT2 is used to construct \math{\Pi_1} then
\math{s=O\left(\frac{1}{\eps^2}d^{\rho+\frac92+\eta}
\log^{\frac32}(\frac d\eps)\right)}.
\end{theorem}

\noindent\textbf{Remarks.}
Several remarks about our results for the $\ell_1$ regression problem are 
in order.
\begin{itemize}
\item
Our 
proof analyzes a more general problem $\min_{x\in {\cal C}}||Xx||_1$, 
where ${\cal C} \subseteq \R^d$ is a convex set.
In order to get the result, we need to preserve norms under sampling, which 
is what Lemma~\ref{lem:L1Chernoff} allows us to do. 
We mention that our methods extend with minor
changes to $\ell_p$ regression, for $p>1$.
This is discussed in Section~\ref{sxn:extensionlp}.
\item
A natural extension of our algorithm to matrix-valued right hand 
sides \math{b} gives a \math{(1+\eps)} approximation in a similar 
running time for the \math{\ell_1}-norm subspace approximation problem.
See Section~\ref{sxn:resultsL1-hyperplane} for details.
\item
We can further improve the efficiency of solving this simple $\ell_1$ 
regression problem, thereby replacing the \math{nd\log n} running time term 
in Theorem~\ref{theorem:FCRsimp} with \math{nd\log(d\eps^{-1}\log n)}, but 
at the expense of a slightly larger sample size \math{s}.
The improved algorithm is essentially the same as the 
\textsf{FastCauchyRegression} algorithm, except with two differences: 
$\Pi_2$ is chosen to be a matrix of i.i.d. Gaussians, for a value 
$r_2 = O(\log (d\eps^{-1}\log n))$; and, 
to accommodate this, the size of \math{s} needs to be increased.
Details are presented in Section~\ref{sxn:resultsL1-l1reg-faster}.
\end{itemize}

%
%

\subsubsection{A Faster Algorithm for $\ell_1$ Regression}
\label{sxn:resultsL1-l1reg-faster}

Here, we present an algorithm that improves the efficiency of our
$\ell_1$ regression algorithm from Section~\ref{sxn:resultsL1-l1reg-main}; 
and we state and prove an associated quality-of-approximation theorem.
See Figure~\ref{l1reg-mainOpt}, which presents the 
\textsf{OptimizedFastCauchyRegression} algorithm.
This algorithm has a somewhat larger sample size \math{s} than our previous 
algorithm, but our main theorem for this algorithm will replace the 
\math{nd\log n} running time term in Theorem~\ref{theorem:FCRsimp} with a 
\math{nd\log(d\eps^{-1}\log n)} term.

\input{alg-L1RegOpt}

The intuition behind the \textsf{OptimizedFastCauchyRegression} algorithm is 
as follows. 
The $(i,j)$-th entry $(U\Pi_2)_{ij}$ will be a $0$-mean Gaussian with 
variance $\norm{U_{(i)}}_2^2$. 
Since the row has $d$-dimensions, the $\ell_2$ norm and $\ell_1$ norm only 
differ by $\sqrt{d}$. 
Hence, at the expense of some factors of \math{d} in the sampling complexity 
\math{s}, we can use sampling probabilities based on the \math{\ell_2} norms.
The nice thing about using \math{\ell_2} norms is that we can use Gaussian
random variables for the entries of $\Pi_2$ rather than Cauchy random variables.
Given the exponential tail of a Gaussian random variable, for a $\Pi_2$ with
fewer columns we can still gurantee that no sampling probability {\it increases}
by more than a logarithmic factor. 
The main difficulty we encounter is that some sampling probabilities may 
{\it decrease} by a larger factor, even though they do not increase by much --
however, one can argue that with large enough 
probability, no row is sampled by the algorithm if its probability shrinks 
by a large factor.
Therefore, the behavior of the algorithm is as if all sampling probabilities 
change by at most a $\poly(d \eps^{-1} \ln n)$ factor, and the result will 
follow.
Here is our main theorem for the \textsf{OptimizedFastCauchyRegression} 
algorithm.
The proof of this theorem may be found in Appendix~\ref{sxn:pfL1faster}.

\begin{theorem}[Optimized Fast Cauchy \math{\ell_1} Regression]
\label{theorem:FCRsimpFaster}
Given are \math{\eps\in(0,1)}, \math{\rho>0}, 
\math{A\in\R^{n\times d}} and \math{b\in\R^{n}}.
\textsf{OptimizedFastCauchyRegression}\math{(A,b)} 
constructs a coreset specified by the diagonal sampling matrix \math{D} and 
a solution vector
\math{\hat x\in\R^d} that minimizes the weighted regression objective
$\norm{D(Ax-b)}_1$.  The solution \math{\hat x} satisfies, with probability
at least \math{1-\frac{1}{d^\rho}-\frac{1}{\log^\rho n}},
\mand{
\norm{A\hat x-b}_1\le\left(\frac{1+\eps}{1-\eps}\right)\norm{Ax-b}_1,
\qquad\forall x\in\R^d.
} 
Further, with probability
\math{1-o(1)}, the entire algorithm to construct \math{\hat x},  runs in time
\mand{
O\left(nd\log(\rho d\eps^{-1}\log n) +\phi(s,d)\right)=
O\left(nd\log(\rho d\eps^{-1}\log n) + \poly(d,\log(d\eps^{-1}\ln n))\right).}
where \math{\phi(s,d)} is the time to solve an \math{\ell_1}-regression 
problem on
\math{s} vectors in \math{d} dimensions, and if 
FCT2 is used to construct \math{\Pi_1} then
\math{s=O\left(\frac{1}{\eps^2}d^{2\rho+6+\eta}
\log^{\frac52}(\frac d\eps)\right)
}
\end{theorem}

Note that our algorithms and results also extend to multiple regression with
\math{b\in\R^{n\times k}}, a fact that will be exploited in the next section.

\subsection{$\ell_1$ norm Subspace Approximation}
\label{sxn:resultsL1-hyperplane}

Finally, we consider the \emph{$\ell_1$ norm subspace approximation problem}:
Given the \math{n} points in the \math{n \times d} matrix \math{A} and a 
parameter $k \in [d-1]$, embed these points into a subspace of dimension 
\math{k} to obtain the embedded points \math{\hat A} such that 
\math{\norm{A-\hat A}_1} is minimized.
(Note that this is the $\ell_1$ analog of the $\ell_2$ problem that is 
solved by the Singular Value Decomposition.)
When \math{k=d-1}, the subspace is a hyperplane, and the task is to find the
hyperplane passing through the origin so as to minimize the sum of 
\math{\ell_1} distances of the points to the hyperplane.
In order to solve this problem with the methods from 
Section~\ref{sxn:resultsL1-l1reg}, we take advantage of the observation made 
in~\cite{bd09} (see also Lemma 18 of~\cite{SW11}) that this problem can be 
reduced to \math{d} related \math{\ell_1} regressions of \math{A} onto each 
of its columns, a problem sometimes called \emph{multiple regression}.
Thus, in Section~\ref{sxn:resultsL1-hyperplane-gen}, we extend our 
\math{\ell_1} ``simple'' regression algorithm to an \math{\ell_1} 
``multiple'' regression algorithm; and then in
Section~\ref{sxn:resultsL1-hyperplane-sub}, we show how this can be used 
to solve the $\ell_1$ norm subspace approximation problem.

\subsubsection{Generalizing to Multiple \math{\ell_1} Regression}
\label{sxn:resultsL1-hyperplane-gen}

The multiple $\ell_1$ regression problem is similar to the simple $\ell_1$ 
regression problem, except that it involves solving for multiple right hand
sides, \emph{i.e.}, both \math{x} and \math{b} become matrices ($W$ and $B$, 
respectively). 
Specifically, let 
\math{A\in\R^{n\times d}} and
\math{B\in\R^{n\times k}}. We wish to find 
\math{W\in\R^{d\times k}} which solves 
\mand{
\min_{W}\norm{AW-B}_1.
} 
Although the optimal \math{W} can clearly be obtained by solving \math{k} 
separate simple $\ell_1$ regressions, with \math{b=B^{(j)}} for \math{j\in[k]},
one can do better.
As with simple regression, we can reformulate the more general constrained 
optimization problem:
\mand{
\min_{Z\in \cl C}\norm{XZ}_1.
}
To recover multiple $\ell_1$ regression, we set
$X  =\left[\begin{matrix}A&-B\end{matrix}\right]$ and
$Z^T=\left[\begin{matrix}W&I_k\end{matrix}\right]^T$,
in which case the constraint set is 
\math{\cl C=\left\{Z=\left[\begin{smallmatrix}W\\I_k\end{smallmatrix}\right]:
W\in\R^{d\times k}\right\}}. 

A detailed inspection of the proof of Theorem~\ref{theorem:FCRsimp} in
Section~\ref{sxn:resultsL1-l1reg}
(see Appendix~\ref{sxn:FCRsimp} for the proof)
reveals that nowhere is it necessary that \math{x} be a vector, 
\emph{i.e.}, the whole proof generalizes to a matrix \math{Z}. 
In particular, the inequalities in (\ref{eq:mainX}) continue to hold, since 
if they hold for every vector \math{x}, then it must hold for a matrix 
\math{Z} because \math{\norm{XZ}_1=\sum_{j\in[k]}\norm{XZ^{(j)}}_1}.
Similarly, 
if Lemma~\ref{lem:preservenorm} continues to hold for 
\emph{vectors} then it will imply the desired result for matrices, and
so the only change in all the algorithms and results
is that the short dimension of
\math{X} changes from \math{d+1} to \math{d+k}. 
Thus, by shrinking \math{\delta} by an additional factor of \math{k}, and
taking a union bound we get a relative error approximation for each 
individual regression.
We refer to this modified algorithm, where a matrix $B$ is input and the 
optimization problem in the last step is modified appropriately, as 
\textsf{FastCauchyRegression}\math{(A,B)}, overloading notation in the 
obvious way.
This discussion is summarized in the following theorem.

\begin{theorem}[Fast Cauchy Multiple \math{\ell_1} Regression]
\label{theorem:FCRmult}
Given \math{\eps\in(0,1)}, \math{\rho>0}, a matrix
\math{A\in\R^{n\times d}} and \math{B\in\R^{n\times k}}, 
\textsf{FastCauchyRegression}\math{(A,B)} 
constructs a coreset specified by the diagonal sampling matrix \math{D} and 
a solution
\math{{\hat W}\in\R^{d \times k}} that minimizes the weighted multiple 
regression objective
$\norm{D(AW-B)}_1$.  The solution \math{\hat W} satisfies, with probability
at least \math{1-\frac{1}{(d+k)^\rho}},
\mand{
\norm{A\hat W^{(j)}-B^{(j)}}_1\le
\left(\frac{1+\eps}{1-\eps}\right)\norm{Ax-B^{(j)}}_1,
\qquad\forall x\in\R^d\hbox{ and } \forall j\in[k].} 
Further, with probability
\math{1-o(1)}, the entire algorithm to construct \math{\hat W},  runs in time
\mand{
O\left(n(d+k)\log n +\phi(s,d,k)\right),
}
where \math{\phi(s,d,k)} is the time to solve \math{k} \math{\ell_1}-regression problem on the same
\math{s} vectors in \math{d} dimensions, and if 
FCT2 is used to construct \math{\Pi_1}, then
\math{s=O\left(\frac{1}{\eps^2}(d+k)^{\rho+\frac{11}{2}+\eta}
\log^{\frac32}(\frac{d+k}{\eps})\right)}.

\end{theorem}

\noindent\textbf{Remarks.}
Several remarks about our results for this $\ell_1$ multiple regression 
problem are in order.
\begin{itemize}
\item
First, we can save an extra factor of \math{(d+k)} in \math{s} in the above 
theorem 
if all we want is a relative error approximation to the entire 
multiple regression and we do not need relative error 
approximations to each individual regression.
\item
Second, when \math{k=O(d)} it is interesting that there is essentially no 
asymptotic overhead in solving this problem other than the increase from 
\math{\phi(s,d)} to \math{\phi(s,d,k)}; in general, by preprocessing the
matrix \math{DA}, solving \math{k} regressions on this same matrix
\math{DA} is much
quicker than solving \math{k} separate regressions.
This should be compared with \math{\ell_2} regression, where solving
\math{k} regressions with the same \math{A} takes \math{O(nd^2+ndk+kd^2)} 
(since the SVD of \math{A} needs to be done only once), versus a time of 
\math{O(nkd^2)} for \math{k} separate $\ell_2$ regressions.
\item
Third, we will use this version of \math{\ell_1} multiple regression 
problem, which is more efficient than solving \math{k} separate 
\math{\ell_1}-simple regressions, to solve the \math{\ell_1}-subspace 
approximation problem. 
See Section~\ref{sxn:resultsL1-hyperplane-sub} for details.
\end{itemize}

\subsubsection{Application to $\ell_1$ norm Subspace Approximation}
\label{sxn:resultsL1-hyperplane-sub}

Here, we will take advantage of the observation made in~\cite{bd09} that 
the $\ell_1$ norm subspace approximation problem can be reduced to 
\math{d} related \math{\ell_1} regressions of \math{A} onto each of its 
columns.
To see this, consider the following \math{\ell_1} regression problem:
\mand{\min_{w:w_j=0}\norm{Aw-A^{(j)}}_1.}
This regression problem is fitting (in the \math{\ell_1} norm) the \math{j}th
column of \math{A} onto the remaining columns. 
Let \math{w_j^*} be an optimal solution. 
Then if we replace \math{A^{(j)}} by \math{Aw_j^*}, the resulting vectors 
will all be in a \math{d-1} dimensional subspace.
Let \math{A_j} be \math{A} with \math{A^{(j)}} replaced by \math{Aw_j^*}.
The crucial observation made in~\cite{bd09} (see also Lemma 18 
of~\cite{SW11}) is that \emph{one of the \math{A_j} is optimal}---and so 
the optimal subspace can be obtained by simply doing a hyperplane fit to the 
embedded points. So,
\mand{
\min_{j\in[d]}\norm{A-A_j}_1=\min_{\rank{\hat A}=d-1}\norm{A-\hat A}_1.
} 
When viewed from this perspective, the \math{\ell_1}-norm subspace 
approximation problem makes the connection between low-rank matrix 
approximation and overconstrained $\ell_1$ regression.
(A similar approach was used in the $\ell_2$ case to obtain 
relative-error low-rank CX and CUR matrix 
decompositions~\cite{DMM08_CURtheory_JRNL,CUR_PNAS}.)
We thus need to perform \math{k} constrained regressions, which can be 
formulated into a single constrained multiple regression problem, which can 
be solved as follows:
Find the matrix \math{W} that solves:
\mand{
\min_{W\in\cl C}\norm{AW}_1,
}
where the  constraint set is
\math{\cl C=\{W\in\R^{d\times d}:W_{ii}=-1\}}.
Since the constraint set effectively places an independent constraint on each
column of \math{W},
after some elementary manipulation, it is easy to see that this regression 
is equivalent to the \math{d} individual regressions to obtain
\math{w_j^*}. Indeed, for an optimal solution
\math{W^*}, we can set \math{w_j^*=W^{*(j)}}.

Thus, using our approximation algorithm for constrained multiple $\ell_1$ 
regression that we described in Section~\ref{sxn:resultsL1-hyperplane-gen}, 
we can build an approximation algorithm for the $\ell_1$-norm subspace 
approximation problem that improves upon the previous best algorithm 
from~\cite{SW11} and~\cite{bd09}.
(The running time of the algorithm of~\cite{SW11} is
$\Omega(nd^{\omega^+} + \poly(d\eps^{-1} \log n))$, where 
$\omega \approx 2.376$ and $\beta > 0$ is any constant.) 
Our improved algorithm is basically our multiple $\ell_1$ regression 
algorithm, \textsf{FastCauchyRegression}\math{(A,B)}, invoked with \math{A} 
and \math{b=\{\}} (NULL).
The algorithm proceeds exactly as outlined in Figure~\ref{l1reg-mainSimp}, 
except for the last step, which instead uses linear programming to solve for 
\math{\hat W} that minimizes \math{\norm{AW}_1} with respect to 
\math{W\in\cl C}. 
(Note that the constraints defining \math{\cl C} are very simple affine 
equality constraints.)
Given \math{\hat W}, we define \math{\hat w_j=\hat W^{(j)}} and compute
\math{j^*=\argmin_{j\in[d]}\norm{A-\hat A_j}} where \math{\hat A_j} is 
\math{A} with the column \math{A^{(j)}} replaced by \math{A\hat w_j}. 
It is easy to now show that \math{\hat A_{j^*}} is a 
\math{(1+\eps)}-approximation to the \math{d-1} dimensional subspace 
approximation problem. 
Indeed, recall that \math{W^*} is optimal and the optimal error is
\math{\norm{AW^{*(j)}}_1} for some \math{j\in[d]}; however,
for any \math{j\in[d]}:
\mand{
\norm{AW^{*(j)}}_1
{\buildrel(a)\over\ge} 
\left(\frac{1-\eps}{1+\eps}\right)\norm{A\hat W^{(j)}}_1
{\buildrel(b)\over\ge} 
\left(\frac{1-\eps}{1+\eps}\right)\norm{A\hat W^{(j^*)}}_1
=\left(\frac{1-\eps}{1+\eps}\right)\norm{A-\hat A_{j^*}}_1,
}
where (a) is from the \math{(1+\eps)}-optimality of the constrained 
multiple regression as analyzed in Appendix~\ref{sxn:FCRsimp} and (b) is 
because \math{j^*} attained minimum error among all \math{j\in[d]}.
This discussion is summarized in the following theorem.

\begin{theorem}
\label{theorem:L1Sub}
Given \math{A\in\R^{n\times d}} (\math{n} points in \math{d} dimensions),
there is a randomized algorithm which outputs a 
$(1+\eps)$-approximation to the $\ell_1$-norm subspace approximation 
problem for these \math{n} points with 
probability at least $1-\frac{1}{d^\rho}$. 
Further, the running time, with probability
\math{1-o(1)}, is 
$$O\left(nd \log n + {\textstyle
\frac{1}{\eps^2}\poly(d,\log\frac {d}{\eps})}\right).$$
\end{theorem}

%% file: alg-L1Basis.tex
\begin{figure}[t]
\begin{center}
\framebox[6.4in]{\parbox{6.2in}{
\textsf{FastL1Basis}$(A)$:
\begin{algorithmic}[1]
\STATE
Let $\Pi_1$ be an $r_1 \times n$ matrix satisfying \r{eq:main},
\emph{e.g.}, as constructed with one of the FCTs of Section~\ref{sxn:main}.
\STATE
Compute $\Pi_1A \in \mathbb{R}^{r_1 \times d}$ and its QR-factorization:
\math{\Pi_1A=QR}, where \math{Q} is an orthogonal matrix, \emph{i.e.}, 
\math{Q^TQ=I}.
\STATE
Return $U=AR^{-1}=A(Q^T\Pione A)^{-1}$
\end{algorithmic}
}}
\end{center}
\caption{Our main algorithm for the fast construction of an $\ell_1$ 
well-conditioned basis of an $n \times d$ matrix $A$.
Note the structural similarities with the algorithm of~\cite{DMMW12_ICML} for
computing quickly approximations to the $\ell_2$ leverage scores and an 
$\ell_2$ well-conditioned basis.
}
\label{alg:basisL1}
\end{figure}

%% file: alg-L1Reg.tex
\begin{figure}
\begin{center}
\framebox[6.4in]{\parbox{6.2in}{
\textsf{FastCauchyRegression}$(A, b)$:
\begin{algorithmic}[1]
\STATE Let \math{X=\left[\begin{matrix}A&-b\end{matrix}\right]\in
\R^{n\times(d+k)}} and construct
$\Pi_1$, an $r_1 \times n$ matrix satisfying 
\r{eq:main} with \math{A} replaced by \math{X}. (If \math{b} is a vector
then \math{k=1}.) 
\STATE 
Compute $X'=\Pi_1X \in \mathbb{R}^{r_1 \times (d+k)}$ and its QR factorization,
\math{\Pione X=QR}.  
(Note that 
\math{\Pione X R^{-1}} has orthonormal columns.)
\STATE Let 
$\Pi_2\in\R^{(d+k) \times r_2}$ be
a matrix of independent Cauchys, with $r_2 = 15\log \frac{2n}{\delta}$. 
\STATE
Let \math{U=X\RR} 
and
construct \math{\Lambda=U\Pi_2\in\R^{n\times r_2}}.
\STATE 
For \math{i\in[n]}, compute
$
\lambda_i=\median_{j\in [r_2]}|\Lambda_{ij}| .
$
\STATE 
For \math{i\in[n]} and 
$s = \frac{63\kappa (d+k)\sqrt{r_1}}{\eps^2}\left((d+k)\log\frac{24\kappa(d+k)
\sqrt{r_1}}{\eps}+
\log\frac2\delta\right)$, 
compute probabilities
$$
\hat p_i=\min\left\{1,s\cdot \frac{\lambda_i}{\sum_{i\in[n]}\lambda_i}
\right\}.
$$
\STATE 
Let $D \in \R^{n \times n}$ be diagonal with independent entries:
$
D_{ii} = \begin{cases}
\frac{1}{\hat p_i}&\text{prob. }\hat p_i;\\
0&\text{prob. }1 - \hat p_i.
\end{cases}
$
\STATE Return $\hat x\in\R^d$ 
that minimizes
$\norm{DAx - Db}_1$ w.r.t. \math{x} (using linear programming). 
\end{algorithmic}
}}
\end{center}
\caption{Algorithm for solving $\ell_1$ regression.
Note that 
in Step 6, we sample rows of \math{A} and \math{b}
so that the expected number of rows sampled is at most \math{s}. 
Instead of this independent sampling (without replacement), we 
could sample exactly \math{s} rows independently \emph{with} 
replacement according to the probabilities 
\math{\hat p_i=\lambda_i/\sum_{i\in[n]}\lambda_i}, and all our results
continue to hold up to small factors.
}
\label{l1reg-mainSimp}
\end{figure}

%% file: alg-L1RegOpt.tex
\begin{figure}
\begin{center}
\framebox[6.4in]{\parbox{6.2in}{
\textsf{OptimizedFastCauchyRegression}$(A, b)$:
\begin{algorithmic}[1]
\STATE Let \math{X=\left[\begin{matrix}A&-b\end{matrix}\right]\in
\R^{n\times(d+k)}} and construct
$\Pi_1$, an $r_1 \times n$ matrix satisfying 
\r{eq:main} with \math{A} replaced by \math{X}.
\STATE 
Compute $X'=\Pi_1X \in \mathbb{R}^{r_1 \times (d+k)}$ and its QR factorization,
\math{\Pione X=QR}.  
(Note that 
\math{\Pione X R^{-1}} has orthonormal columns.)
\STATE Set the parameters
\eqan{
s& =& \frac{210\kappa^2 \sqrt{r_1(d+k)}}{\eps^2}\left((d+k)\log\frac{24
\kappa(d+k)\sqrt{r_1}}{\eps}+
\log\frac2\delta\right)\\
r_2&=&2\log\left(2sq\sqrt{r_1}\log^{2\rho+1/2} n\right)=O\left(\log\left(\rho(d+k)\epsilon^{-1}\log n\right)\right)
}
\STATE Let 
$\Pi_2\in\R^{(d+k) \times r_2}$ be
a matrix of independent standard Gaussians.
\STATE
Construct \math{\Lambda=X\RR\Pi_2\in\R^{n\times r_2}}.
\STATE 
For \math{i\in[n]}, compute
$
\hat\lambda_i=\median_{j\in [r_2]}|\Lambda_{ij}|
$
\STATE 
For \math{i\in[n]}
compute probabilities
$
\hat p_i=\min\{1,s\cdot \hat\lambda_i\}.
$
\STATE 
Let $D \in \R^{n \times n}$ be diagonal with independent entries:
$
D_{ii} = \begin{cases}
\frac{1}{\hat p_i}&\text{prob. }\hat p_i;\\
0&\text{prob. }1 - \hat p_i.
\end{cases}
$
\STATE 
Return $\hat x\in\R^d$ 
that minimizes
$\norm{DAx - Db}_1$ w.r.t. \math{x} (using linear programming). 
\end{algorithmic}
}}
\end{center}
\caption{An optimized version of our main algorithm for solving $\ell_1$ 
regression.
Note that for this algorithm $\Pi_2$ consists of independent Gaussian random 
variables and achieves the desired running time at the cost of a
larger corset size, increased by a factor of 
\math{\poly(d\epsilon^{-1}\log n)}.
}
\label{l1reg-mainOpt}
\end{figure}

%% file: resultsLp.tex
\section{Extensions to $\ell_p$, for \math{p>1}}
\label{sxn:extensionlp}

In this section, we describe extensions of our methods to $\ell_p$, for
\math{p>1}.  We will first (in Section~\ref{sxn:extensionlp-conditioning})
discuss $\ell_p$ norm conditioning and connect it to ellipsoidal rounding,
followed by a fast rounding algorithm for general centrally symmetric convex
sets (in Section~\ref{sxn:extensionlp-rounding}); and we will then (in
Section~\ref{sxn:extensionlp-basis}) show how to obtain quickly a
well-conditioned basis for the $\ell_p$ norm, for any $p \in [1,\infty)$ and (in
Section~\ref{sxn:extensionlp-regression}) show how this basis can be used for
improved $\ell_p$ regression.  These results will generalize our results for
$\ell_1$ from Sections~\ref{sxn:resultsL1-basis} and~\ref{sxn:resultsL1-l1reg},
respectively, to general $\ell_p$.

\subsection{$\ell_p$ norm Conditioning and Ellipsoidal Rounding}
\label{sxn:extensionlp-conditioning}

As with $\ell_2$ regression, $\ell_p$ regression problems are easier to
solve when they are well-conditioned.  Thus, we start with the definition of the
$\ell_p$ norm condition number $\kappa_p$ of a matrix $A$.

\begin{definition}[$\ell_p$ norm conditioning]
\label{def:lpnormcond}
Given an $n \times d$ matrix $A$, let
\begin{equation*}
  \sigma_p^{\max}(A) = \max_{\|x\|_2 \leq 1} \|A x\|_p \text{ and } \sigma_p^{\min}(A) = \min_{\|x\|_2 \geq 1} \|A x\|_p.
\end{equation*}
Then, we denote by $\kappa_p(A)$ the \emph{$\ell_p$ norm condition number of
  $A$}, defined to be:
\begin{equation*}
  \kappa_p(A) = \sigma_p^{\max}(A) / \sigma_p^{\min}(A).
\end{equation*}
For simplicity, we will use $\kappa_p$, $\sigma_p^{\min}$, and 
$\sigma_p^{\max}$ when the underlying matrix is clear.
\end{definition}

\noindent
There is a strong connection between the $\ell_p$ norm condition number and the
concept of an $(\alpha, \beta, p)$-conditioning developed by Dasgupta et al.~\cite{DDHKM09_lp_SICOMP}.

\begin{definition}[$(\alpha, \beta, p)$-conditioning (from \cite{DDHKM09_lp_SICOMP})]
\label{def:lpbasis}
Given an $n \times d$ matrix $A$ and $p\in[1,\infty]$, let $\|\cdot\|_q$ be the dual norm
of $\|\cdot\|_p$, i.e., $1/p+1/q=1$.  Then $A$ is \emph{$(\alpha,\beta,p)$-conditioned} if (1) $\|A\|_p \leq
\alpha$, and (2) for all $z \in \mathbb{R}^{d}$, $\|z\|_q \leq \beta \|A
z\|_p$. Define $\bar{\kappa}_p(A)$ as the minimum value of $\alpha \beta$ such
that $A$ is $(\alpha, \beta, p)$-conditioned. We say that $A$ is
\emph{$\ell_p$ well-conditioned} if $\bar{\kappa}_p(A) = \mathcal{O}(\poly(d))$, independent
of~$n$.
\end{definition}

\noindent
The following lemma characterizes the relationship between these two 
quantities.

\begin{lemma}
  \label{lemma:kappa_equiv}
  Given an $n \times d$ matrix $A$ and $p \in [1, \infty]$, we always have
  \begin{equation*}
    d^{-|1/2-1/p|} \kappa_p(A) \leq \bar{\kappa}_p(A) \leq d^{\max \{1/2, 1/p\}} \kappa_p(A).
  \end{equation*}
\end{lemma}

\begin{proof}
To see the connection, recall that 
\begin{align*}
\|A\|_p = \left(\sum_{j=1}^d \|A e_j\|_p^p\right)^{1/p} 
        \leq \left(\sum_{j=1}^d (\sigma^{\max}_p \| e_j \|_2)^p\right)^{1/p} 
        = d^{1/p} \sigma^{\max}_p ,
\end{align*}
and that
\begin{equation*}
  \|A x\|_p \geq \sigma^{\min}_p \|x\|_2 \geq d^{\min \{1/p-1/2, 0\}}  \sigma^{\min}_p \|x\|_q, \quad \forall x \in \mathbb{R}^n.
\end{equation*}
Thus, $A$ is $(d^{1/p} \sigma^{\max}_p, 1/( d^{\min \{1/p-1/2, 0\}}
\sigma^{\min}_p), p)$-conditioned and $\bar{\kappa}_p(A) \leq d^{\max \{1/2,
  1/p\}} \kappa_p(A)$. On the other hand, if $A$ is $(\alpha, \beta,
p)$-conditioned, we have, for all $x \in \mathbb{R}^d$,
\begin{equation*}
  \|A x\|_p \leq \|A\|_p \|x\|_q \leq d^{\max \{1/2-1/p, 0\}} \alpha \cdot \|x\|_2,
\end{equation*}
and
\begin{equation*}
  \|A x\|_p \geq \|x\|_q/\beta \geq d^{\min \{1/2-1/p, 0\}} / \beta \cdot \|x\|_2.
\end{equation*}
Thus, $\kappa_p(A) \leq d^{|1/p-1/2|} \alpha \beta$.
\end{proof}

\noindent
Although it is easier to describe sampling algorithms in terms of
$\bar{\kappa}_p$, after we show the equivalence between $\kappa_p$ and
$\bar{\kappa}_p$, it will be easier for us to discuss conditioning algorithms in
terms of $\kappa_p$, which naturally connects to ellipsoidal rounding algorithms.

\begin{definition}
  Let $\mathcal{C} \subseteq \mathbb{R}^d$ be a convex set that is
  full-dimensional, closed, bounded, and centrally symmetric with respect to the
  origin. 
An ellipsoid $\mathcal{E} = \{ x \in \mathbb{R}^d \,|\, \|R x\|_2 \leq 1 \}$ 
is a \emph{$\kappa$-rounding} of $\mathcal{C}$ if it satisfies 
$ \mathcal{E}/\kappa \subseteq \mathcal{C} \subseteq \mathcal{E} $, for some 
$\kappa \geq 1$, where $\mathcal{E}/\kappa$ means shrinking $\mathcal{E}$ by 
a factor of $1/\kappa$. 
\end{definition}

\noindent
To see the connection between rounding and conditioning, let $\mathcal{C} = \{ x \in \mathbb{R}^d \,|\, \|A x\|_p
\leq 1\} $ and assume that we have a $\kappa$-rounding of $\mathcal{C}$:
$\mathcal{E} = \{ x \,|\, \|R x\|_2 \leq 1 \}$. This implies
\begin{equation*}
  \|R x\|_2 \leq \|A x\|_p \leq \kappa \|R x\|_2, \quad \forall x \in \mathbb{R}^d.
\end{equation*}
If we let $y = R x$, then we get
\begin{equation*}
  \|y\|_2 \leq \|A \RR y\|_p \leq \kappa \|y\|_2, \quad \forall y \in \mathbb{R}^d.
\end{equation*}
Therefore, we have $\kappa_p(A\RR) \leq \kappa$. So a $\kappa$-rounding of
$\mathcal{C}$ leads to a $\kappa$-conditioning of $A$.

\subsection{Fast Ellipsoidal Rounding}
\label{sxn:extensionlp-rounding}

Here, we provide a deterministic algorithm to compute a $2d$-rounding of a centrally
symmetric convex set in $\mathbb{R}^d$ that is described by a separation oracle.
Recall the well-known result due to John~\cite{john1948extremum} that for a
centrally symmetric convex set $\mathcal{C}$ there exists a $d^{1/2}$-rounding
and that such rounding is given by the L{\"o}wner-John (LJ) ellipsoid of
$\mathcal{C}$, \emph{i.e.}, the minimal-volume ellipsoid containing
$\mathcal{C}$.  However, finding this $d^{1/2}$-rounding is a hard problem.  To
state algorithmic results, suppose that $\mathcal{C}$ is described by a
separation oracle and that we are provided an ellipsoid $\mathcal{E}_0$ that
gives an $L$-rounding for some $L \geq 1$.  In this case, the best known
algorithmic result of which we are aware is that we can find a
$(d(d+1))^{1/2}$-rounding in polynomial time, in particular, in $O(d^4 \log L)$
calls to the oracle; see Lov{\'a}sz~\cite[Theorem 2.4.1]{lovasz}. This result
was used by Clarkson~\cite{Cla05} and by
Dasgupta \emph{et al.}~\cite{DDHKM09_lp_SICOMP}. Here, we follow the same construction, but we
show that it is much faster to find a (slightly worse) $2d$-rounding. The proof
of this theorem may be found in Appendix~\ref{sxn:app-rounding1pf}.

\begin{theorem}[Fast Ellipsoidal Rounding]
  \label{thm:rounding}
  Given a centrally symmetric convex set $\mathcal{C} \subseteq \mathbb{R}^d$
  centered at the origin and described by a separation oracle, and an ellipsoid
  $\mathcal{E}_0$ centered at the origin such that $\mathcal{E}_0/L
  \subseteq \mathcal{C} \subseteq \mathcal{E}_0$ for some $L \geq 1$, it takes
  at most $3.15 d^2 \log L$ calls to the oracle and additional $O(d^4 \log L)$ 
  time to find a $2d$-rounding of $\mathcal{C}$.
\end{theorem}

\noindent
Applying Theorem \ref{thm:rounding} to the convex set $\mathcal{C} = \{ x \,|\,
\|A x\|_p \leq 1\}$, with the separation oracle described via a subgradient of
$\|A x\|_p$ and the initial rounding provided by the ``$R$'' matrix from the QR
decomposition of $A$, we improve the running time of the algorithm used by
Clarkson~\cite{Cla05} and by Dasgupta \emph{et al.}~\cite{DDHKM09_lp_SICOMP} from
$\mathcal{O}(n d^5 \log n)$ to $\mathcal{O}(n d^3 \log n)$ while maintaining an
$\mathcal{O}(d)$-conditioning. The proof of this theorem may be found in
Appendix~\ref{sxn:app-rounding2pf}.

\begin{theorem}
  \label{thm:lp_cond_2d}
  Given an $n \times d$ matrix $A$ with full column rank, it takes at most
  $\mathcal{O}(n d^3 \log n)$ time to find a matrix $R \in \mathbb{R}^{d \times
    d}$ such that $\kappa_p(A R^{-1}) \leq 2 d$.
\end{theorem}

\subsection{Fast Construction of an $\ell_p$ Well-conditioned Basis}
\label{sxn:extensionlp-basis}

Here, we consider the construction of a basis that is well-conditioned for
$\ell_p$.
To obtain results for general $\ell_p$ that are analogous to those we obtained
for $\ell_1$, we will extend the FCT2 construction from
Section~\ref{sec:FCT2}, combined with Theorem \ref{thm:rounding}. 

Our main algorithm for constructing a $p$-well-conditioned basis, the
\textsf{FastLpBasis} algorithm, is summarized in Figure~\ref{alg:basisLp}.  The
algorithm first applies block-wise embeddings in the $\ell_2$ norm, similar to
the construction of FCT2; it then uses the algorithm of Theorem
\ref{thm:rounding} to compute a $(2 d)$-rounding of a special convex set
$\tilde{\mathcal{C}}$ and obtain the matrix $R$. It is thus a generalization of
our \textsf{FastL1Basis} algorithm of Section~\ref{sxn:resultsL1-basis}, and it
follows the same high-level structure laid out by the algorithm
of~\cite{DMMW12_ICML} for computing approximations to the $\ell_2$ leverage
scores and an $\ell_2$ well-conditioned basis.

\input{alg-LpBasis}

The next theorem is our main result for the \textsf{FastLpBasis} algorithm.  It
improves the running time of the algorithm of
Theorem~\ref{thm:lp_cond_2d}, at the cost of slightly worse conditioning
quality.  However, these worse factors will only contribute to a low-order
additive $\poly(d)$ term in the running time of our $\ell_p$ regression
application in Section~\ref{sxn:extensionlp-regression}.  The proof of this
theorem may be found in Appendix~\ref{sxn:app-onepasspf}.

\begin{theorem}[Fast $\ell_p$ Well-conditioned Basis]
\label{thm:basisLp}
For any \math{A\in\R^{n\times d}} with full column rank, the basis $A R^{-1}$
constructed by \textsf{FastLpBasis}$(A)$ (Figure~\ref{alg:basisLp}), with
probability at least \math{1-1/n}, is $\ell_p$ well-conditioned with $\kappa_p(A
R^{-1}) = \mathcal{O}(d t^{|1/p-1/2|})$.  The time to compute $R$ is $O(nd \log
n)$.
\end{theorem}

\noindent
When \math{d > \log n}, $\kappa_p(A R^{-1}) = \mathcal{O}(d^{1 + 3 \cdot
  |1/p-1/2|})$ and hence $\bar{\kappa}_p(A R^{-1}) = \mathcal{O}(d^{1 + 3 \cdot
  |1/p-1/2| + \max\{1/p, 1/2\}})$ by Lemma \ref{lemma:kappa_equiv}. Note that,
even for the case when $p = 1$, we have $\bar{\kappa}_p(A R^{-1}) =
\mathcal{O}(d^{7/2})$, which is slightly better than FCT2 (see
Corollary~\ref{thm:basisL1-FCT2}). However, we have to solve a rounding problem
of size $ns/t \times d$ in the step 2 of \textsf{FastLpBasis}, which requires
storage and work depending on $n$.

\subsection{Fast $\ell_p$ Regression}
\label{sxn:extensionlp-regression}

Here, we show that the overconstrained $\ell_p$ regression problem can be 
solved with a generalization of the algorithms of 
Section~\ref{sxn:resultsL1-l1reg} for solving $\ell_1$ regression; we will 
call this generalization the \textsf{FastLpRegression} algorithm.
In particular, as with the algorithm for $\ell_1$ regression, this
\textsf{FastLpRegression} algorithm for the 
$\ell_p$ regression problem uses an 
$\ell_p$ well-conditioned basis and samples rows of $A$ with probabilities 
proportional to the \math{\ell_p} norms of the 
rows of the corresponding well-conditioned 
basis (which are the $\ell_p$ analogs of the $\ell_2$ leverage scores). 
As with the \textsf{FastCauchyRegression}
, this entails using---for speed---a 
second random projection $\Pi_2$ applied to $A\RR$---on the right---to 
estimate the row norms.
This allows fast estimation of the $\ell_2$ norms of the rows of $A\RR$, 
which provides an estimate of the $\ell_p$ norms of those rows, up to a 
factor of $d^{|1/2 - 1/p|}$.
We use these norm estimates, \emph{e.g.}, as in the above algorithms or in 
the sampling algorithm of~\cite{DDHKM09_lp_SICOMP}.
As discussed for the running time bound of~\cite{DDHKM09_lp_SICOMP},
Theorem 7, this algorithm samples a number of rows proportional to 
$\bar{\kappa}_p^p(A \RR) d$. This factor, together with a sample complexity
increase of $(d^{|1/2-1/p|})^{p} = d^{|p/2 - 1|}$ needed to compensate for error
due to using $\Pi_2$, gives a sample complexity increase for the
\textsf{FastLpRegression} algorithm
while the leading term in the complexity (for $n\gg d$) is reduced from
$O(nd^5\log n)$ to $O(nd\log n)$.
We modify Theorem 7 of \cite{DDHKM09_lp_SICOMP} 
to obtain the following theorem.

\begin{theorem}[Fast \math{\ell_p} Regression]
\label{thm:lprunning}
Given \math{\eps\in(0,1)}, \math{A\in\R^{n\times d}}, and 
\math{b\in\R^{n}}, 
there is a random sampling algorithm 
(the \textsf{FastLpRegression} algorithm described above) 
for $\ell_p$ regression that constructs a coreset specified by a diagonal 
sampling matrix \math{D}, and a solution vector \math{\hat x\in\R^d} that 
minimizes the weighted regression objective $\norm{D(Ax-b)}_p$.  
The solution \math{\hat x} satisfies, with probability at least \math{1/2}, 
the relative error bound that
\math{\norm{A\hat x-b}_p\le (1+\eps)\norm{Ax-b}_p}
for all $x\in\R^d$.
Further, with probability \math{1-o(1)}, the entire algorithm to construct 
\math{\hat x}  runs in time
\mand{
O\left(nd\log n +\phi_p(s,d)\right)=
O\left(nd\log n+{\textstyle\frac{1}{\eps^2}\poly(d,\log\frac d\eps)}
\right),
}
where \math{s=O(\eps^{-2}d^k\log(1/\eps))}
with  $k = p + 1 + 4 |p/2-1|+ \max\{p/2, 1\}$, and
\math{\phi_p(s,d)} is the time to solve an \math{\ell_p} regression 
problem on \math{s} vectors in \math{d} dimensions.
\end{theorem}

%% file: alg-LpBasis.tex
\begin{figure}[t]
\begin{center}
\framebox[6.4in]{\parbox{6.2in}{
\textsf{FastLpBasis}$(A)$:
\begin{algorithmic}[1]
  \STATE Let $s = \Theta(d + \log n)$, $t = \Theta(s d^2)$, and $G$ be an $s
  \times t$ Fast Johnson-Lindenstrauss matrix, the same as the matrix $G$ in the
  FCT2 construction.
  
  \STATE Partition $A$ along its rows into sub-matrices of size $t \times d$,
  denoted by $A_1, \ldots, A_N$, compute $\tilde{A}_i = G A_i$ for $i=1,\ldots,N$,
  and define
  \begin{equation*}
     \tilde{\mathcal{C}} = \left\{ x \,\left|\, \left(\sum_{i=1}^N \|\tilde{A}_i x\|_2^p\right)^{1/p} \leq
       1 \right. \right\}, \text{ and } \tilde{A} = {\tiny \begin{pmatrix} \tilde{A}_1 \\
         \vdots \\ \tilde{A}_N
       \end{pmatrix}}.
  \end{equation*}

  \STATE Apply the algorithm of Theorem~\ref{thm:rounding} to obtain a $(2
  d)$-rounding of $\tilde{\mathcal{C}}$: $\mathcal{E} = \{ x \,|\, \|R x\|_2
  \leq 1 \}$.

  \STATE Output $A R^{-1}$.
\end{algorithmic}
}}
\end{center}
\caption{Our main algorithm for the fast construction of an $\ell_p$
  well-conditioned basis of an $n \times d$ matrix $A$.  Note the structural
  similarities with our \textsf{FastL1Basis} algorithm of
  Figure~\ref{alg:basisL1} for computing quickly an $\ell_1$ well-conditioned
  basis.  }
\label{alg:basisLp}
\end{figure}

%% file: numerical-long.tex
\section{Numerical Implementation and Empirical Evaluation}
\label{sec:numer}

In this section, we describe the results of our empirical evaluation.
We have implemented and evaluated 
the Fast Cauchy Transforms (both FCT1 and FCT2) as well as
the Cauchy transform (CT) of~\cite{SW11}. 
For completeness, we have also compared our method against two $\ell_2$-based 
transforms: the Gaussian Transform (GT) and a version of the FJLT.
Ideally, the evaluation would be based on the evaluating the distortion of 
the embedding, \emph{i.e.}, evaluating the smallest $\kappa$ such that
\begin{equation*}
  \|A x\|_1 \leq \| \Pi A x \|_1 \leq \kappa \|A x\|_1, \quad \forall x \in \mathbb{R}^d, 
\end{equation*}
where $\Pi \in \mathbb{R}^{r \times n}$ is one of the Cauchy transforms. 
Due to the non-convexity, there seems not to be a way to compute, tractably 
and accurately, the value of this $\kappa$. 
Instead, we evaluate both $\ell_1$-based transforms (CT, FCT1, and FCT2) and 
$\ell_2$-based transforms (GT and FJLT) based on how they perform in 
computing well-conditioned bases and approximating $\ell_1$ 
regression~problems. 

\subsection{Evaluating the Quality of $\ell_1$ Well-conditioned Bases}

We first describe our methodology.
Given a ``tall and skinny'' matrix $A \in \mathbb{R}^{n \times d}$ with full 
column rank, as in Section~\ref{sxn:resultsL1-basis}, we compute 
well-conditioned bases of $A$:
$U = A \RR = A (Q^T \Pi A)^{-1}$, where $\Pi$ is one of those transforms, and where 
$Q$ and $R$ are from the QR decomposition of $\Pi A$. 
Our empirical evaluation is based on the metric 
$\bar{\kappa}_1(U)$.
Note that $\bar{\kappa}_1$ is scale-invariant: if $U$ is 
$(\alpha, \beta)$-conditioned with $\bar{\kappa}_1(U) = \alpha \beta$, $\gamma U$ is 
$( \alpha \gamma, \beta/\gamma)$-conditioned, and hence
$\bar{\kappa}_1(\gamma U) =\alpha \gamma \beta / \gamma =\alpha \beta = \bar{\kappa}_1(U)$. 
This saves us from determining the scaling constants when implementing CT, 
FCT1, and FCT2. 
While computing $\alpha = \|U\|_1$ is trivial, computing 
$\beta = 1 / (\min_{\|z\|_\infty = 1} \|U z\|_1)$ is not as easy: it 
requires solving $d$ linear programs:
\begin{equation*}
  \beta = \frac{1}{\underset{j=1,\ldots,d}{\min} \underset{\tiny \begin{array}{c}
        \|z\|_\infty \leq 1 \\
        z_j = 1
      \end{array}}{\min} \|U z\|_1}.
\end{equation*}
Note that this essentially limits the size of the test problems in our 
empirical evaluation: although we have applied our algorithms to much 
larger problems, we must solve these linear programs if we want to provide 
a meaningful comparison by comparing our fast $\ell_1$-based algorithms 
with an ``exact'' answer. 
Another factor limiting the size of our test problems is more subtle and 
is a motivation for our comparison with $\ell_2$-based algorithms.
Consider a basis induced by the Gaussian transform: $U = A(Q^T G A)^{-1}$, 
where $G \in \mathbb{R}^{O(d) \times n}$ is a matrix whose entries 
are i.i.d.\ Gaussian.
We know that $\kappa_2(U) = O(1)$ with high probability. 
In such case, we have
\begin{equation*}
  \|U\|_1 = \sum_{j=1}^d \|U e_j\|_1 \leq \sum_{j=1}^d n^{1/2} \|U e_j\|_2 \leq n^{1/2} d \cdot \sigma^{\max}_2(U),
\end{equation*}
and
\begin{equation*}
  \|U z\|_1 \geq \|U z\|_2 \geq \sigma^{\min}_2(U) \|z\|_2 \geq \sigma^{\min}_2(U) \|z\|_\infty.
\end{equation*}
Hence $\bar{\kappa}_1(U) \leq n^{1/2} d \cdot \sigma^{\max}_2(U)/\sigma^{\min}_2(U) =
O(n^{1/2} d)$. 
Similar results apply to the FJLTs that work on an entire subspace of 
vectors, \emph{e.g.}, 
the Subsampled Randomized Hadamard Transform (SRHT)~\cite{tropp2011improved}. 
In our empirical evaluation, we use SRHT as our 
implementation of FJLT, but we note that similar running times hold for
other variants of the FJLT~\cite{AMT10}.
Table \ref{tab:cond} lists the running time and worst-case performance of 
each transform on $\ell_1$ conditioning, clearly showing the 
cost-performance trade-offs. 
For example, comparing the condition number of GT or FJLT, $O(n^{1/2} d)$, 
with the condition number of CT, $O(d^{5/2} \log^{3/2} d)$, we will need 
$n > O(d^3 \log^3 d)$ to see the advantage of CT over $\ell_2$-based 
algorithms (\emph{e.g.}, $n$ should be at least at the scale of $10^8$ when $d$ is 
$100$). 
To observe the advantage of FCT1 and FCT2 over $\ell_2$-based transforms, $n$ 
should be relatively even larger. 

\begin{table}
  \centering
  \begin{tabular}{c|c|c}
    & time & $\bar{\kappa}_1$ \\
    \hline
    CT & $O(n d^2 \log d)$ & $O(d^{5/2} \log^{3/2} d)$ \\
    FCT1 & $O(n d \log d)$ & $O(d^{11/2} \log^{9/2} d)$ \\
    FCT2 & $O(n d \log d)$ & $O(d^{7/2+\eta} \log^{3/2} d)$ \\
    \hline
    GT & $O(n d^2)$ & $O(n^{1/2} d)$ \\
    FJLT & $O(n d \log n)$ & $O(n^{1/2} d)$ \\
  \end{tabular}
  \caption{Summary of time complexity and $\ell_1$ conditioning performance 
for $\ell_1$-based and $\ell_2$-based transforms used in our empirical 
evaluation.}
  \label{tab:cond}
\end{table}

Motivated by these observations, we create two sets of test problems. 
The first set contains matrices of size 
$2^{18} \times 4$ and the second set contains 
matrices of size $2^{16} \times 16$. 
We choose the number of rows to be powers of $2$ to implement FCT2 and FJLT
in a straightforward way. 
Based on our theoretical analysis, we expect $\ell_1$-based algorithms should 
work better on the first test set than $\ell_2$-based algorithms, at least on 
some worst-case test problems; and that this advantage should disappear on 
the second test set. 
For each of these two sizes, we generate four test matrices:
$A_1$ is randomly generated ill-conditioned matrix with slightly 
heterogeneous leverage scores; 
$A_2$ is randomly generated ill-conditioned matrix with strongly 
heterogeneous leverage scores; and
$A_3$ and $A_4$ are two ``real'' matrices chosen to illustrate the 
performance of our algorithms on real-world data.
In more detail, the test matrices are as follows:
\begin{itemize}
\item $A_1 = D_1 G_1 D_2 G_2$, where $D_1 \in \mathbb{R}^{n \times n}$ is a
  diagonal matrix whose diagonals are linearly spaced between $1$ and $10^4$,
  $G_1 \in \mathbb{R}^{n \times d}$ is a Gaussian matrix, $D_2 \in \mathbb{R}^{d
    \times d}$ is a diagonal matrix whose diagonals are linearly spaced between
  $1$ and $10^4$, and $G_2 \in \mathbb{R}^{d \times d}$ is a Gaussian matrix.
  $A_1$ is chosen in this way so that it is ill-conditioned (due to the choice of $D_2$) and 
  its bottom rows tend to have high leverage scores (due to the choice of $D_1$).
\item $A_2 =
  \left(
    \begin{smallmatrix}
      1 & & & & & \\
      & 1 & & & & \\
      & & \ddots & & & \\
      & & & 1 & \cdots & 1 
    \end{smallmatrix} \right)^T G$, where $G \in \mathbb{R}^{d \times d}$ is a
  Gaussian matrix. The first $d-1$ rows tend to have very high leverage scores 
  because missing any of them would lead to rank deficiency, while the rest $n-d+1$ 
  rows are the same from each other and hence they tend to have very low leverage scores.
  $A_2$ is also ill-conditioned because we have $A_2^T A_2 = G^T \left(
    \begin{smallmatrix}
      1 & & & \\
      & \ddots & & \\
      & & 1 & \\
      & & & w 
    \end{smallmatrix} \right)G$, where $w = (n - d + 1)^2$ is very large.
\item $A_3$, the leading submatrix of the SNP matrix used by Paschou et
  al.~\cite{Paschou10b}. The SNP matrix is of size $492516 \times
  2250$, from the Human Genome Diversity Panel 
  and the HapMap Phase 3
  dataset. See \cite{Paschou10b} for more descriptions of the data.
\item $A_4$, the leading submatrix of the TinyImages matrix created by 
  Torralba et al.~\cite{torralba200880}. The original images are in RGB 
  format. We convert them to grayscale intensity images, resulting a matrix 
  of size $8e7 \times 1024$.
\end{itemize}

To implement FCT1 and FCT2 for our empirical evaluations, we have to fix 
several parameters in Theorems \ref{theorem:FCT} and \ref{theorem:FCT2}, 
finding a compromise between theory and practice. 
We choose $r_1 = \lceil 2 d \log d \rceil$ except $r_1 = 2 d$ for GT. 
We choose $s = \lceil 2 d \log d \rceil$ and $t = 2 d^2$ for FCT1, and 
$s = 2^{\lceil 2 \log_2 (2 d \log d) \rceil}$ for FCT2. 
Although those settings don't follow Theorems \ref{theorem:FCT} and
\ref{theorem:FCT2} very closely, they seem to be good for practical use. 
Since all the transforms are randomized algorithms that may fail with 
certain probabilities, for each test matrix and each transform, we take 
$50$ independent runs and show the first and the third quartiles of 
$\bar{\kappa}_1$ in Tables \ref{tab:precond_small} and \ref{tab:precond_medium}.

\begin{table}
  \centering
  \begin{tabular}{c|cccc}
    & $A_1$ & $A_2$ & $A_3$ & $A_4$ \\
    \hline
    $\bar{\kappa}_1(A_i)$ & 1.93e+04 & 7.67e+05 & 8.58 & 112\\
    \hline
    \texttt{CT} & [10.8, 39.1] & [10.4, 41.7] & [10.2, 33] & [8.89, 42.8]\\
    \texttt{FCT1} & [9.36, 21.2] & [15.4, 58.6] & [10.9, 38.9] & [11.3, 40.8]\\
    \texttt{FCT2} & [12.3, 32.1] & [17.3, 76.1] & [10.9, 43] & [11.3, 42.1]\\
    \hline
    \texttt{GT} & [6.1, 8.81] & [855, 1.47e+03] & [5.89, 8.29] & [6.9, 9.17]\\
    \texttt{FJLT} & [5.45, 6.29] & [658, 989] & [5.52, 6.62] & [6.18, 7.53]
  \end{tabular}
\caption{$\ell_1$-norm conditioning, $\bar{\kappa}_1(U)$, on 
matrices of size $2^{18} \times 4$. 
We compute the first and the third quartiles of the $\ell_1$-norm 
conditioning number in $50$ independent runs for each matrix and each 
algorithm. 
The size is chosen to demonstrate the difference between $\ell_1$-based and 
$\ell_2$-based conditioning algorithms and the superiority of the 
$\ell_1$-based algorithms in the asymptotic regime. 
GT and FJLT don't work well on $A_2$, resulting condition numbers close to 
the worst-case bound of $n^{1/2} d = 2048$. 
CT, FCT1, and FCT2 perform consistently across all matrices.}
  \label{tab:precond_small}
\end{table}

\begin{table}
  \centering
  \begin{tabular}{c|cccc}
    & $A_1$ & $A_2$ & $A_3$ & $A_4$ \\
    \hline
    $\bar{\kappa}_1(A_i)$ & 4.21e+05 & 2.39e+06 & 36.5 & 484\\
    \hline
    \texttt{CT} & [90.2, 423] & [386, 1.44e+03] & [110, 633] & [150, 1e+03]\\
    \texttt{FCT1} & [113, 473] & [198, 1.1e+03] & [114, 765] & [127, 684]\\
    \texttt{FCT2} & [134, 585] & [237, 866] & [106, 429] & [104, 589]\\
    \hline
    \texttt{GT} & [27.4, 31] & [678, 959] & [28.8, 32.3] & [29.4, 33.5]\\
    \texttt{FJLT} & [19.9, 21.2] & [403, 481] & [21.4, 23.1] & [21.8, 23.2]
  \end{tabular}
\caption{$\ell_1$-norm conditioning, $\bar{\kappa}_1(U)$, on 
matrices of size $2^{16} \times 16$. 
We compute the first and the third quartiles of the $\ell_1$-norm 
conditioning number in $50$ independent runs for each matrix and each 
algorithm. 
The size is chosen to demonstrate that $\ell_2$-based conditioning algorithms 
can be as good as or even better than $\ell_1$-based conditioning algorithms. 
GT and FJLT still don't work well on $A_2$, but they become comparable to 
$\ell_1$-based algorithms. 
Although still performing consistently across all matrices, $\ell_1$-based 
algorithms perform much worse than in the first test set due to the 
increase of $d$ and decrease of $n$.}
  \label{tab:precond_medium}
\end{table}

The empirical results, described in detail in Tables~\ref{tab:precond_small} 
and \ref{tab:precond_medium}, conform with our expectations.
The specifically designed $\ell_1$-based algorithms perform consistently 
across all test matrices, while the performance of $\ell_2$-based algorithms 
is quite problem-dependent. 
Interestingly, though, the $\ell_2$-based methods often perform reasonably 
well: at root, the reason is that for many input the $\ell_2$ leverage 
scores are not too much different than the $\ell_1$ leverage scores.
That being said, the matrix $A_2$ clearly indicates that $\ell_2$-based 
methods can fail for ``worst-case'' input; while the $\ell_1$-based methods 
perform well for this input.

On the first test set,
$\ell_1$-based algorithms are comparable to $\ell_2$-based algorithms on 
$A_1$, $A_3$, and $A_4$ but much better on $A_2$. 
The differences among $\ell_1$-based algorithms are small. 
In terms of conditioning quality, CT leads FCT1 and FCT2 by a small amount
on average; but when we take running times into account, FCT1 and FCT2 are 
clearly more favorable choices in this asymptotic regime. 
On the second test set, $\ell_1$-based algorithms become worse than 
$\ell_2$-based on $A_1$, $A_3$, and $A_4$ due to the increase of $d$ and the 
decrease of $n$. 
All the algorithms perform similarly on $A_2$; but $\ell_1$-based 
algorithms, involving Cauchy random variables, have larger variance than 
$\ell_2$-based algorithms.


\begin{figure}
  \centering
  \begin{tabular}{cc}
    \includegraphics[width=0.45\textwidth, clip, trim=1.5cm 7cm 2cm 7cm]{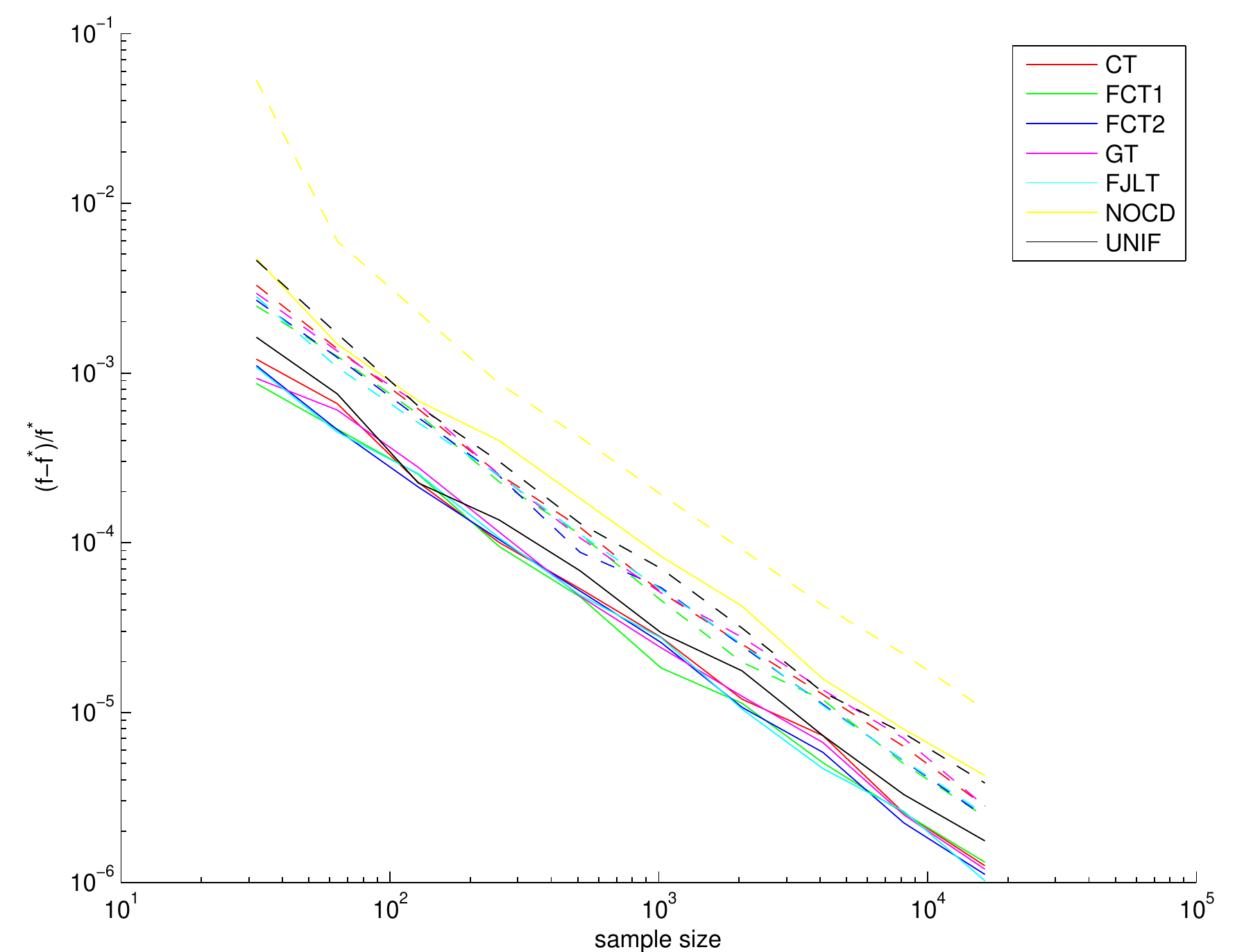} &
  \includegraphics[width=0.45\textwidth, clip, trim=1.5cm 7cm 2cm 7cm]{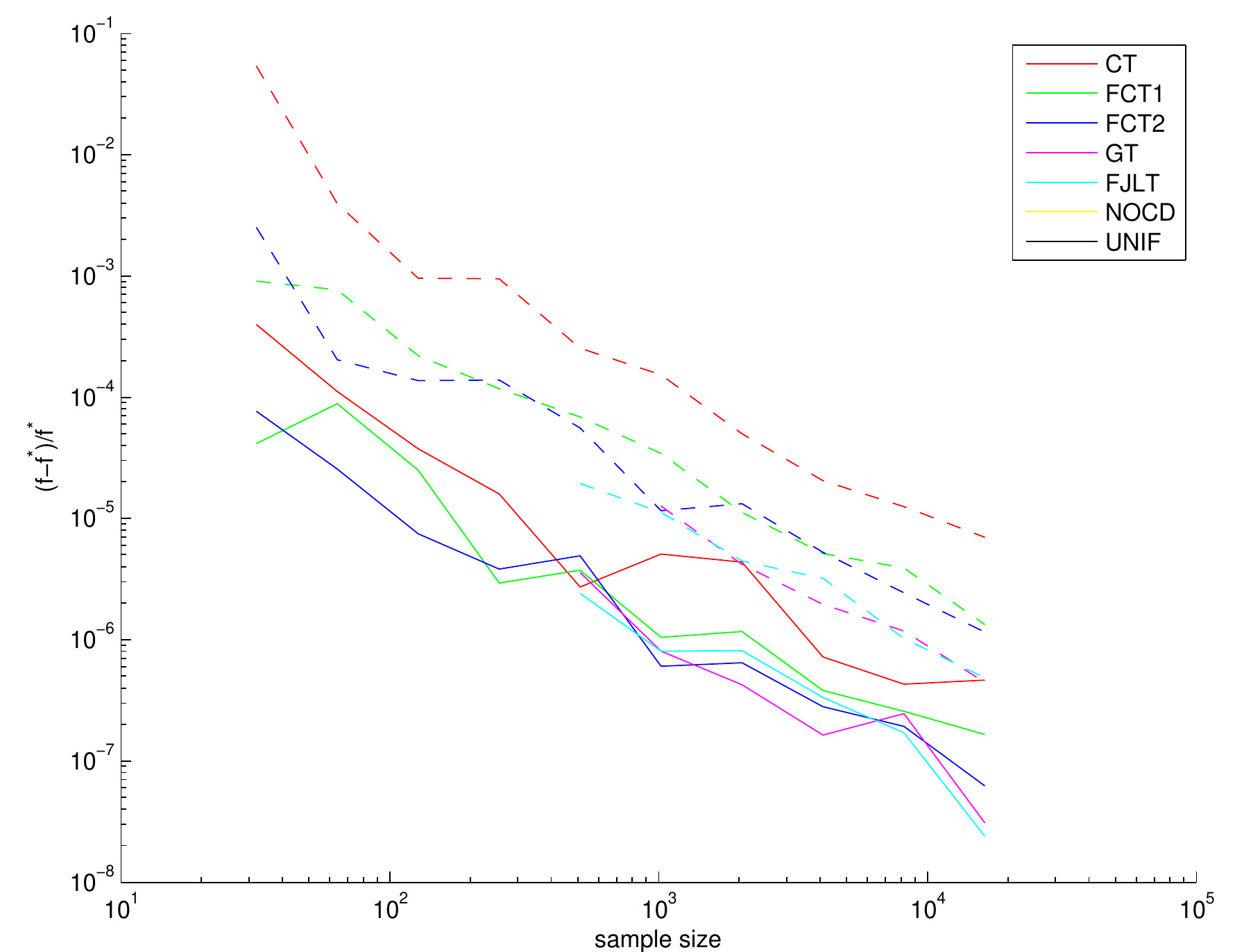} \\
  $A_1$ & $A_2$ \\[1em]
\includegraphics[width=0.45\textwidth, clip, trim=1.5cm 7cm 2cm 7cm]{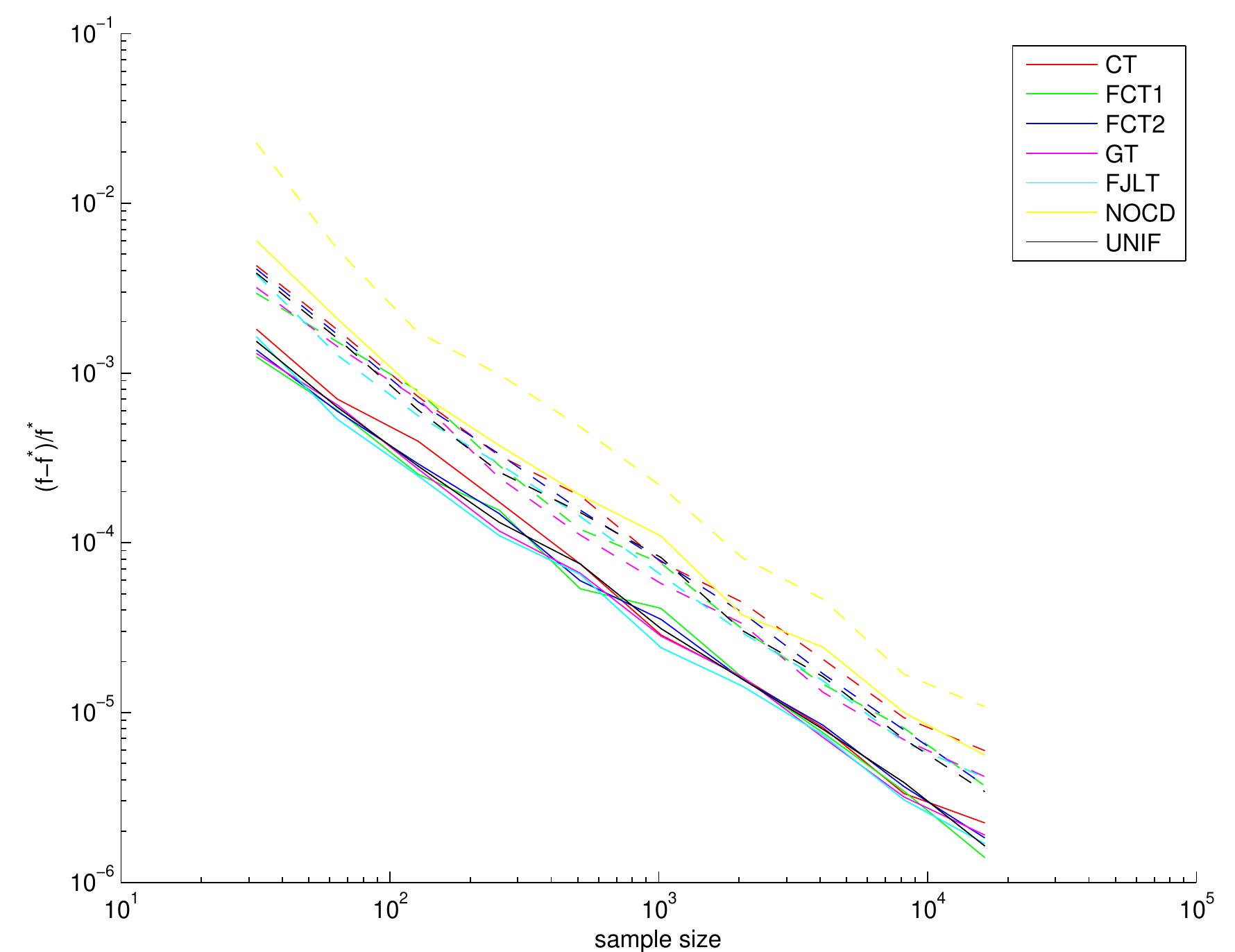} & 
  \includegraphics[width=0.45\textwidth, clip, trim=1.5cm 7cm 2cm 7cm]{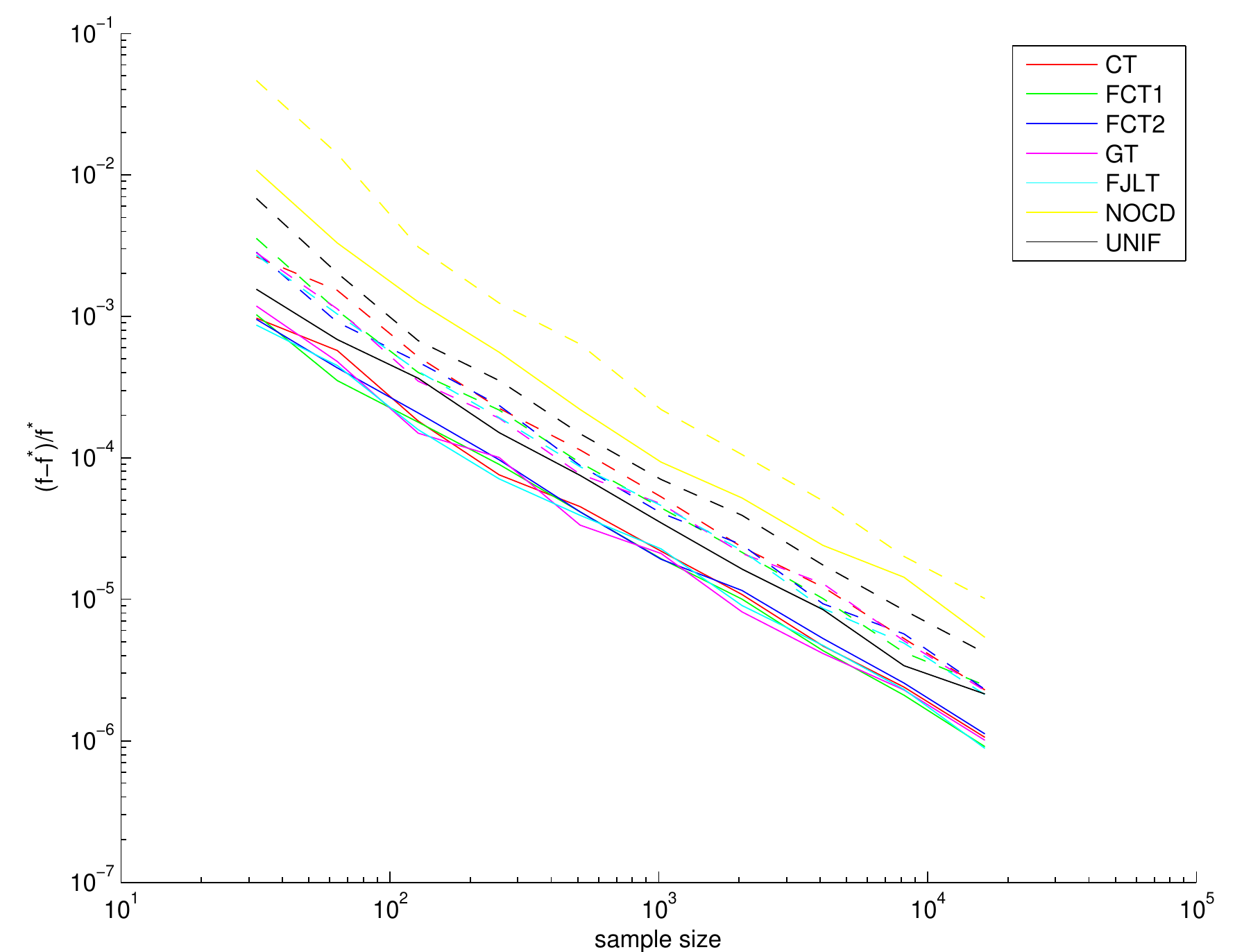} \\
  $A_3$ & $A_4$ 
  \end{tabular}
  \caption{The first and the third quartiles of relative errors in objective
    value. The problem size is $2^{18} \times 7$. The first quartiles are drawn
    in solid lines while the third drawn in dashed lines. If the subsampled
    problem is rank-deficient, we set corresponding relative error to
    $\infty$. If a quartile is larger than $100$, we remove it from the
    plot. There are few differences among those algorithms on $A_1$, $A_3$, and
    $A_4$. UNIF and NOCD are clearly inferior to algorithms that explore both
    conditioning and leverage score-based sampling. UNIF and NOCD also failed on
    $A_2$ completely. GT and FJLT failed on $A_2$ when the sample size is
    smaller than $512$. CT works slightly worse than FCT1 and FCT2 on these
    tests. One interesting fact from the result is that we see $\eps \sim
    1/s$ instead of $1/s^{1/2}$. }
  \label{fig:sample}
\end{figure}

\subsection{Application to $\ell_1$ Regression}

Next, we embed these transforms into fast approximation of 
$\ell_1$ regression problems to see how they affect the accuracy of 
approximation. 
We implement the \textsf{FastCauchyRegression} algorithm of
Section~\ref{sxn:resultsL1-l1reg}, except that 
we compute the row norms of $U$ exactly instead of estimating them. 
Although this takes $O(n d^2)$ time, it is free from errors introduced by 
estimating the row norms of $U$, and thus it permits a more direct 
evaluation of the regression algorithm. 
Unpublished results indicate that using approximations to the $\ell_1$ 
leverage scores, as is done at the beginning of the 
\textsf{FastCauchyRegression} algorithm, leads to very similar 
quality-of-approximation results.

We generate a matrix $A$ of size $2^{18} \times 7$ and generate the
right-hand sides $b = A x_{\text{exact}} + \eps$, where
$x_{\text{exact}}$ is a Gaussian vector, and $\eps$ is a random vector
whose entries are independently sampled from the Laplace distribution and scaled
such that $\|\eps\|_2 / \| A x_{\text{exact}} \|_2 = 0.1$. Then, for each
row $i$, with probability $0.001$ we replace $b_i$ by $100 \|\eps\|_2$ to
simulate corruption in measurements. 
On this kind of problems, $\ell_1$ regression should give very accurate 
estimate, while $\ell_2$ regression won't work well. 
For completeness, we also add uniform sampling (UNIF) and no conditioning 
(NOCD) into the evaluation. 
Instead of determining the sample size from a given tolerance, we accept the 
sample size as a direct input; and we choose sample sizes from $2^5$ to 
$2^{14}$. 

The results are shown in Figure \ref{fig:sample}, where we draw the
first and the third quartiles of the relative errors in objective value from
$50$ independent runs. If the subsampled matrix is rank-deficient, we set
corresponding relative error to $\infty$ to indicate a failure. We remove
relative errors that are larger than $100$ from the plot in order to show more
details. 
As expected, we can see that UNIF and NOCD are certainly not among reliable
choices; they failed (either generating rank-deficient subsampled problems 
or relative errors larger than $100$) completely on $A_2$.
In addition, GT and FJLT failed partially on the same test. 
Empirically, there is not much difference among $\ell_1$-based algorithms: 
CT works slightly worse than FCT1 and FCT2 on these tests, which certainly 
makes FCT1 and FCT2 more favorable.
(One interesting observation is that we find that, in these tests 
at least, the relative error is proportional to $1/s$ instead of $1/s^{1/2}$. 
At this time, we don't have theory to support this observation.)
This coupled with the fact that $\ell_1$ leverage scores can be 
approximated more quickly with FCT1 and FCT2 suggests the use of these 
transforms in larger-scale applications of $\ell_1$ regression.

\subsection{Evaluation on a Large-scale $\ell_1$ Regression Problem}

Here, we continue to demonstrate the capability of sampling-based algorithms 
in large-scale applications by solving a large-scale $\ell_1$ regression 
problem with imbalanced and corrupted measurements. 
The problem is of size $5.24e9 \times 15$, generated in the following way:
\begin{enumerate}
\item The true signal $x^*$ is a standard Gaussian vector.
\item Each row of the design matrix $A$ is a canonical vector, which means that
  we only estimate a single entry of $x^*$ in each measurement. The number of
  measurements on the $i$-th entry of $x^*$ is twice as large as that on the
  $(i+1)$-th entry, $i=1,\ldots,14$. We have $2.62$ billion measurements on the
  first entry while only $0.16$ million measurements on the last. Imbalanced
  measurements apparently create difficulties for sampling-based algorithms.
\item The response vector is given by 
  \begin{equation*}
    b_i =
    \begin{cases}
      1000 \eps_i & \text{with probability } 0.001 \\
      a_i^T x^* + \eps_i & \text{otherwise}
    \end{cases}, \quad i=1,\ldots,
  \end{equation*}
  where $a_i$ is the $i$-th row of $A$ and $\{\eps_i\}$ are i.i.d.\ samples
  drawn from the Laplace distribution. $0.1\%$ measurements are corrupted to
  simulate noisy real-world data. Due to these corrupted measurements, $\ell_2$
  regression won't give us accurate estimate, and  $\ell_1$ regression is 
  certainly a more robust alternative.
\end{enumerate}
Since the problem is separable, we know that an optimal solution is simply given
by the median of responses corresponding to each entry. 

The experiments were performed on a Hadoop cluster with $40$ cores.  Similar to
our previous test, we implemented and compared Cauchy-conditioned sampling (CT),
Gaussian-conditioned sampling (GT), un-conditioned sampling (NOCD), and uniform
sampling (UNIF). Since $A$ only has $2 n$ non-zeros, CT takes $O(n d \log d)$ 
time instead of $O(n d^2 \log d)$, which makes it the fastest
among CT, FCT1, and FCT2 on this particular problem. 
Moreover, even if $A$ is dense, data at this scale are usually stored on 
secondary storage, and thus time spent on scanning the data typically 
dominates the overall running time. 
Therefore, we only implemented CT for this test. 
Note that the purpose of this test is not to compare CT, FCT1, and FCT2 
(which we did above), but to reveal some inherent differences among 
$\ell_1$ conditioned sampling (CT, FCT1, and FCT2),
$\ell_2$ conditioned sampling (GT and FJLT), and other sampling algorithms 
(NOCD and UNIF). 
For each algorithm, we sample approximately $100000$ ($0.019\%$) rows
and repeat the sampling $100$ times, resulting $100$ approximate solutions. Note
that those $100$ approximate solutions can be computed simultaneously in a
single~pass.

We first check the overall performance of these sampling algorithms, measured by
relative errors in $1$-, $2$-, and $\infty$-norms. The results are shown in
Table \ref{tab:err}.
\begin{table}
  \centering
  \begin{tabular}{c|ccc}
    & $\frac{\|x-x^*\|_1}{\|x^*\|_1}$ & $\frac{\|x-x^*\|_2}{\|x^*\|_2}$ & $\frac{\|x-x^*\|_\infty}{\|x^*\|_\infty}$ \\
    \hline
    CT & [0.008, 0.0115] & [0.00895, 0.0146] & [0.0113, 0.0211]\\
    GT & [0.0126, 0.0168] & [0.0152, 0.0232] & [0.0184, 0.0366]\\
    NOCD & [0.0823, 22.1] & [0.126, 70.8] & [0.193, 134]\\
    UNIF & [0.0572, 0.0951] & [0.089, 0.166] & [0.129, 0.254]\\
    \end{tabular}
    \caption{The first and the third quartiles of relative errors in $1$-, $2$-, and $\infty$-norms. CT clearly performs the best. GT follows closely. NOCD generates large errors, while UNIF works but it is about a magnitude worse than CT.}
  \label{tab:err}
\end{table}
Since the algorithms are all randomized, we show the first and the third
quartiles of the relative errors in $100$ independent runs. We see that CT
clearly performs the best, followed by GT. UNIF works but it is about a
magnitude worse than CT. NOCD is close to UNIF at the first quartile, but makes
very large errors at the third. Without conditioning, NOCD is more likely to
sample outliers because the response from a corrupted measurement is much larger
than that from a normal measurement. However, those corrupted measurements
contain no information about $x^*$, which leads to NOCD's poor performance.
UNIF treats all the measurements the same, but the measurements are
imbalanced. Although we sample $100000$ measurements, the expected number of
measurements on the last entry is only $3.05$, which downgrades UNIF's overall
performance.

\begin{figure}
  \centering
    \includegraphics[width=0.618\textwidth, clip, trim=1.5cm 7cm 2cm 7cm]{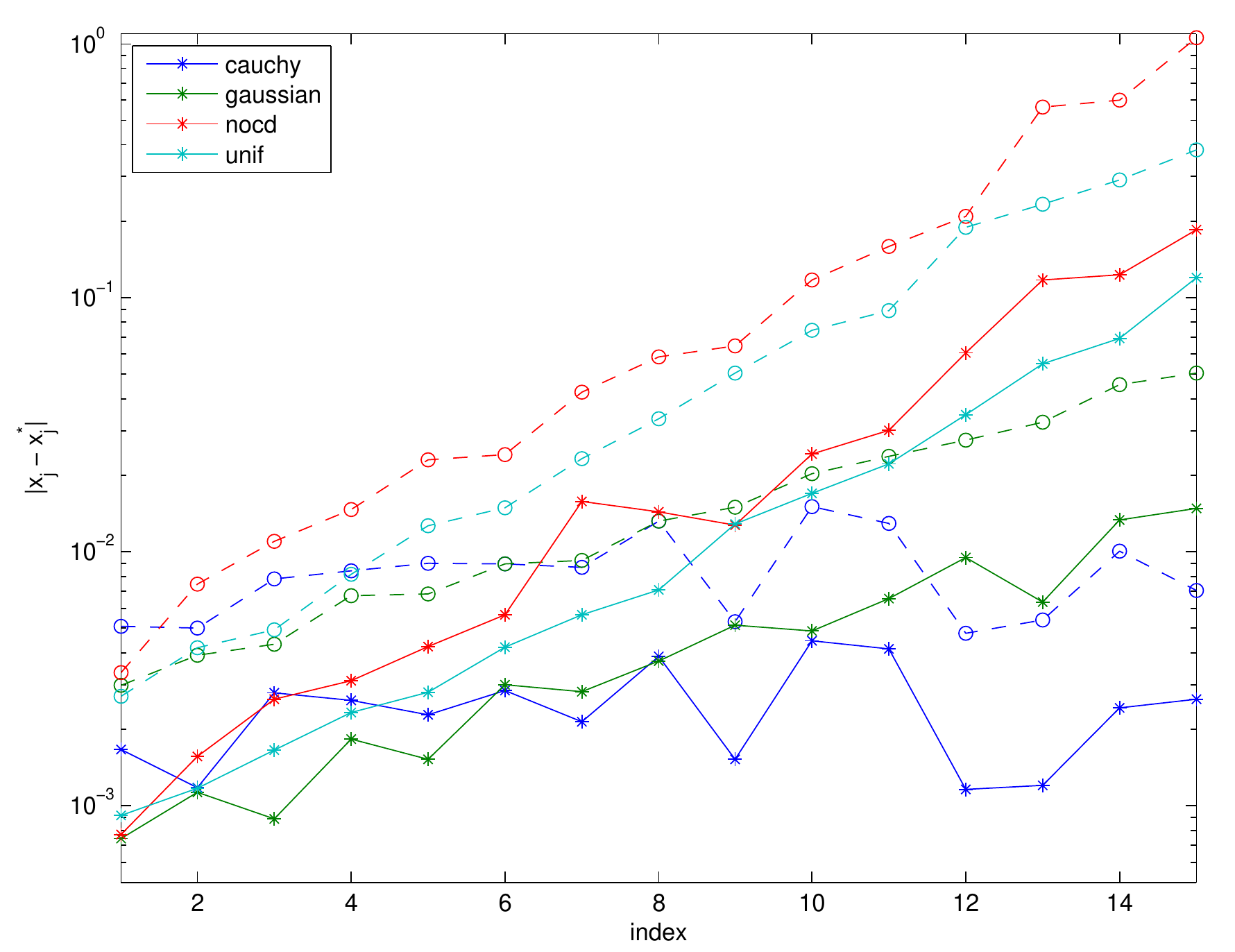}
    \caption{The first (solid) and the third (dashed) quartiles of entry-wise
      absolute errors for our large-scale $\ell_1$ regression empirical 
      evaluation.  See the text for details.
      }
  \label{fig:ele_err}
\end{figure}

We continue to analyze entry-wise errors. Figure \ref{fig:ele_err} draws the
first and the third quartiles of entry-wise absolute errors, which clearly
reveals the differences among $\ell_1$ conditioned sampling,
$\ell_2$ conditioned sampling, and other sampling algorithms. 
While UNIF samples uniformly row-wise, CT tends to sample uniformly
entry-wise. Although not as good as other algorithms on the first entry, CT
maintains the same error level across all the entries, delivering the best
overall performance. The $\ell_2$-based GT sits between CT and UNIF.  $\ell_2$
conditioning can help detect imbalanced measurements to a certain extent and
adjust the sampling weights accordingly, but it is still biased towards the
measurements on the first several entries.

To summarize, we have shown that $\ell_1$ conditioned sampling indeed works 
on large-scale $\ell_1$ regression problems and its performance looks 
promising. 
We obtained about two accurate digits ($0.01$ relative error) on a problem 
of size $5.24e9 \times 15$ by passing over the data twice and sampling only 
$100000$ $(0.019\%)$ rows in a judicious manner.

%% file: conc.tex
\section{Conclusion}
\label{sxn:conc}

We have introduced the Fast Cauchy Transform, an $\ell_1$-based analog of 
fast Hadamard-based random projections. 
We have also demonstrated that this fast $\ell_1$-based random projection 
can be used to develop algorithms with improved running times for a range 
of $\ell_1$-based problems; we have provided extensions of these results to $\ell_p$; and we have provided the first implementation
and empirical evaluation of an $\ell_1$-based random projection.
Our empirical evaluation clearly demonstrates that for large and very 
rectangular problems, for which low-precision solutions are acceptable, 
our implementation follows our theory quite well; 
and it also points to interesting connections between $\ell_1$-based 
projections and $\ell_2$-based projections in practical settings.
Understanding these connections theoretically, exploiting other properties such as sparsity, and using these ideas 
to develop improved algorithms for high-precision solutions to large-scale $\ell_1$-based 
problems, are important future directions raised by our work.

%% file: appdx-techlemmas.tex
\section{Proofs of Technical Cauchy Lemmas}
\label{sec:proofs-technical}

\subsection{Proof of Lemma~\ref{lem:tail} (Cauchy Upper Tail Inequality)}

The proof uses similar techniques to the  
bounds due to Indyk~\cite{Indyk06} for sums
of independent clipped half-Cauchy random variables.
Fix \math{M>0} (we will choose \math{M} later) and define the events
\mand{
F_{i}=\{|C_{i}|\le M\},
}
and \math{F=\cap_{i\in[m]}F_{i}}.
Note that
\math{F\cap F_{i}=F}.
Using the pdf of a Cauchy and because \math{\tan^{-1}x\le x}, we have that:
\mand{\Prob[F_{i}]=\frac{2}{\pi}\tan^{-1}\left(M\right)
=1-\frac{2}{\pi}\tan^{-1}\left(\frac{1}{M}\right)
\ge 
1-\frac{2}{\pi M}  .
}
By a union bound,
\math{\Prob[F]\ge 1-\frac{2m}{\pi M}}. Further,
\math{
\Prob[F|F_{i}]\Prob[F_{i}]=\Prob[F\cap F_i]=\Prob[F]},
hence 
\math{\Prob[F|F_{i}]=\Prob[F]/\Prob[F_{i}]}.
We now bound \math{\Exp\left[|C_{i}|\ \bigl|\ F\right]}.
First, observe that
\eqan{
\Exp\left[|C_{i}|\ \bigl|\ F_{i}\right]
&=& 
\Exp\left[|C_{i}|\ \bigl|\ F_{i}\cap F\right]\Prob[F|F_{i}]+
\Exp\left[|C_{i}|\ \bigl|\ F_{i}\cap \bar F\right]\Prob[\bar F|F_{i}]
\nonumber\\
&\ge&\Exp\left[|C_{i}|\ \bigl|\ F_{i}\cap F\right]\Prob[F|F_{i}].
}
Next, since \math{F_{i}\cap F=F}, we have that 
\mand{\Exp\left[|C_{i}|\ \bigl| F\right]\le
\frac{\Exp\left[|C_{i}|\ \bigl|\ F_{i}\right]}{\Prob[F|F_{i}]}
=\frac{\Exp\left[|C_{i}|\ \bigl|\ F_{i}\right]\Prob[F_i]}{\Prob[F]}.
}
Finally, by
using the pdf of a Cauchy,
\math{
\Exp\left[|C_{i}|\bigl|F_{i}\right]
={\frac{1}{\pi}\log(1+M^2)}/{\Prob[F_{i}]}
}, and so
\mand{
\Exp\left[|C_{i}|\ \bigl| F\right]
\le
\frac{\frac{1}{\pi}\log(1+M^2)}{\Prob[F]}
\le
\frac{\frac{1}{\pi}\log(1+M^2)}{1-2m/\pi M}
.
}
We conclude that
\mand{
\Exp[X|F]=\sum_{i\in[m]}\gamma_i\Exp\left[|C_{i}|\ \bigl| F\right]
\le \frac{\gamma}{\pi}\cdot\frac{\log(1+M^2)}{1-2m/\pi M}.
}
By Markov's inequality and because \math{\Prob[X\ge \gamma t|\bar F]\le 1},
we have:
\eqan{
\Prob[X\ge \gamma t]
&=&
\Prob[X\ge\gamma  t|F]\Prob[F]+\Prob[X\ge\gamma  t|\bar F](1-\Prob[F])\\
&\le&
\frac{1}{\pi t}\cdot\frac{\log(1+M^2)}{1-2m/\pi M}
+\frac{2 m}{\pi M}.
}
The result follows by setting \math{M=2mt}.

\subsection{Proof of Lemma~\ref{lem:lower} (Cauchy Lower Tail Inequality)}

To bound the lower tail, we will use  Lemma~\ref{lem:bernstein}.
By homogeneity, it suffices to prove the result for \math{\gamma=1}.
Let \math{Z_i=\gamma_i\min(|C_i|,M)}. Clearly \math{Z_i\le \gamma_i|C_i|} 
and so 
defining \math{Z=\sum_i Z_i}, we have that
\math{Z\le X} and \math{\Prob[X\le 1-t]\le\Prob[Z\le 1-t]}.
Thus, we have that
\mand{
\Prob[Z\le 1-t]
=
\Prob[Z\le \Exp[Z]-(\Exp[Z]-1+t)]
\le
\expon\left(\frac{-(\Exp[Z]-1+t)^2}{2\sum_{i}\Exp[Z_i^2]}\right),
}
where the last step holds by Lemma~\ref{lem:bernstein} for
\math{1-t<\Exp[Z]}.
Using the distribution of the half-Cauchy, one can verify using standard 
techniques that 
by choosing \math{M\approx 1.6768},
\math{\Exp[Z_i]=\gamma_i} and
\math{\Exp[Z_i^2]\le\frac32\gamma_i^2}, so
\math{\sum_{i}\Exp[Z_i]=1} and
\math{\sum_{i}\Exp[Z_i^2]\le\frac{3}{2}\sum_i\gamma_i^2\le\frac{3}{2\beta^2}}.
It follows that 
\math{\Prob[Z\le 1-t]\le\expon\left(-t^2/
{\textstyle\frac{3}{\beta^2}}\right),}
and the result follows.

\subsection{Proof of Lemma~\ref{lem:L1Chernoff} (\math{\ell_1} Sampling Lemma)}

First, observe that
\math{\norm{DZx}_1=\sum_{i\in[n]}D_{ii}|Z_{(i)}x|}, and since
\math{\Exp[D_{ii}]=1}, \math{\Exp[\norm{DZx}_1]=\sum_{i\in[n]}|Z_{(i)}x|
=\norm{Zx}_1}.
Next, observe that 
\mand{
\sum_{i\in[n]}D_{ii}|Z_{(i)}x|-\sum_{i\in[n]}|Z_{(i)}x|
=
\sum_{\hat p_i<1}D_{ii}|Z_{(i)}x|-\sum_{\hat p_i<1}|Z_{(i)}x|  ,
}
because when \math{\hat p_i=1}, that row must be sampled, and so does not
contribute to the deviation. So, we only need to analyze the
RHS of the above equation.
From now on, we only consider those \math{i} with 
\math{\hat p_i<1}, in which case
\math{\hat p_i=s\cdot t_i}, where 
\math{t_i\ge a\norm{Z_{(i)}}_1/\norm{Z}_1}.
Let \math{Q_i} be the (positive)
random variable \math{D_{ii}|Z_{(i)}x|}; either \math{Q_i=0} or
\mand{
Q_i
=
\frac{|Z_{(i)}x|}{\hat p_i}
\le
\frac{\norm{Z_{(i)}}_1\norm{x}_\infty}{\hat p_i}
=
\frac{\norm{Z_{(i)}}_1\norm{x}_\infty}{st_i}
\le
\frac{1}{a s}\norm{Z}_1\norm{x}_\infty
=
\frac{\gamma}{s},
}
where we defined \math{\gamma=\frac{1}{a}\norm{Z}_1\norm{x}_\infty}.
We can also obtain a bound for \math{\sum_{\hat p_i<1}\var[Q_i]}:
\mand{
\sum_{\hat p_i<1}\var[Q_i]=\sum_{\hat p_i<1}\var[Q_i]\le\sum_{\hat p_i<1}
\Exp[Q_i^2]
=
\sum_{\hat p_i<1}\frac{|Z_{(i)}x|^2}{\hat p_i}
=
\sum_{\hat p_i<1}
Q_i|Z_ {(i)}x|
\le\frac{\gamma}{s}\norm{Zx}_1,
}
where, in the last inequality, we used the upper bound for \math{Q_i} and
we further upper bounded by summing over all \math{i\in[n]}.
Let \math{Q=\sum_i Q_i} with \math{Q_i\le \gamma}; the 
standard Bernstein bound states that
\mand{
\Prob\left[
\left|Q-\Exp[Q]\right|>\eps\right]
\le
2\expon\left(\frac{-\eps^2}{2\sum_{i}\var[Q_i]+\frac23\eps\gamma}
\right)  .
}
Plugging in our bounds for
\math{\sum_{i}\var[Q_i]} and \math{\gamma}, we deduce that 
\mand{
\Prob\Bigl[\bigl|\norm{DZx}_1-\norm{Zx}_1\bigr|>\eps\norm{Zx}_1\Bigr]
\le
2\expon\left(
\frac{-\eps^2\norm{Zx}_1^2}{\frac{2\gamma}{s}\norm{Zx}_1+
\frac{2\eps\gamma}{3s}\norm{Zx}_1}\right).
}
The lemma follows after some simple algebraic manipulations.

%% file: appdx-pfFCTlcm.tex
\section{Proof of Theorem~\ref{theorem:FCT2} (Fast Cauchy Transform (FCT1))}
\label{sxn:FCT2}

\paragraph{Preliminaries.}
Before presenting the proof, we describe the main idea.
It follows a similar line of reasoning to \cite{SW11}, and it uses an 
``uncertainty principle'' (which we state as Lemma~\ref{lem:uncertainty} 
below). 

The uncertainty principle we prove follows from the fact that the
concatenation of the Hadamard matrix with the identity matrix is a 
dictionary of {\it low coherence}. For background, and similar
arguments to those we use in Lemma~\ref{lem:uncertainty} below,
see Section 4 of \cite{Indyk_uncertainty}. In particular, see
Claim 4.1 and Lemma 4.2 of that section. 

To prove the upper bound, we use the existence of a 
$(d,1)$-conditioned basis \math{U} and apply \math{\Pi_1} to this 
basis to show that 
\math{\norm{\Pi_1Ux}_1} cannot expand too much, which in turn means that 
\math{\norm{\Pi_1Ax}_1} cannot expand too much (for any \math{x}).
To prove the lower bound, we show that the inequality holds with 
exponentially high probability, for a particular \math{y}; and we then 
use a suitable \math{\gamma}-net to obtain the result for all \math{y}.

\paragraph{Main Proof.}
We now proceed with the proof of Theorem~\ref{theorem:FCT2}.
We will first prove an upper bound (Proposition~\ref{prop:1}) 
and then a lower bound (Proposition~\ref{prop:2}); the theorem 
follows by combining Propositions~\ref{prop:1} and~\ref{prop:2}.

\begin{proposition}\label{prop:1}
With probability at least 
\math{1-\delta}, for all \math{x\in\R^d},
$\mynorm{\Pione Ax}_1 \le \kappa \mynorm{Ax}_1$, where 
\math{\kappa=O(\frac{d\sqrt{s}}{\delta}\log (r_1 d))}.
\end{proposition}
\begin{proof}
Let \math{U\in\R^{n\times d}} be a $(d,1)$-conditioned basis
(see Definition~\ref{def:l1basis} below)
for the column space of $A$, which  implies that 
for some $Z\in\R^{d\times d}$ we can write $A=UZ$.
Since 
$\mynorm{\Pione Ax}_1 \le \kappa \mynorm{Ax}_1$
if and only if
$\mynorm{\Pione UZx}_1 \le \kappa \mynorm{UZx}_1$, 
it suffices to prove the proposition for
\math{U}.
By construction of 
\math{U}, for any \math{x\in\R^d},
$\mynorm{x}_\infty \le \mynorm{Ux}_1$, and
so
$$\mynorm{\Pione Ux}_1 \le \mynorm{\Pione U}_1 \mynorm{x}_\infty \le  
\mynorm{\Pione U}_{1} \mynorm{Ux}_1.$$
Thus it is enough to show that
$ \mynorm{\Pione U}_{1}\le\kappa$.
We have
\[
 \mynorm{\Pione U}_{1} = 4\mynorm{BC\tH  U}_1
=4\sum_{j\in [d]} \mynorm{BC\tH U^{(j)} }_1
=4\sum_{j\in [d]} \mynorm{BC\hat{U}^{(j)}}_1  ,
\]
where $\hat{U} \equiv \tH U$.
We will need bounds for $\mynorm{\hat{U}^{(j)}}_1$ for $j\in [d]$, 
and $\mynorm{\hat{U}}_1$.
For any vector $y\in\R^n$, we represent \math{y} by its 
\math{n/s} blocks of size 
\math{s}, so \math{z_i\in\R^s} and \math{y^T=[z_1^T,z_2^T,\ldots,z_{n/s}^T]}.
Recall that
$G_s\equiv \left[\begin{smallmatrix} H_{s}\\ I_s\end{smallmatrix}\right]$, 
and observe that  
\math{\mynorm{G_s}_2=\sqrt{2}}.
By explicit calculation,
\mand{
\mynorm{\tH y}_1
	= \sum_{i\in [n/s]} \mynorm{G_sz_i}_1.
}
Since \math{\mynorm{G_sz_i}_1\le\sqrt{2s}\mynorm{G_sz_i}_2\le
\sqrt{4s}\mynorm{z_i}_2\le
\sqrt{4s}\mynorm{z_i}_1}, it follows that  
\[
\mynorm{\tH y}_1
	\le \sqrt{4s}\sum_{i\in [n/s]} \mynorm{z_i}_1
	= \sqrt{4s}\mynorm{y}_1.
\]
Applying this to $y=U^{(j)}$ for $j\in [d]$ yields
\begin{equation}\label{eq:hAnorm}
\mynorm{
\hat{U}^{(j)}}_1\le\sqrt{4s} \mynorm{
{U}^{(j)}}_1, \hbox{ and}\qquad
\mynorm{\hat{U}}_1= \sum_{j\in [d]} \mynorm{\hat{U}^{(j)}}_1
	\le \sqrt{4s}\mynorm{U}_1
\le d\sqrt{4s},
\end{equation}
since \math{\mynorm{U}_1\le d} because $U$ is $(d,1)$-conditioned.

The $(i,j)$ entry of $BC\hat{U}$ is 
\math{
\sum_{k}B_{ik}C_{kk}{\hat U}_{kj},
}
which is a Cauchy scaled by
\math{\gamma_{ij}=\sum_{k}|B_{ik} \hat U_{kj}|}. So,
\mand{
\norm{BC\hat{U}}_1=\sum_{i\in[r_1],j\in[d]}\gamma_{ij}\tilde C_{ij},
}
where \math{\tilde C_{ij}} are \emph{dependent} Cauchy random variables. 
Using \math{\sum_{i}B_{ik}=1}, we obtain:
\mand{\sum_{i,j}\gamma_{ij}=
\sum_{i,j,k}|B_{ik} \hat U_{kj}|
=
\sum_{j,k}|\hat U_{kj}|\sum_{i}B_{ik}
=
\sum_{j,k}|\hat U_{kj}|
=
\norm{\hat U}_1 .
}
Hence, we can
apply Lemma~\ref{lem:tail} with \math{\gamma=\norm{\hat U}_1} and
\math{m=r_1 d} to obtain
\mand{
\Prob\left[\mynorm{BC\hat{U}}_1> t\mynorm{{\hat U}}_1\right]
\le \frac{(\log(r_1 d)+\log t)}{t}\left(1+o(1)\right).
}
Setting the RHS to \math{\delta}, it suffices that  
\math{t=O(\frac1\delta\log(d r_1))}. 
Thus,
with  probability at least \math{1-\delta},
\mand{\mynorm{BC\hat{U}}_1=
O\left(\frac1\delta\log(d r_1)\mynorm{{\hat U}}_1\right)
=
O\left(\frac{d}{\delta}\sqrt{s}\log(d r_1)\right).
}
\end{proof}

Before we prove the lower bound, we need the following lemma 
which is derived using a sparsity result for matrices with unit norm 
rows and low ``coherence,'' as measured by the maximum magnitude of 
the inner product between distinct rows (\math{G_s} is a matrix with 
low coherence). This result mimics results in
\cite{donoho89,donoho01,gribonval03,tropp04,Indyk_uncertainty}.

\begin{lemma}\label{lem:uncertainty}
For $G=\left[\begin{smallmatrix} H_{s}\\ I_s\end{smallmatrix}\right]$
and any $z\in\R^{s}$,
$\mynorm{Gz}_1 \ge \frac{1}{2} s^{1/4}\mynorm{z}_2$.
\end{lemma}

\begin{proof}
We can assume $\mynorm{z}_2=1$, and so $\mynorm{Gz}_2^2 = 2$.
Let \math{G_{(S')}} be \math{k} rows of \math{G}, with \math{\kappa} of them
coming 
from \math{H_{s}} and \math{k-\kappa} from \math{I_s}.
\math{G_{(S')}G_{(S')}^T=I+\Lambda} where \math{\Lambda} is a symmetric
\math{2\times 2} 
block matrix \math{\left[\begin{smallmatrix} 
\bm{0}&\frac{1}{\sqrt{s}}Q\\ 
\frac{1}{\sqrt{s}}Q^T&\bm{0}\\ 
\end{smallmatrix}
\right]} where the entries in \math{Q\in\R^{\kappa\times(k-\kappa)}}
are \math{\pm1}, and so \math{\mynorm{Q}_2\le\sqrt{\kappa(k-\kappa)}
\le \frac12k}. 
\mand{\mynorm{G_{(S')}}_2^2=\mynorm{G_{(S')}G_{(S')}^T}_2
\le 1+\mynorm{\Lambda}_2=1+\frac{1}{\sqrt{s}}\mynorm{Q}_2\le 1+
\frac{k}{2\sqrt{s}}.}
Now, given any \math{z}, we set \math{k=2\beta\sqrt{s}} 
with \math{\beta=\frac{2}{5}}, and choose
\math{G_{(S')}} to be the rows corresponding to the \math{k} components
of \math{Gz} having largest magnitude, with \math{G_{(S)}} being the rows
with indices in \math{[s]\setminus S'}. Then
\math{\mynorm{G_{(S')}z}_2^2\le 1+\beta}, and so the entry in 
\math{G_{(S')}z} with smallest magnitude has magnitude at most
\math{a=\sqrt{(1+\beta)/k}=\sqrt{(1+\beta)/2\beta}s^{-1/4}}.
We now consider \math{G_{(S)}z}. Since $\mynorm{Gz}_2^2 = 2$,
$\mynorm{G_{(S)}z}_2^2 \ge 1-\beta$; further, all components  
have magnitude at most \math{a}
(as all the components of \math{G_{(S)}z} have smaller magnitude than those of
\math{G_{(S')}z}). \math{\mynorm{G_{(S)}z}_1} is minimized by concentrating all
the entries into as few components as possible. Since the number of 
non-zero components is at least 
\math{(1-\beta)/a^2=2\beta s^{1/2}(1-\beta)/(1+\beta)}, giving these 
entries the maximum possible magnitude results in
\mand{\mynorm{G_{(S)}z}_1\ge a\times\frac{(1-\beta)}{a^2}= 
(1-\beta)\sqrt{2\beta(1+\beta)}s^{1/4}
\ge 0.63 s^{1/4}} 
(where we used \math{\beta=\frac25}). We are done 
because
\math{\mynorm{Gz}_1\ge \mynorm{G_{(S)}z}_1}
\end{proof}
We now prove the lower bound. We assume that
Proposition~\ref{prop:1} holds for \math{\Pione},
which is true with probability at least
\math{1-\delta} for \math{\kappa} as defined in Proposition~\ref{prop:1}.
Then, by a union bound,
both Propositions~\ref{prop:1} and~\ref{prop:2} hold with 
probability at least 
\mand{1-\delta-\expon\left(-\frac{r_1}{48}+d\log (2d\kappa)\right)-
\expon\left(-\frac{s^{1/2}}{8r_1^2}+\log r_1+d\log (2d\kappa)\right)} 
(\math{\delta} and \math{\kappa} are from 
Proposition~\ref{prop:1}). 
Since \math{s^{1/2}=r_1^3},
by choosing \math{r_1=\alpha d\log \frac{d}{\delta}} for
large enough \math{\alpha}, the final probability of failure is 
at most \math{2\delta},
because 
\math{\kappa=O(\frac{d\sqrt{s}}{\delta}\log(r_1 d))=O(\poly(d))}.

\begin{proposition}\label{prop:2}
Assume Proposition~\ref{prop:1} holds. Then, for all \math{x\in\R^d},
$\mynorm{\Pione Ax}_1 \ge 
\mynorm{Ax}_1$ holds with probability
at least 
\mand{1-\expon\left(-\frac{r_1}{48}+d\log (2d\kappa)\right)-
\expon\left(-\frac{s^{1/2}}{8r_1^2}+\log r_1+d\log (2d\kappa)\right)}.
\end{proposition}

\begin{proof}
First we will show a result for fixed $y\in\R^n$, summarized in the 
next lemma.
\begin{lemma}\label{lem:fixedy}
$\displaystyle \Prob\left[\norm{\Pione y}_1 < 2\norm{y}_1\right]
\le \expon\left(-{\frac{r_1}{48}}\right)+\expon\left(-\frac{s^{1/2}}{8r_1^2}+\log r_1\right)$
\end{lemma}
Given this lemma, the proposition follows 
by putting a $\gamma$-net $\Gamma$ on the range of \math{A} (observe that
the range of \math{A} has 
dimension at most \math{d}). This argument follows the same line as in
Sections 3 and 4 of \cite{SW11}. Specifically,
let \math{L} be any fixed 
(at most) \math{d} dimensional subspace of \math{\R^n}
(in our case, \math{L} is the range of \math{A}). 
Consider the 
\math{\gamma}-net
on \math{L} with cubes of side \math{\gamma/d}.
There are \math{(2d/\gamma)^d} such cubes required to cover the
hyper-cube \math{\norm{y}_\infty\le 1}; and,
 for any two points \math{y_1,y_2} inside the same 
\math{\gamma/d}-cube, \math{\mynorm{y_1-y_2}_1\le\gamma}. 
From each of the \math{\gamma/d}-cubes, select a fixed representative point
which we will generically refer to as \math{y^*}; select the
representative to have 
\math{\mynorm{y^*}_1=1} if possible.
By a union bound and Lemma~\ref{lem:fixedy}, 
\mand{\Prob\left[\min_{y^*}\norm{\Pione y^*}_1/\norm{y^*}_1 <
2\right]
\le (2d/\gamma)^d 
\left(\expon\left(-{\frac{r_1}{48}}\right)+\expon\left(-\frac{s^{1/2}}{8r_1^2}+\log r_1\right)\right).}
We will thus condition on the high probability event that
\math{\norm{\Pione y^*}_1\ge 2\norm{y^*}} for all \math{y^*}.
For any \math{y\in L}
with \math{\norm{y}_1=1},
let \math{y^*} denote the representative point for the cube in which 
\math{y} resides (\math{\norm{y^*}_1=1} as well). Then 
\math{\norm{y-y^*}\le \gamma}.
\mand{\norm{\Pione y}_1=\norm{\Pione y^*+\Pione(y-y^*)}_1
\ge\norm{\Pione y^*}_1-\norm{\Pione(y-y^*)}_1
\ge 2{\norm{y^*}_1}-\kappa\norm{y-y^*}_1,
}
where  the last inequality holds using Proposition~\ref{prop:1}.
By choosing \math{\gamma=1/\kappa} and
recalling that \math{\norm{y^*}_1=1}, we have that
\math{\norm{\Pione y}_1\ge 1}, with probability at least
\mand{1-\expon(d\log (2d\kappa))\left(\expon\left(-{\frac{r_1}{48}}\right)+\expon\left(-\frac{s^{1/2}}{8r_1^2}+\log r_1\right)\right).}

All that remains is to prove Lemma~\ref{lem:fixedy}.
As in the proof of Proposition~\ref{prop:1},
we represent 
any vector $y\in\R^n$ by its \math{n/s} blocks of size 
\math{s}, so \math{z_i\in\R^s} and \math{y^T=[z_1^T,z_2^T,\ldots,z_{n/s}^T]}.
Let \math{g=\tH y},
\mand{
g=
\left[
\begin{smallmatrix}
G_sz_1\\
G_sz_2\\
\vdots\\
G_sz_{n/s}
\end{smallmatrix}
\right].
}
We have that \math{\norm{g}_2^2=\sum_i\norm{G_sz_i}_2^2=2
\sum_i\norm{z_i}_2^2=2\norm{y}_2^2}, and 
\mand{
\norm{g}_1=\sum_i\norm{G_sz_i}_1
\ge
\frac12s^{1/4}\sum_i\norm{z_i}_2
\ge
\frac12s^{1/4}\left(\sum_i\norm{z_i}_2^2\right)^{1/2}
=
\frac12s^{1/4}\norm{y}_2.
}
We conclude that \math{\norm{g}_1\ge\frac{1}{2\sqrt{2}}s^{1/4}\norm{g}_2}, 
which intuitively means that \math{g} is ``spread out.''
We now analyze \math{\norm{BCg}_1}. (Recall that 
\math{\Pione y=4BCg}, where \math{g=\tH y}).
\mand{
(BCg)_j=\sum_{i=1}^{2n}B_{ji}C_{ii}g_i
}
is a Cauchy random variable \math{\tilde C_j} scaled by
\math{\gamma_j=\sum_{i=1}^{2n}B_{ji}|g_i|}. Further, because each
column of \math{B} has exactly one non-zero element, the 
\math{\tilde C_j} for \math{j\in[r_1]} are independent.
Thus, the random variables 
\math{\norm{BCg}_1} and \math{\sum_{j\in[r_1]}|\tilde C_j|\gamma_j} 
have the same distribution. 
To apply Lemma~\ref{lem:lower},
we need to bound \math{\sum_j\gamma_j} and \math{\sum_j\gamma_j^2}.
First,
\mand{
\sum_{j\in[r_1]}\gamma_j=\sum_{j\in[r_1]}\sum_{i\in[n]}B_{ji}|g_i|
=\sum_{i\in[n]}|g_i|\sum_{j\in[r_1]}B_{ji}=\sum_{i=1}^{2n}|g_i|=\norm{g}_1,
}
where the last inequality is because \math{B^{(i)}} is a standard basis
vector. To bound \math{\sum_j\gamma_j^2}, we will show that 
\math{\gamma_j} is nearly uniform. Since 
\math{\gamma_j} is a weighted sum of independent Bernoulli random
variables (because \math{B_{ji}} and \math{B_{jk}} are independent
for \math{i\not=k}), we can use Lemma~\ref{lem:bernoulli} with 
\math{\xi_i=|g|_i} and \math{1-p=1-1/r_1\le 1}, and so 
\math{\sum_{i}\xi_i=\norm{g}_1} and
\math{\sum_{i}\xi_i^2=\norm{g}^2_2}; setting \math{t=1/r_1} in  
Lemma~\ref{lem:bernoulli}:
\mand{
\Prob
\left[\gamma_j\ge\frac{2\norm{g}_1}{r_1}\right]
\le
\expon\left(-\frac{\norm{g}_1^2}{2\norm{g}_2^2r_1^2}\right)
\le
\expon\left(-\frac{s^{1/2}}{8r_1^2}\right).
}
By a union bound, none of the \math{\gamma_j} exceed 
\math{2\norm{g}_1/r_1} with probability at most 
\math{r_1\expon\left(-{s^{1/2}}/{8r_1^2}\right)}. We assume this high 
probability event, in which case
\math{\sum_{j}\gamma_j^2\le 4\norm{g_1}^2_1/r_1}.
We can now apply Lemma ~\ref{lem:lower} with \math{\beta^2=r_1/4} and 
\math{t=\frac12} to obtain
\mand{
\Prob\left[\sum_{j\in[r_1]}|\tilde C_j|\gamma_j\le{\textstyle \frac12}\norm{g}_1
\right]\le\expon\left(-{\frac{r_1}{48}}\right).
}
By a union bound, \math{\norm{BCg}_1\ge\frac12\norm{g}_1} with probability at least 
\math{1-\expon\left(-{\frac{r_1}{48}}\right)-\expon\left(-\frac{s^{1/2}}{8r_1^2}+\log r_1\right)}.
Scaling both sides by 
\math{4} gives the lemma.
%
\end{proof}

\paragraph{Running Time.} The running time follows from the time to 
compute the product \math{H_s x} for a Hadamard matrix \math{H_s},
which is \math{O(s\log s)} time. The time to compute \math{\tH y} is dominated
by \math{n/s} computations of \math{H_s z_i}, which is a total of
\math{O(\frac{n}{s}\cdot s \log s)= O(n\log s)} time.
Since \math{C} is diagonal, pre-multiplying by \math{C} is 
\math{O(n)} and further pre-multiplying by 
\math{B} takes time \math{O(nnz(B))}, the number of non-zero elements in 
\math{B} (which is \math{2n}). Thus the total time is
\math{O(n\log s +n)=O(n\log r_1)} as desired.

%% file: appdx-pfFCTfjl.tex
\section{Proof of Theorem~\ref{theorem:FCT} (Fast Cauchy Transform (FCT2))}
\label{sxn:FCTFJLT}

\paragraph{Preliminaries.}
We will need results from prior work, which we paraphrase in our notation.
\begin{definition}[Definition 2.1 of \cite{AL08}]\label{def:JLP}
For  $\eps\in(0,\frac12]$,
a distribution 
on $s \times t$ real matrices $G$ $(s \leq t)$ has
the \emph{Manhattan Johnson-Lindenstrauss property} (MJLP) if for any (fixed)
 vector 
$x \in \mathbb{R}^t$, the inequalities 
\eqan{ 
&(1-\eps)\norm{x}_2\le\norm{Gx}_2\le(1+\eps)\norm{x}_2&\\
&c_3{\sqrt{s}}(1-\eps)\norm{x}_2
\le\norm{Gx}_1\le c_3{\sqrt{s}}(1+\eps)\norm{x}_2&\\
}
holds with probability 
at least \math{1-c_1 e^{-c_2 k \eps^2}} (w.r.t. \math{G}),
for global constants $c_1, c_2 ,c_3> 0$. 
\end{definition}

\noindent\textbf{Remark.}
This is the standard Johnson-Lindenstrauss property with the additional
requirement on \math{\norm{Gx}_1}. Essentially it says that 
\math{Gx} is a nearly uniform, so that 
\math{\norm{Gx}_1\approx\sqrt{s}\norm{Gx}_2\approx\sqrt{s}\norm{x}_2}.

\begin{lemma}[Theorem 2.2 of \cite{AL08}]\label{thm:al}
Let $\eta > 0$ be an arbitrarily small constant. For any $s, t$ satisfying
$s \leq t^{1/2-\eta}$, there exists an algorithm that 
 constructs a random $s \times t$ matrix
$G$ that is sampled from an MJLP distribution
with \math{c_3=\sqrt{\frac{2}{\pi}}}. 
Further, the time to compute $Gx$
for any $x \in \mathbb{R}^t$ is $O(t \log s)$. 
\end{lemma}

We will need these lemmas to get a result for
how an arbitrary subspace \math{L} behaves under the action of \math{G}, 
extending Lemma~\ref{thm:al} to every \math{x\in L}, not just a fixed \math{x}. 
In the next lemma, the 2-norm bound can be derived 
using Lemma~\ref{thm:al} (above) 
and Theorem 19 of \cite{KN12}
by placing a \math{\gamma}-net 
on \math{L} and bounding the
size of this \math{\gamma}-net.  
(See Lemma 4 of \cite{ahk06}.) 
The Manhattan norm bound is then 
derived using a second \math{\gamma}-net 
argument together with an application of 
the 2-norm bound. 
The constants \math{c_1,\ c_2} and 
\math{c_3} in this lemma are from
Definition~\ref{def:JLP}; and the \math{G} 
in Lemmas~\ref{thm:al} and~\ref{thm:subspace},
with the constants \math{c_1,\ c_2} and 
\math{c_3} from Definition~\ref{def:JLP}, 
is the same \math{G} used in our 
FCT2 construction for \math{\tH}. 
We present the complete proof of Lemma~\ref{thm:subspace} in 
Appendix~\ref{sxn:app-pf-thm-subspace}.

\begin{lemma}
\label{thm:subspace}
Let \math{L} be any (fixed) \math{d} dimensional subspace of \math{\R^t},
 and  $G$ an $s \times t$ matrix sampled from a distribution
having the MJLP property.
Given $\eps\in(0,\frac13]$, let \math{s=36(k+\frac{8d}{c_3\eps}+
\log (2c_1))/c_2\eps^2=O(\frac{k}{\eps^2}+\frac{d}{\eps^3})}.
Then,
with probability at least $1-e^{-k}$, for every \math{x\in\R^t},
\eqan{ 
&\sqrt{1-\eps}\norm{x}_2\le\norm{Gx}_2\le\sqrt{1+\eps}\norm{x}_2&\\
&c_3{\sqrt{s}}(1-\eps)\norm{x}_2
\le\norm{Gx}_1\le c_3{\sqrt{s}}(1+\eps)\norm{x}_2&\\
}
\end{lemma}

We also need a result on 
how the matrix of Cauchy random variables \math{C} 
behaves when it hits a vector \math{y}.
The next theorem is Theorem 5 of \cite{SW11}.
For completeness and also to fix some minor errors in the proof 
of~\cite{SW11}, we give a proof of Theorem~\ref{thm:sw} in 
Appendix~\ref{sxn:app-pf-thm-sw}.

\begin{theorem}\label{thm:sw}
Let $L$ be an arbitrary (fixed)
subspace of $\mathbb{R}^n$ having dimension at most $d$, and  $C$ an 
$r_1 \times n$
matrix of i.i.d. Cauchy random variables with 
\math{r_1=c\cdot  d\log \frac{d}{\delta}} for large enough constant
\math{c}. Then,  with
probability at least \math{1-\delta}, and for all 
$y \in L$, 
$$\|y\|_1 \leq {\textstyle\frac{4}{r_1}}\|Cy\|_1 \leq \kappa'\|y\|_1,$$
where \math{\kappa'= 
O(\frac{d}{\delta}\log(r_1 d))}. 
\end{theorem}

\noindent
Note that for \math{\delta} fixed to some
small error probability, \math{r_1=O(d\log d)},
and the product
 \math{Cy} in the theorem above can be computed in 
time \math{O(r_1 n)=O(n d\log d).} 

\paragraph{Main Proof.}
We now proceed with the proof of Theorem~\ref{theorem:FCT}.
We need to analyze the product \math{C\tH A x} for all \math{x\in\R^d}.
Let \math{y=Ax\in\R^n}, so that \math{y\in \colsp A\equiv \{Az\mid z\in\R^d\}},
and the column space $\colsp A$ is a $d$-dimensional subspace of $\R^n$.
Partition the coordinate set $[n]$ into $n/t$ contiguous groups of $t$ 
coordinates. We will work with the block representation of \math{y}, as 
defined by this partition,
\emph{i.e.}, with
\math{y^T=[z_1^T,z_2^T,\ldots,z_{n/t}^T]}, where 
\math{z_i=A_{(\{i\})}x} and where \math{A_{(\{i\})}} is the block of 
\math{t}
rows in
\math{A} corresponding to the indices in \math{z_i}. Then,
\mand{
\tH y=
\left[
\begin{smallmatrix}
Gz_1\\
Gz_2\\
\vdots\\
Gz_{n/t}
\end{smallmatrix}
\right].
}
The vector \math{z_{i}\in\colsp A_{(\{i\})}},
noting that \math{\colsp A_{(\{i\})}} is
a subspace of
\math{\R^t} of dimension at most $d$. Let \math{U_i\in\R^{t\times d}} 
be an orthonormal basis for \math{\colsp A_{(\{i\})}},
and let \math{z_i=U_iw_i}. Setting \math{\eps=\frac12}
in Lemma~\ref{thm:subspace}, and recalling that 
\math{G} is \math{s\times t},
\math{k} in Lemma~\ref{thm:subspace} can be expressed as
\math{k=\frac{c_2 s}{144}-\frac{16d}{c_3}-\log (2c_2)}.
Applying a union bound,
we have that 
for all \math{i\in[n/t]}
with probability at least 
\math{1-2c_1\cdot\frac{n}{t}\cdot\expon(-\frac{c_2s}{144}+\frac{16d}{c_3})},
that for all \math{y=Ax} (and corresponding
\math{z_i\in\R^t}), it holds that
\eqan{
&\sqrt{\textstyle\frac12}\norm{z_i}_2\le\norm{Gz_i}_2\le
\sqrt{\textstyle\frac32}\norm{z_i}_2&\\
&{\textstyle\frac12}c_3\sqrt{s}\norm{z_i}_2\le\norm{Gz_i}_1\le
{\textstyle\frac32}c_3\sqrt{s}\norm{z_i}_2&.
}
We will condition on this event, which occurs with high probability for
the given parameters.
We can now bound \math{\norm{\tH y}_1=\sum_{i\in[n/t]} \|Gz_i\|_1 } as follows.
\eqar{
\norm{\tH y}_1&=&
\sum_{i\in[n/t]} \|Gz_i\|_1 \le
{\textstyle\frac32} c_3\sqrt{s}\sum_{i\in[n/t]}\|z_i\|_2
\leq  
{\textstyle\frac32} c_3\sqrt{s}
\sum_{i\in[n/t]} \|z_i\|_1
=
{\textstyle\frac32} c_3\sqrt{s}
\norm{y}_1;
\label{eq:Hupper}
\\
\norm{\tH y}_1&=&
\sum_{i\in[n/t]} \|Gz_i\|_1\ge 
{\textstyle\frac12} c_3\sqrt{s}\sum_{i\in[n/t]}\|z_i\|_2
\geq 
{\textstyle\frac12} c_3\sqrt{\textstyle\frac{s}{t}} 
\sum_{i\in[n/t]}\|z_i\|_1
={\textstyle \frac12} c_3\sqrt{\textstyle\frac{s}{t}} \norm{y}_1;
\label{eq:Hlower}
}
Since \math{\colsp \tH A} has dimension at most \math{d},
we can apply Theorem \ref{thm:sw} to it.
We have that with probability at least $1-\delta$, 
for all $x \in \mathbb{R}^d$,
$$\| \tilde{H}Ax\|_1 \leq \frac{4}{r_1} \|C\tilde{H}Ax\|_1 
\leq 
\kappa' \| \tilde{H} Ax\|_1,$$
where \math{\kappa'=O(\frac{d}{\delta}\log(r_1d))} from
 Theorem~\ref{thm:sw}.
Recall that $\Pione\equiv  \frac{8}{r_1} \sqrt{\frac{\pi t}{2s}} C \tH$.
We now use \r{eq:Hupper} and \r{eq:Hlower} with \math{y=Ax}, and after 
multiplying by \math{\frac{2}{c_3}\sqrt{\frac{t}{s}}} and setting 
\math{c_3=\sqrt{2/\pi}}, we conclude that 
for all $x \in \mathbb{R}^d$,
\mld{
\|Ax\|_1 \leq \|\Pione Ax\|_1 \leq 3\kappa' \sqrt{t}\|Ax\|_1
\label{eq:floatingbound}
}
holds with probability at least $1-\delta-2c_1\cdot
\frac{n}{t}\expon(-\frac{c_2 s}{144}+\frac{16d}{c_3})\ge 1-2\delta$
(by choosing \math{s\ge\frac{144}{c_2}(\frac{16d}{c_3}
+\log\frac{2c_1n}{\delta t})}).
 The theorem follows because
\math{\log n\le d} and hence \math{\kappa'=O(\frac{d}{\delta}\log d)},
\math{s=O(d+\log\frac1\delta)} and \math{t=O(s^{2+\eta})}.

\paragraph{Running Time.}
We now evaluate the time to compute \math{\Pione y} for \math{y\in\R^n}.
We first compute \math{\tH y} which requires \math{n/t} computations of 
\math{Gz_i}. Since \math{s=t^{1/2-\eta/2}}, we can invoke Lemma~\ref{thm:al}.
The time to compute all \math{Gz_i} is \math{\frac{n}{t}\cdot t\log s=n\log s}.
Since \math{\tH y} is \math{(n s/t)\times 1}, it takes 
\math{O(r_1ns/t)} 
time to compute \math{C\tH y}, which concludes the computation.
The total running time is \math{O(n\log s+nr_1s/t)}.
Using \math{\log n\le d}, \math{s=O(d)}, \math{t=s^{2+\eta}}, 
\math{r_1=O(d\log \frac{d}{\delta})} we need total time
\math{O(n\log \frac{d}{\delta})}.
To compute \math{\Pione A}, we need to compute
\math{\Pione A^{(j)}} for 
\math{d} vectors \math{A^{(j)}}, resulting in a total run time
\math{O(nd\log \frac{d}{\delta})}.

%% file: appdx-pfL1basis.tex
\section{Proof of Theorem \ref{thm:basisL1} (Fast $\ell_1$ Well-conditioned Basis)}
\label{sxn:pfL1basis}

Clearly \math{U=AR^{-1}} is in the range of \math{A} 
and has the same null-space
otherwise \math{\Pione A}
 would not preserve lengths to relative error. Therefore
\math{U} is a basis for the range of \math{A}. Consider 
any \math{x\in\R^d}. 
The first claim of the theorem
follows from the following derivations:
\eqan{
&&\norm{U}_1=\norm{AR^{-1}}_1 \mathop{\buildrel (a)\over\le}
\norm{\Pione AR^{-1}}_1 \mathop\le \sqrt{r_1}\norm{\Pione AR^{-1}}_2{\buildrel (b) \over =}d\sqrt{r_1}; \mbox{and} \\
&&\norm{x}_\infty\mathop\le\norm{x}_2\mathop{\buildrel (b) \over =}
\norm{\Pione AR^{-1}x}_2\le \norm{\Pione AR^{-1}x}_1{\buildrel (c) \over \le} 
\kappa\norm{AR^{-1}x}_1=\kappa\norm{Ux}_1.
}
(a) follows from the lower bound in \r{eq:main}, because it holds for
every column of \math{AR^{-1}}; (b) follows because by the construction of
\math{R}, \math{\Pione AR^{-1}} has \math{d} orthonormal columns;
finally, (c) follows from the upper bound in \r{eq:main}.

Finally, to obtain the Corollary, if \math{\Pione} 
satisfying \r{eq:main} is constructed 
using 
Theorem~\ref{theorem:FCT} 
with small fixed probability of failure \math{\delta}, then 
\math{r_1=O(d\log d)} and \math{\kappa=O(d^{2+\eta}\log d)}. 
The running time to compute \math{\RR} is obtained by summing 
\math{O(nd\log d)} (to compute \math{\Pione A}) and
\math{O(r_1d^2)=O(d^3\log d)} (to obtain \math{R^{-1}\in\R^{d\times d}}).

%% file: appdx-pfL1fast.tex
\section{Proof of Theorem~\ref{theorem:FCRsimp} (Fast Cauchy \math{\ell_1} Regression)}
\label{sxn:FCRsimp}

For \math{X\in\R^{n\times q}}, we analyze the more general constrained 
\math{\ell_1} regression problem, 
\math{
\min_{x\in \cl C}\norm{Xx}_1,
}
where $\cl C \subseteq \R^q$ is a convex set,
and we show that 
\math{\hat x\in\cl C} constructed by our algorithm
is a \math{(1+\eps)}-approximation for this more general problem:
\mand{
\norm{X\hat x}_1\le(1+\eps)
\min_{x\in \cl C}\norm{Xx^*}_1.
}
(That is, we actually prove a somewhat stronger result than we state in 
Theorem~\ref{theorem:FCRsimp}.
This more general problem involves calling our main algorithm with 
\math{b=\{\}} (NULL), and then incorporating the constraint that 
\math{x\in\cl C} into the optimization problem that is solved as a black 
box in the last step.
Of course, if the constraint set is not a polytope, then the last step may 
involve more sophisticated techniques than linear programming.)
The classic \math{\ell_1} regression is obtained by setting
\math{X=\left[\begin{matrix}A&-b\end{matrix}\right]} (\math{q=d+1})
with  constraint \math{\cl C=\{x:e_{d+1}^Tx=1\}}, which corresponds to 
setting \math{x_{d+1}=1}.

The main ingredients in our
proof follow a similar line
to those in \cite{Cla05,DDHKM09_lp_SICOMP,SW11}. 
We use the notation from Figure~\ref{l1reg-mainSimp}. 
From Step 1, \math{\Pi_1} 
satisfies \r{eq:main} with \math{A\gets X}, \emph{i.e.},
 \math{\Pione} preserves the 
\math{\ell_1}-norm of vectors in the range of 
\math{X}:
\mld{
\norm{Xx}_1\le\norm{\Pione Xx}_1\le\kappa\norm{Xx}_1.\label{eq:mainX}
}
Let \math{\cl C'=\{y=\RR x:x\in\cl C\}} be a linear transform of the constraint
set.
We start with a basic lemma that allow us to use \math{U} instead of~\math{X}.
This lemma says that if we can construct a sampling matrix 
\math{D} for \math{U} under the constraint \math{\cl C'}
such that solving the
down-sampled problem for \math{U} gives a \math{(1+\eps)}-approximation,
 then that \emph{same} sampling matrix works for \math{X} under the
constraint \math{\cl C}. 

\begin{lemma}\label{lem:UnotX}
Let \math{U=X\RR} and \math{D} any diagonal sampling matrix as 
in Figure~\ref{l1reg-mainSimp}. Suppose that for any \math{\hat y\in\cl C'} 
that 
minimizes
\math{\norm{DUy}}, \math{\hat y} is
a \math{(1+\eps)}-approximation for the problem
\math{\min_{y\in\cl C'}\norm{Uy}}.  Let 
\math{\hat x} be any solution
to \math{\min_{x\in\cl C}\norm{DXx}}. Then \math{\hat x} is
a \math{(1+\eps)}-approximation for the problem
\math{\min_{x\in\cl C}\norm{Xx}}.
\end{lemma}
\begin{proof}
Select \math{\hat y=\RR\hat x\in\cl C'}. For any 
\math{y\in\cl C'}, there is some \math{x\in \cl C} with 
\math{y=\RR x}, and we have:
\mand{
\norm{DUy}_1
=
\norm{DU\RR x}_1
=
\norm{DX x}_1
{\buildrel (a)\over \ge}
\norm{DX \hat x}_1
=
\norm{DU \RR \hat x}_1
=
\norm{DU\hat y}_1,
}
where (a) is by the optimality of \math{\hat x}.
So, \math{\hat y} minimizes \math{\norm{DUy}_1}, hence for 
all \math{y\in\cl C'},
\math{\norm{U\hat y}_1\le (1+\eps)\norm{Uy}_1}. Now consider 
any \math{x\in \cl C} and let \math{y=\RR x\in\cl C'}. Then,
\mand{
\norm{X\hat x}_1
=
\norm{U\RR\hat x}_1
=
\norm{U\hat y}_1
\le
(1+\eps)\norm{U y}_1
=
(1+\eps)
\norm{U\RR x}_1
=
(1+\eps)
\norm{X x}_1.
}
\end{proof}

\noindent\textbf{Remarks.} 
We emphasize that our proof accommodates an \emph{arbitrary} constraint set 
\math{\cl C}.  
For the classical \math{\ell_1} regression problem, 
that is of interest to us in Theorem~\ref{theorem:FCRsimp},
we only need \math{\cl C} to be specified by a single linear constraint.
In the remaining, we will work with \math{U} and show that our algorithm 
generates a coreset that works, regardless of the constraint set \math{\cl C'}.

By Theorem~\ref{thm:basisL1}, \math{U=X\RR} is an 
\math{(\alpha,\beta)}-conditioned basis for the range of \math{X}, where
\mand{\alpha\le q\sqrt{r_1}, \qquad\text{ and }\beta\le\kappa.} 
So,
\math{\norm{U}_1\le q\sqrt{r_1}} and for all \math{y\in\R^{q}},
\math{\norm{y}_\infty\le\kappa\norm{Uy}_1}.
We next show that \math{\lambda_i} estimates \math{\norm{U_{(i)}}_1}.
The following lemma is a straightforward application of a Chernoff bound
 to independent half-Cauchys (see also Claims 1 and 2 and 
Lemmas 1 and 2 in \cite{Indyk06}).
\begin{lemma}\label{lem:medCauchy}
Let \math{Z_1,\ldots,Z_{r_2}} be \math{r_2} independent Cauchys. Then, 
\math{\frac12\le\median\{|Z_1|,\ldots,|Z_{r_2}|\}\le\frac32}
with probability at least \math{1-2e^{-cr_2}}, where 
\math{c\ge 2(\tan^{-1}(\frac15))^2\ge0.07} is a constant.
\end{lemma}
Fix \math{i} and for \math{j\in[r_2]} define the random variables
\math{Z_j=\Lambda_{ij}} to apply Lemma~\ref{lem:medCauchy}.
Observe that for \math{j\in[r_2]},
\math{Z_j=\Lambda_{ij}=U_{(i)}\Pi_2^{(j)}} 
are independent 
Cauchy random variables
scaled by \math{\norm{U_{(i)}}_1}. Applying
Lemma~\ref{lem:medCauchy} with \math{\lambda_i=
\median_{j\in r_2}|\Lambda_{ij}|}, we have that with 
probability at least \math{1-2e^{-cr_2}}
\mld{
{\textstyle\frac12}\norm{U_{(i)}}_1\le\lambda_i\le{\textstyle\frac32}
\norm{U_{(i)}}_1.\label{app:eq:l1-1}
}
By a union bound, these inequalities hold for all 
\math{i\in[n]} with probability at least 
\math{1-2ne^{-cr_2}\ge 1-\delta} for \math{r_2\ge \frac{1}{c}\log\frac{2n}{\delta}}
(since \math{\frac{1}{c}\le 15}, our algorithm satisfies this condition).
Next we show that if the sampling matrix
preserves the \math{\ell_1}-norm of every vector in the range of \math{U}, 
then
we are done.
\begin{lemma}\label{lem:preservenorm}
Given \math{D} with \math{n} columns,  Suppose that for all 
\math{y\in\R^{q}},
\mld{(1-\eps)\norm{Uy}_1\le\norm{DUy}_1\le(1+\eps)\norm{Uy}_1,
\label{eq:precondition}}
and suppose that \math{\hat y} is a solution to 
\math{\min_{y\in\cl C'}\norm{DUy}}. Then, for all \math{y\in\cl C'},
\mand{
\norm{U\hat y}_1
\le\left(\frac{1+\eps}{1-\eps}\right)\norm{Uy}_1.
} 
\end{lemma}
\begin{proof}
Since \math{D} preserves norms, for any \math{y\in\cl C'} we have that:
\mand{
\norm{U\hat y}_1
\le
\frac{1}{1-\eps}\norm{DU\hat y}_1
{\buildrel (a) \over \le}
\frac{1}{1-\eps}\norm{DU y}_1
\le
\frac{1+\eps}{1-\eps}\norm{U y}_1.
}
(a) is by the optimality of \math{\hat y}.
\end{proof}
The remainder of the proof is to 
show that \math{D} from our algorithm in Figure~\ref{l1reg-mainSimp}
satisfies the pre-condition of Lemma~\ref{lem:preservenorm}, namely
\r{eq:precondition}.
We need two ingredients. 
The first is the \math{\ell_1}-sampling lemma 
Lemma~\ref{lem:L1Chernoff}.
The second ingredient is a standard \math{\gamma}-net argument.

We are going to apply Lemma~\ref{lem:L1Chernoff} with
\math{Z=U}. From \r{app:eq:l1-1} (which holds for all \math{i\in[n]} with 
probability at least \math{1-\delta}), 
\math{\lambda_i/\sum_{i\in[n]}\lambda_i\ge\frac13\norm{U_{(i)}}_1/\norm{U}_1}, 
and so we can apply   Lemma~\ref{lem:L1Chernoff} with 
\math{a=\frac13}. Since
\math{U} is \math{(\alpha,\beta)}-conditioned,
\math{\norm{Uy}_1\ge\frac{1}{\beta}\norm{y}_\infty}, and
\math{\norm{U}_1\le\alpha}, and so we have that with probability at least 
\math{1-\delta},
\mand{
(1-\eps)\norm{Uy}_1
\le
\norm{DUy}_1
\le
(1+\eps)\norm{Uy}_1,
}
where \math{\delta\le2\expon\left(\frac{-s\eps^2}{(6+2\eps)\alpha\beta}\right)}, and \math{\alpha\beta\le\kappa q\sqrt{r_1}}.
If \math{y=\bm0} then the bounds trivially hold; 
by rescaling, it thus suffices to show the bound for all 
\math{y\in\R^{q}} with \math{\norm{y}_\infty= 1}.
We now show this using a standard
\math{\gamma}-net argument. Consider the uniform lattice on 
\math{\R^{q}} specified by \math{T=\frac{\gamma}{q}\mathbb{Z}^{q}}
(we assume that \math{q/\gamma} is a positive integer for simplicity).
Let \math{H=\{z:
\norm{z}_\infty\le 1\}\cap T} be the restriction of this grid to 
only its points within the hypercube of points with
 \math{\ell_\infty}-norm equal to 1;  
\math{|H|\le(\frac{2q}{\gamma})^{q}}. 
Consider any \math{y} with \math{\norm{y}_\infty=1}, and 
let \math{h} be the closest grid point in \math{H} to \math{y}. Then
\mld{
y=h+\frac{\gamma}{q}\sum_{i=1}^{q}\zeta_ie_i,\label{eq:net}}
where \math{0\le|\zeta_i|\le 1}.
Observe that \math{e_i\in H}. By a union bound, for every 
\math{h\in H}, with probability at least \math{1-\delta}, 
\mand{
(1-\eps)\norm{Uh}_1
\le
\norm{DUh}_1
\le
(1+\eps)\norm{Uh}_1,
}
where \math{\delta\le 2|H|\expon\left(\frac{-s\eps^2}{(6+2\eps)\alpha\beta}\right)}. We condition on this high probability event. Then,
\eqan{
\norm{DUy}_1=
\Norm{DUh+\frac{\gamma}{q}\sum_{i=1}^{q}\zeta_iDUe_i}_1
&\le&
\norm{DUh}_1+\frac{\gamma}{q}\sum_{i=1}^{q}|\zeta_i|\norm{DUe_i}_1
\\
&\le&
(1+\eps)\left(
\norm{Uh}_1+\frac{\gamma}{q}\sum_{i=1}^{q}\norm{Ue_i}_1
\right).
}
Applying \math{U} to both sides of \r{eq:net} and using the triangle inequality,
\math{\norm{Uh}_1\le\norm{Uy}_1+
\frac{\gamma}{q}\sum_{i=1}^{q}|\zeta_i|\norm{Ue_i}_1
}.
We conclude that 
\mand{
\norm{DUy}_1\le
(1+\eps)\left(
\norm{Uy}_1+\frac{2\gamma}{q}\sum_{i=1}^{q}\norm{Ue_i}_1
\right)
\le(1+\eps)\norm{Uy}_1
\left(1+\frac{2\gamma\alpha\beta}{q}\right),
}
where we used
\math{\norm{Uy}_1\ge\frac1\beta\norm{y}_\infty
=\frac1\beta} (since \math{\norm{y}_\infty=1}) and
\math{\sum_{i=1}^{q}\norm{Ue_i}_1=\norm{U}_1\le\alpha}.
In an analogous way, we get the lower bound:
\eqan{
\norm{DUy}_1=
\Norm{DUh+\frac{\gamma}{q}\sum_{i=1}^{q}\zeta_iDUe_i}_1
&\ge&
\norm{DUh}_1-\frac{\gamma}{q}\sum_{i=1}^{q}|\zeta_i|\norm{DUe_i}_1
\\
&\ge&
(1-\eps)
\norm{Uh}_1-\frac{\gamma(1+\epsilon)}{q}\sum_{i=1}^{q}\norm{Ue_i}_1\\
&=&
(1-\eps)
\norm{Uh}_1-\frac{\gamma(1+\epsilon)}{q}\norm{U}_1.
}
Again, applying \math{U} to \r{eq:net} and using the triangle
inequality gives
\math{\norm{Uh}_1\ge \norm{Uy}_1-\frac{\gamma}
{q}\norm{U}_1}. Further, 
\math{\norm{Uy}_1\le\norm{U}_1\norm{y}_\infty\le\alpha}, and so
we have
\mand{
\norm{DUy}_1\ge
(1-\eps)\left(
\norm{Uy}_1-\frac{2\gamma}{q(1-\epsilon)}\norm{U}_1
\right)
\ge(1-\eps)\norm{Uy}_1
\left(1-\frac{4\gamma\alpha\beta}{q}\right),
}
where we assume \math{\epsilon\le\frac12}.
Setting \math{\gamma= q\eps/(4\alpha\beta)},  using 
\math{(1+\eps)^2\le 1+3\eps} and
\math{(1-\eps)^2\ge 1-3\eps}
(for \math{\eps <\frac12}), and rescaling \math{\eps} 
by dividing by 3,
we obtain that with probability at least \math{1-\delta},
\mand{
(1-\eps)\norm{Uy}_1\le\norm{DUy}_1\le(1+\eps)\norm{Uy}_1.
}
where \math{\delta\le 2|H|\expon\left(\frac{-s\eps^2}{9(6+2\eps/3)\alpha\beta}\right)}, and \math{|H|\le(\frac{24\alpha\beta}{\eps})^{q}}.
Solving for \math{s} using \math{\alpha\le q\sqrt{r_1}} and 
\math{\beta\le\kappa}, and simplifying a little, we require
\mand{
s\ge\frac{63\kappa q\sqrt{r_1}}{\eps^2}\left(q\log\frac{24
\kappa q\sqrt{r_1}}{\eps}+
\log\frac2\delta\right).
}
The total success probability is \math{1-2\delta}, which results from 
a union bound applied to the 
two random processes involving \math{\Pi_2} and \math{D}.
The Theorem now follows by setting \math{\delta=1/d^\rho} for a constant 
\math{\rho}.
This concludes the proof of the correctness.

\paragraph{Running Time.}
Set \math{\delta=\frac{1}{3q^\rho}}, for \math{\rho=O(1)}.
We compute the running time as follows. In Step~2, 
if we use Theorem~\ref{theorem:FCT}
for \math{\Pione}  
(which succeeds with probability \math{1-\delta}), the time to
compute \math{\Pione X} is
\math{O(nq\log q)} and \math{r_1=O(q\log q)}
and \math{\kappa=O\left(q^{\rho+2}\log q)\right)}
(\math{r_1} and \math{\kappa} affect the running time of later steps);
We need to compute an orthogonal factorization in 
\math{O(r_1q^2)} and then compute \math{\RR} in \math{O(q^3)} for a 
total run time of Step 2 that is 
\math{O((n+q^2)q\log q)}. 
In Step 3, \math{r_2=O(\log n)} by our choice of $r_2$, so the time to compute 
\math{\Lambda=X\RR\Pi_2} is in \math{O(nqr_2+r_2q^2)=
O(nq\log n+q^2 \log n)}, where $O(nq \log n + q^2 \log n)$ is the time needed to compute $R^{-1} \Pi_2$
followed by $X \cdot (R^{-1} \Pi_2)$. Note that $q^2 \log n \leq nq \log n$. 

Since computation of the median of 
\math{r_2} elements is in \math{O(r_2)}, computing all \math{\lambda_i}
takes \math{O(nr_2)=O(n\log n)} time. Thus, the running time for
Steps 1-5 is \math{O(nq\log n)+q^3\log q}.

In Step~6, 
\math{s=O\left(\eps^{-2}
q^{\rho+\frac92}\log^{\frac52}(\frac q\eps)\right)}.
It takes \math{O(n)} time to sample the diagonal matrix
\math{D} and then \math{O(qS)} time to
construct \math{DA} and \math{Db}, where \math{S} is the number of non-zero
entries in \math{D}. Lastly Step 8 takes \math{\phi(S,d)=\Omega(dS)}
time to solve the \math{\ell_1} regression on the smaller problem with
\math{s} rows in \math{d} dimensions. The total running time is thus:
\mand{
O\left(nq\log n +\phi(S,q)\right).
}
where \math{\Exp[S]\le s=O\left(\eps^{-2}
q^{\rho+\frac92}\log^{\frac52}(\frac q\eps)\right)} and \math{S} is very 
tightly concentrated around \math{s} (via a standard Bernstein bound) because 
it is the sum of independent binomial variables; 
specifically, with probability at least \math{1-e^{-\frac38s}},
\math{S\le 2s}.
Hence
\math{S=O(s)} with probability \math{1-o(1)}.
 The probability of
success is \math{1-3\delta=1-\frac{1}{q^\rho}} (union bound).
Since \math{s=\eps^{-2}q^{\rho+\frac92}\poly(\log\frac{q}{\eps})}, and
since standard algorithms for linear programming give
\math{\phi(S,q)=Sq^{O(1)}}, we have the result claimed in the theorem
for \math{q=O(d)}.

%% file: appdx-pfL1faster.tex
\def\Good{\textsf{Good}}
\def\BadRow{\textsf{BadRow}}
\def\AllBounded{\textsf{AllBounded}}
\section{Proof of Theorem~\ref{theorem:FCRsimpFaster} 
(Optimized Fast Cauchy \math{\ell_1} Regression)}
\label{sxn:pfL1faster}

As in the proof of Theorem~\ref{theorem:FCRsimp} in Section ~\ref{sxn:FCRsimp},
given is \math{X\in\R^{n\times q}} and the constraint set \math{\cl C}.
We condition on \math{\Pione\in\R^{r_1\times n}}
satisfying (\ref{eq:main}) and Theorem \ref{thm:basisL1}. 
So, \math{U=XR^{-1}} is 
\math{(q\sqrt{r_1},\kappa)}-conditioned where \math{r_1} and \math{\kappa}
depend on \math{n,q,\delta} (this holds 
with probability at least \math{1-\delta}). Thus,
\math{\norm{U}_1\le q\sqrt{r_1}} and 
\math{\norm{x}_\infty\le\kappa\norm{Ux}_1} for any \math{x\in\R^q}. It follows that
\mand{
\frac{q}{\kappa}\le\norm{U}_1\le q\sqrt{r_1}.
}
(The lower bound follows from
\math{\sum_{i\in[q]}\norm{e_i}_\infty\le\sum_{i\in[q]}\norm{Ue_i}_1}, where
\math{e_i} are standard basis vectors.)
In the proof of Theorem~\ref{theorem:FCRsimp}, we proved the 
following result. Given \emph{weights}
\math{t_i}, with 
\mld{
t_i\ge a\cdot\frac{\norm{U_{(i)}}_1}{\norm{U}_1} \qquad \forall i\in[n],
\label{eq:app:leverage1}
}
define leverage probabilities 
$$\ell_i = \min \left (1, s \cdot  t_i \right ),$$
with 
\mld{
s\ge\frac{63\kappa q\sqrt{r_1}}{a \eps^2}\cdot
\left(q\log\frac{4
q\sqrt{r_1}\max(q\sqrt{r_1},\kappa)}{\eps}+
\log\frac2\delta\right).\label{app:eq:s}
}
and construct the random diagonal sampling matrix 
\math{D} with \math{D_{ii}=1/ \ell_i} with probability \math{ \ell_i}
and zero otherwise.
Then, with probability at least \math{1-\delta}, solving the coreset problem
 given by
\math{D X} and the constraints \math{\cl C} gives a \math{(1+\eps)/(1-\eps)}
approximate solution to the full \math{L_1} regression with 
\math{X} and \math{\cl C}.
In the proof of Theorem~\ref{theorem:FCRsimp}, 
the purpose of \math{\Pi_2} was to allow us to construct weights
\math{t_i} quickly such that with high probability,
\math{\frac{1}{a}\le 3}.
Here, we show that our faster way to get weights, results in only a 
\math{\poly(d\epsilon^{-1}\log n)} 
factor increase in \math{\frac{1}{a}}.

Recall \math{\Lambda=U\Pi_2}, where \math{\Pi_2\in\R^{n\times r_2}} is a matrix
of \math{i.i.d.} standard Gaussian random variables,
and \math{\hat\lambda_i=\median_{j\in[r_2]}|\Lambda_{ij}|}, with
and
\mand{
\hat p_i = \min \left (1, s \cdot \hat \lambda_i \right ).
}
For \math{j\in[r_2]}, the \math{\Lambda_{ij}} are i.i.d. zero mean Gaussians
with variance \math{\norm{U_{(i)}}_2^2}, so
\math{ |\Lambda_{ij}|} are i.i.d half Gaussians.
We need a result from \cite{Indyk06}.
\begin{lemma}\label{lemma:indyk}(Lemma 2 of \cite{Indyk06}). 
Let \math{x_1,\ldots,x_r} be  i.i.d. with continuous distribution 
function \math{F} and \math{\lambda=\median_{i\in[r]}x_i}. Then,
\eqan{
\Prob\left[\lambda\ge F^{-1}({\textstyle\frac12}-\epsilon)\right]
\ge {1-\expon({-2\epsilon^2r})},&
\\
\Prob\left[\lambda\le F^{-1}({\textstyle\frac12}+\epsilon)
\right]
\ge {1-\expon({-2\epsilon^2r})},&
}
\end{lemma}
For the half Gaussian with variance \math{\sigma^2},
\math{F(x)=2\phi(x/\sigma)-1} where \math{\phi} is the standard Gaussian
distribution function. Choosing \math{\epsilon=\frac14} and 
using Lemma~\ref{lemma:indyk}, with probability at least
\math{1-\expon({-r_2/2})},
\mand{
\hat\lambda_i\ge 0.3\cdot\norm{U_{(i)}}_2
}
where we used
\math{0.3<\phi^{-1}({\textstyle\frac58})}.
Using \math{\norm{U_{(i)}}_2\ge\norm{U_{(i)}}_1/\sqrt{q}} and
\math{\norm{U}_1\ge q/\kappa},
it
follows that
\mld{
\hat \lambda_i
\ge\frac{0.3}{\sqrt q}\norm{U_{(i)}}_1
\ge
\frac{0.3\sqrt{q}}{\kappa}\frac{\norm{U_{(i)}}_1}{\norm{U}_1}
\label{eq:app:leverage2}
}
holds with probability at least \math{1-2\expon({-r_2/2})} for any
particular \math{i}.
If we required these bounds to hold for all \math{i\in[n]}, then to apply the
union bound succesfully, we would need to set \math{r_2=\Omega(\log n)},
which is too costly. We want \math{r_2=O(\log(d\epsilon^{-1}\log n))},
so we need a more subtle argument.
We choose \math{s} as in \r{app:eq:s} with \math{a=0.3\sqrt{q}/\kappa}.

Let \math{t_i=\norm{U_{(i)}}_1} and  
\math{p_i=\min(1,s\cdot t_i)} 
be sampling probabilities obtained from the 
the exact \math{L_1} leverage scores for
\math{U}. For these sampling probabilities, \math{a} in  
\r{eq:app:leverage1} is larger which would imply that a smaller \math{s} 
is needed. 
Nevertheless,  any larger value of \math{s} will also
work, and so the  same value of 
\math{s} with \math{a=0.3\sqrt{q}/\kappa} will work with 
the weights~\math{t_i}. Note that since 
\math{\Pione} is fixed, \math{p_i} is not a random variable, but 
\math{\hat p_i} is a random variable depending on~\math{\Pi_2}.
Fix \math{r_2} and generate \math{\Pi_2} and thence 
\math{\hat \lambda_i,\hat p_i}.

We define a set of indices \math{T\subseteq[n]} as those \math{i} for which
\math{\hat \lambda_i\ge(0.3\sqrt{q}/\kappa)\norm{U_{(i)}}_1/\norm{U}_1}.
These are the indices for which \math{\Pi_2} `worked'. Essentially, these
are the large leverage scores. The intuition behind our 
argument is that even though
there may be some indices for which \math{\Pi_2} did not work, there are enough
large leverage scores for which \math{\Pi_2} did work that the probability of
these faulty indices coming into play is miniscule.

To make this argument, we 
define hybrid weights \math{w_i} to equal 
\math{\hat \lambda_i} for \math{i\in T} and \math{t_i} for \math{i\not\in T}.
By construction,
\mand{
w_i\ge\frac{0.3\sqrt{q}}{\kappa}\frac{\norm{U_{(i)}}_1}{\norm{U}_1},
}
and so the same \math{s} works for constructing sampling probabilities
\math{q_i=\min(1,s\cdot w_i)}. 
Note that in the algorithm, we do not
actually construct (or know) \math{p_i,q_i}; 
they are just used here as a hypothetical
set of sampling probabilities which help us to analyze the 
performance of the actual sampling probabilities we use, which are
\math{\hat p_i}. The important property about the \math{q_i} is that 
for \math{i\in T}, \math{q_i=\hat p_i}.

We call a set of rows that are sampled and rescaled according to a set of
probabilities a {\it good} coreset if the coreset
 solution from this sample is
 a $(1+\eps)$-approximation to the full \math{L_1} regression.
The sampling probabilities \math{q_i} give a good coreset 
with probability at least \math{1-\delta}.
We now define several events over three random processes: \math{\Pi_2}, 
sampling a coreset according to \math{\hat p_i} and sampling a coreset
according to \math{q_i}. The last two random processes depend on the outcome of
\math{\Pi_2}.
\begin{itemize}
\item {\sf AllBounded} is the event
$\{\hat\lambda_i\le C\sqrt{\log n} \cdot \norm{U_{(i)}}_2\ \forall i\in[n]\}$
(we will choose \math{C} later).
We show 
$$\Pr_{\Pi_2}[{\sf AllBounded}] \geq 1-\frac{1}{n^{\frac12C^2-1}}.$$
Indeed,
\math{\hat\lambda_i} is the median of \math{r_2} i.i.d. zero mean
 Gaussians \math{x_1,\ldots,x_{r_2}} with variance \math{\norm{U_{(i)}}_2^2},
where
\math{\Prob\left[x_i>C\sqrt{\log n}\norm{U_{(i)}}_2\right]
=1-\phi(C\sqrt{\log n})
\le 1/(\sqrt{\pi} n^{C^2/2})} (by the properties of the Gaussian distribution).
Define \math{z_i=1} if \math{x_i > C\sqrt{\log n}} and 0 otherwise.
Then \math{\hat\lambda_i>C\sqrt{\log n}} if and only if \math{\sum_{i\in[r_2]}
z_i>r_2/2}. We have \math{\Exp\left[\sum_{i\in[r_2]}
z_i\right]\le r_2/(\sqrt{\pi} n^{C^2/2})} and the result follows by a Markov
bound and a union bound over \math{i\in[n]}.

\item Let \math{\Good(q)} be the event that the coreset sampled according to
probabilities $q_i$
are good. 

\item Let \math{\Good(\hat p)} be the event that the coreset sampled
according to $\hat p_i$ are good. 

\item Let {\sf BadRow} be the event that
either of the two coresets above contains a row $i \notin T$.
\end{itemize}
In what follows, we consider probabilities with respect to the joint
distribution of \math{\Pi_2} and the randomness of choosing the coresets 
according to \math{q_i} and \math{\hat p_i}.
\eqan{
\Prob[\Good(q)]
&\le&
\Prob[\BadRow]+\Prob[\Good(q)|\neg\BadRow](1-\Prob[\BadRow])
\\
&=&
\Prob[\BadRow]+\Prob[\Good(\hat p)|\neg\BadRow](1-\Prob[\BadRow])\\
&\le&
\Prob[\BadRow]+\Prob[\Good(\hat p)],
}
where the second step 
follows because conditioning on \math{\neg\BadRow}, the sampling probabilities
\math{\hat p_i} and \math{q_i} are identical (by construction).
Thus,
\eqan{
\Prob[\Good(\hat p)] 
&\geq& \Prob[{\Good(q)}] - \Prob[{\BadRow}]\\
&\ge& 1-\delta -\Prob[{\BadRow}],
}
because we know that the sampling probabilities \math{q_i} satisfy the 
conditions to get a good coreset with probability at least \math{1-\delta}.
To get an upper bound on 
\math{\Prob[\BadRow]}, observe that
\eqan{
\Prob[{\BadRow}]&\le&
\Pr[\neg {\AllBounded}]+
\Pr[{\BadRow} | {\AllBounded}]\\
&\le&
\frac{1}{n^{\frac12C^2-1}}+\Pr[{\BadRow} | {\AllBounded}].
}
To conclude, we obtain a bound on 
$\Pr[{\BadRow} | {\AllBounded}]$. 
Condition on  
$\Pi_2$ and that
{\sf AllBounded} holds.  This fixes $T$ and $\hat p_i$ and also 
means that \math{\hat p_i\le C\sqrt{\log n}q_i}. Hence,
\mld{\Pr[{\BadRow} | {\AllBounded}]
\le \sum_{i \notin T} (\hat p_i+q_i) 
= (1+C\sqrt{\log n}) \cdot \sum_{i \notin T} q_i
= (1+C\sqrt{\log n}) \cdot \sum_{i \notin T} p_i,\label{app:eq:badrow}}
where the last equality is because \math{q_i=p_i} for \math{i\not\in T}.
So the bound is determined by the sum of the leverage scores over the
indices for which \math{\Pi_2} did not work. This is the quantification of
our intuition that the algorithm will work as long as \math{\Pi_2} preserves
enough of the large leverage scores.
We need to bound \math{\sum_{i \notin T} p_i}, where \math{T} is a random
set of indices depending on \math{\Pi_2}.
We will use a Markov bound to bound \math{\sum_{i \notin T} p_i} 
with high probability.
We have
\eqan{
\Exp_{\Pi_2}\left[\sum_{i \notin T} p_i\right]
&\ge&
\Exp_{\Pi_2}\left[\sum_{i \notin T} p_i \ \mid \ {\AllBounded}\right]
\Prob[\AllBounded]\\
&\ge&\Exp_{\Pi_2}\left[\sum_{i \notin T} p_i \ \mid \ {\AllBounded}\right]
\cdot\left(1-\frac{1}{n^{\frac12 C^2-1}}\right)
}
Since \math{\Prob[i\not\in T]\le\expon({-r_2/2})},
\mand{
\Exp_{\Pi_2}\left[\sum_{i \notin T} p_i\right]
=
\sum_{i\in[n]} p_i\cdot
\Prob[i\not\in T]\le e^{-r_2/2}\sum_{i\in[n]} p_i.
}
But, \math{\sum_{i\in[n]} p_i\le s\sum_{i\in[n]}t_i=s\norm{U}_1\le
sq\sqrt{r_1}}, where the last step follows from the conditioning of
\math{U} which is assumed. Putting all this together,
\mand{
\Exp_{\Pi_2}\left[\sum_{i \notin T} p_i \ \mid \ {\AllBounded}\right]
\le
\frac{\displaystyle sq\sqrt{r_1}e^{-r_2/2}}{1-{n^{1-\frac12 C^2}}}.
\le
2sq\sqrt{r_1}e^{-r_2/2},
}
where the last expression follows by setting \math{C=2} in which case
\math{1-1/n\ge \frac12}.
Now, recalling $\rho > 0$ is given as in the theorem statement, if we set 
\mand{
r_2=2\log\left(2sq\sqrt{r_1}\log^{2\rho+1/2} n\right)=
O\left(\log(d\epsilon^{-1}\log n)\right),
}
then, \math{\Exp_{\Pi_2}\left[\sum_{i \notin T} p_i \ \mid 
\ {\AllBounded}\right]\le 1/\log^{2\rho+1/2} n}. Applying a Markov bound and
conditioning on \math{\AllBounded}, with probability at most
\math{1/\log^\rho n}, the bound
\math{\sum_{i \notin T} p_i>1/\log^{\rho+1/2}n} holds.
Condition on this bad event not happening, in which case, 
using \r{app:eq:badrow},
\mand{\Pr[{\BadRow} | {\AllBounded}]
\le
\frac{1+2\sqrt{\log n}}{\log^{\rho+3/2}n}
=O(1/\log^\rho n),}
where we used \math{C=2}.
Using a union bound over this bad event not happening, we finally have that 
\mand{
\Prob[\BadRow]\le
\frac{1}{n}+\frac{1}{\log^{\rho+1/2} n}+\frac{1+2\sqrt{\log n}}{\log^{\rho + 3/2}n},
}
from which \math{\Prob[\Good(\hat p)]\ge 1-\delta-O(\log^{-\rho}n)}.
This completes the proof.

%% file: appdx-pfEllRnd.tex
\section{Proof of the Fast Ellipsoidal Rounding Theorems}
\label{sxn:app-roundingpf}

\subsection{Proof of Theorem~\ref{thm:rounding} (Fast Ellipsoidal Rounding)}
\label{sxn:app-rounding1pf}

For completeness, we state the following lemma which is 
from~\cite{todd1982minimum} and which we will use in the
proof of this theorem.

\begin{lemma}
  \label{lemma:todd}
  (Todd~\cite{todd1982minimum}) Given an ellipsoid $\mathcal{E} = \{ u \in
  \mathbb{R}^d \,|\, u^T E^{-1} u \leq 1 \}$ where $E \in \mathbb{R}^{d \times
    d}$ is symmetric positive-definite and $\mathcal{K} = \{ u \in \mathbb{R}^d
  \,|\, - \beta (g^T E g)^{1/2} \leq g^T u \leq \beta (g^T E g)^{1/2} \}$ for
  some $g \in \mathbb{R}^d$, the minimum-volume ellipsoid that contains
  $\mathcal{E} \cap \mathcal{K}$ is given by
  \begin{equation*}
    \mathcal{E}_+ =
    \begin{cases}
      \mathcal{E} & \text{if } \beta \geq d^{-1/2} \\
      \{ u \in \mathbb{R}^d \,|\, u^T E_+^{-1} u \leq 1 \} & \text{if } 0 < \beta < d^{-1/2},
    \end{cases}
  \end{equation*}
  where
  \begin{align*}
    E_+ &= \delta \left( E - \sigma \frac{(E g) (E g)^T}{g^T E g} \right), \\
    \quad
    \delta &= \frac{d (1-\beta^2)}{d-1}, \quad \sigma = \frac{1-d \beta^2}{1 - \beta^2}.
  \end{align*}
  When $\beta < d^{-1/2}$, we have
  \begin{equation*}
    \frac{|\mathcal{E}_+|}{|\mathcal{E}|} 
    = d^{1/2} \left( \frac{d}{d-1} \right)^{(d-1)/2} \beta (1 - \beta^2)^{(d-1)/2}.
  \end{equation*}
\end{lemma}

Now we proceed with the main part of the proof.
  We construct a sequence of ellipsoids $\mathcal{E}_1,\mathcal{E}_2,\ldots$,
  all centered at the origin, such that $\mathcal{E}_k \supseteq \mathcal{C}$
  and $|\mathcal{E}_{k}|/ |\mathcal{E}_{k-1}| < e^{3/8}/2,\ k=1,2,\ldots$, and
  thus this sequence must terminate in 
  $$\log( L^{-d} )/\log(e^{3/8}/2) < 3.15 d \log L$$ 
  steps. Suppose we have $\mathcal{E}_k \supseteq \mathcal{C}$ centered
  at the origin. Determine all the extreme points of $\mathcal{E}_k$ along its
  axes. Let these points be $\pm x_{k, i},\ i=1,\ldots,d$, and then check
  whether $\frac{1}{2 \sqrt{d}} x_{k, i} \in \mathcal{C}$ for $i=1,\ldots,d$. If
  all these points are in $\mathcal{C}$, so is their convex hull, denoted by
  $\mathcal{H}$. Apparently, $\frac{1}{2 \sqrt{d}} \mathcal{E}_k$ is the LJ
  ellipsoid of $\mathcal{H}$, and hence shrinking $\frac{1}{2 \sqrt{d}}
  \mathcal{E}_k$ by a factor $\frac{1}{\sqrt{d}}$ makes it contained in
  $\mathcal{H} \subseteq \mathcal{C}$. We have $\frac{1}{2 d} \mathcal{E}_k
  \subseteq \mathcal{C} \subseteq \mathcal{E}_k$. Now suppose that $\frac{1}{2
    \sqrt{d}} x_{k, i_k} \notin \mathcal{C}$ for some $i_k$ and the separation
  oracle returns $\mathcal{K}_k = \{ x \in \mathbb{R}^d \,|\, -1 \leq g_k^T x
  \leq 1\}$ such that $\mathcal{C} \subseteq \mathcal{K}_k$ but $\frac{1}{2
    \sqrt{d}} x_{k, i_k} \notin \mathcal{K}_k$. Let $\mathcal{E}_{k+1}$ be the
  LJ ellipsoid of $\mathcal{E}_k \cap \mathcal{K}_k \supseteq \mathcal{C}$,
  which must be centered at the origin.  
  Lemma~\ref{lemma:todd} gives analytic
  formulas of $\mathcal{E}_{k+1}$ and
  $|\mathcal{E}_{k+1}|/|\mathcal{E}_k|$. 
  Adopting the notation from Lemma~\ref{lemma:todd}, 
  let $\mathcal{E}_k = \{ x \in \mathbb{R}^d \,|\, x^T E_k^{-1} x \leq 1\}$ 
  and we~have
  \begin{align*}
    (g_k^T E_k g_k)^{1/2} &= \left[ g_k^T \left(\sum_{i=1}^d x_{k,i} x_{k,i}^T \right) g_k \right]^{1/2} \\
    &\geq |g_k^T x_{k, i_k}| > 2 \sqrt{d}.
  \end{align*}
  The last inequality comes from the fact that $\frac{1}{2 \sqrt{d}} x_{k, i_k}
  \notin \mathcal{K}_k$.  
  Therefore $\beta = (g_k^T E_k g_k)^{-1/2} < \frac{1}{2 \sqrt{d}}$,~and
  \begin{equation*}
    \frac{|\mathcal{E}_{k+1}|}{|\mathcal{E}_{k}|} < \frac{1}{2} \left( 1 + \frac{3}{4d - 4} \right)^{(d-1)/2} < e^{3/8}/2.
  \end{equation*}
  Thus, our construction is valid.  For each step, it takes at most $d$ calls to
  the separation oracle. Therefore, we need at most $3.15 d^2 \log L$ calls to
  find a $2d$-rounding of $\mathcal{C}$. Computing the extreme points of
  $\mathcal{E}_k$ requires an eigendecomposition, which takes $O(d^3)$
  time. Hence the total cost to find a $2 d$-rounding is $3.15 d^2 \log L$ calls
  and additional $O(d^4 \log L)$ time. We note that rank-one updates
  can be used for computing the eigendecomposition of $\mathcal{E}_k$ for
  efficiency. See Gu and Eisenstat~\cite{gu1994stable}.

\subsection{Proof of Theorem~\ref{thm:lp_cond_2d}}
\label{sxn:app-rounding2pf}

This is a direct consequence of Theorem \ref{thm:rounding}.  We present the
proof for the case $p < 2$. The proof for the case $p > 2$ is similar. Let
$\mathcal{C} = \{ x \in \mathbb{R}^d \,|\, \|A x\|_p \leq 1\}$. For any $z
\notin \mathcal{C}$, define $\mathcal{K}(z) = \{ x \in \mathbb{R}^d \,|\, -1
\leq g(z)^T x \leq 1 \}$, where $g(z)$ is a subgradient of $\|A x\|_p$ at $x = z$. 
We have $\mathcal{K}(z) \supseteq \mathcal{C}$ and $z
\notin \mathcal{K}(z)$, which gives the separation oracle. Let $A = Q R_0$ be
$A$'s QR factorization. We have,
  \begin{align*}
    \|R_0 x\|_2 &= \|A x\|_2 \leq  \|A x\|_p \leq n^{1/p-1/2} \|A x\|_2 \\
    &= n^{1/p-1/2} \|R_0 x\|_2, \quad \forall x \in \mathbb{R}^d,
  \end{align*}
  which means $\mathcal{E}_0 = \mathcal{E}(0, R_0^{-1})$ gives an
  $n^{1/p-1/2}$-rounding of $\mathcal{C}$. 
  Applying Theorem \ref{thm:rounding},
  we can find a $2 d$-rounding of $\mathcal{C}$ in at most $3.15 d^2 \log
  (n^{1/p-1/2})$ calls to the separation oracle. Let $\mathcal{E} =
  \mathcal{E}(0, E)$ be the ellipsoid that gives such rounding. We have
  \begin{equation*}
    \|y\|_2 \leq \| A E y \|_p \leq 2 d \|y\|_2, \quad \forall y \in \mathbb{R}^d.
  \end{equation*}
  The QR factorization takes $O(n d^2)$ time. Each call to the
  separation oracle takes $O(n d)$ time. Computing the extreme points
  of an ellipsoid takes $O(d^3)$ time. In total, we need
  $O(n d^3 \log n)$~time.

%% file: appdx-pfLpCond.tex
\section{Proof of Theorem~\ref{thm:lprunning}}
\label{sxn:app-onepasspf}

The tool we need to verify the \textsf{FastLpBasis} algorithm is simply the
equivalence of vector norms. We present the proof for the case $p < 2$. The proof
for the case $p > 2$ is similar. Adopt the notation from the
\textsf{FastLpBasis} algorithm. $G$ is chosen such that, with a constant
probability,
\begin{equation*}
  \theta_1 \|A_i x\|_2 \leq \|\tilde{A}_i x\|_2 \leq \theta_2 \|A_i x\|_2, \quad i=1,\ldots,N,
\end{equation*}
where $\theta_1 > 0$ and $\theta_2 > 0$ are constants. Conditioning on this event, we have
\begin{equation}
  \label{eq:init_rounding}
  t^{1/p-1/2}/\theta_1 \cdot \tilde{\mathcal{C}} \subseteq \mathcal{C} \subseteq 1/\theta_2 \cdot \tilde{\mathcal{C}},
\end{equation}
where $\mathcal{C} = \{ x \,|\, \|A x\|_p \leq 1\}$, because for all $x \in
\mathbb{R}^d$,
\begin{align*}
  \|A x\|_p^p &= \sum_{i=1}^N \|A_i x\|_p^p \leq t^{1-p/2} \sum_{i=1}^N \|A_i x\|_2^p 
  \leq t^{1-p/2} /\theta_1^p \cdot \sum_{i=1}^N \|\tilde{A}_i x\|_2^p  ,
\end{align*}
and
\begin{align*}
  \|A x\|_p^p &= \sum_{i=1}^N \|A_i x\|_p^p \geq \sum_{i=1}^N \|A_i x\|_2^p \geq
  1/\theta_2^p \cdot \sum_{i=1}^N \|\tilde{A}_i x\|_2^p.
\end{align*}
Let $R_0$ be the $R$ matrix from the QR decomposition of $\tilde{A}$ and define
$\mathcal{E}_0 = \{x \,|\, s^{1/p-1/2} \|R_0 x\|_2 \leq 1\}$. We show that $\mathcal{E}_0$
gives an $(N s)^{1/p-1/2}$-rounding of $\tilde{C}$. For all $x \in
\mathbb{R}^d$, we have
\begin{align*}
  \left( \sum_{i=1}^N \|\tilde{A}_i x\|_2^p \right)^{1/p} \leq N^{1/p-1/2} \left( \sum_{i=1}^N \|\tilde{A}_i x\|_2^2 \right)^{1/2} = N^{1/p-1/2} \|R_0 x\|_2
\end{align*}
and
\begin{align*}
  \left(\sum_{i=1}^N \|\tilde{A}_i x\|_2^p \right)^{1/p} \geq s^{1/2-1/p} \left( \sum_{i=1}^N \|\tilde{A}_i x\|_p^p \right)^{1/p} = s^{1/2-1/p} \| \tilde{A} x\|_p \geq s^{1/2-1/p} \|\tilde{A} x\|_2 = s^{1/2-1/p} \|R_0 x\|_2.
\end{align*}
Hence $\mathcal{E}_0$ gives an $(N s)^{1/p-1/2}$-rounding of
$\tilde{\mathcal{C}}$. Then we compute a $(2 d)$-rounding of
$\tilde{\mathcal{C}}$ and obtain the matrix $R$. The running time is
$\mathcal{O}(Ns d^3 \log(N s) ) = \mathcal{O}(n d \log n)$ since $N s = ns/t =
n/d^2$. Then by \eqref{eq:init_rounding}, we know $\kappa_p(A R^{-1}) =
\mathcal{O}(d t^{1/p-1/2})$.

%% file: appdx-pfSWfixed.tex
\section{Proof of Theorem~\ref{thm:sw}}
\label{sxn:app-pf-thm-sw}

\emph{Upper Bound.} First we prove the upper bound.
Let \math{U} be a \math{\ell_1} \math{(d,1)}-conditioned basis for 
\math{L} (see Section~\ref{sxn:resultsL1-basis}). Therefore, 
\math{\norm{U}_1\le d}, \math{\norm{x}_\infty\le\norm{Ux}_1} for all
\math{x\in\R^d}, and for any \math{y\in L}, \math{y=Ux} for some \math{x}. 
Let \math{y\in L}; we have,
 $$\norm{Ry}=\mynorm{R Ux}_1 \le \mynorm{R U}_1 \mynorm{x}_\infty \le  
\mynorm{R U}_{1} \mynorm{Ux}_1=\mynorm{R U}_{1} \mynorm{y}_1  .
$$
Thus, it suffices to prove an upper bound on \math{\mynorm{RU}_1}.
\math{(RU)_{ij}=\sum_{k}R_{ik} U_{kj}} is a Cauchy scaled by 
\math{\gamma_{ij}=\norm{U^{(j)}}}. So 
\math{\mynorm{RU}_1} is a sum of \math{r_1 d}
scaled, dependent half-Cauchys with 
sum of scalings \math{\gamma=\sum_{i,j}\norm{U^{(j)}}=r_1\norm{U}_1}.
By Lemma~\ref{lem:tail},
\mand{
\Prob[\norm{RU}_1>tr_1\norm{U}_1]
\le
\frac{(\log(r_1 d)+\log t)}{t}\left(1+o(1)\right)
.
}
It suffices to set \math{t=O(\frac1p\log(r_1 d))} for the RHS to be 
at least \math{1-\delta}. Since \math{\norm{U}_1\le d}, 
with probability at least 
\math{1-\delta}, \math{\norm{RU}_1=O(\frac{r_1 d}{\delta} \log(r_1 d))}. Multiplying
both sides by \math{C=4/r_1} gives the upper bound.

\vskip\baselineskip

\noindent\emph{Lower Bound.}
The lower bound is essentially following the proof of the lower
bound in Theorem 5 of \cite{SW11}, and so we only provide a sparse 
sketch of the proof.
Consider an arbitrary, fixed \math{y}. The product
\math{CRy} is distributed as a Cauchy random vector whose components are
independent and scaled by \math{C\norm{y}_1}. 
Therefore
\mand{
\norm{CRy}_1=C\norm{y}_1\sum_{i=1}^{r_1}|X_i|,}
where \math{X_i} are i.i.d. Cauchy random variables. 
We now apply 
Lemma~\ref{lem:lower} with 
\math{\gamma=r_1C\norm{y}_1}, \math{\beta^2=r_1} and setting 
\math{t=\frac12}, to obtain
\mand{
\Prob\left[\norm{CRy}_1\le \frac12 r_1 C\norm{y}_1\right]
\le \expon\left(-r_1/12\right).
}
Since \math{C=4/r_1}, we have 
\math{\Prob\left[\norm{CRy}_1\le 2\norm{y}_1\right]
\le \expon\left(-r_1/12\right).}
The result now follows 
by putting a $\gamma$-net $\Gamma$ on  \math{L} for sufficiently
small \math{\gamma}.
This argument follows the same line as the end of 
Section 3 of \cite{SW11}.

It suffices to show the result for \math{\norm{y}_1= 1}.
Consider the 
\math{\gamma}-net
on \math{L} with cubes of side \math{\gamma/d}.
There are \math{(2d/\gamma)^d} such cubes required to cover the
hyper-cube \math{\norm{y}_\infty\le 1}; and,
 for any two points \math{y_1,y_2} inside the same 
\math{\gamma/d}-cube, \math{\mynorm{y_1-y_2}_1\le\gamma}. 
From each of the \math{\gamma/d}-cubes, select a fixed representative point
which we will generically refer to as \math{y^*}; select the
representative to have 
\math{\mynorm{y^*}_1= 1} if possible.
By a union bound 
\mand{\Prob\left[\min_{y^*}\norm{CR y^*}_1/\norm{y^*}_1 <
2\right]
\le (2d/\gamma)^d\expon(-r_1/12).}
We will thus condition on the high probability event that
\math{\norm{CR y^*}_1\ge \norm{y^*}_1} for all \math{y^*}.
We will also condition on the upper bound holding (which is 
true with probability at least \math{1-\delta}).
For any \math{y\in L}
with \math{\norm{y}_1=1},
let \math{y^*} denote the representative point for the cube in which 
\math{y} resides (by construction, \math{\norm{y^*}_1=1} as well). Then 
\math{\norm{y-y^*}\le \gamma} and \math{y-y^*\in L} since \math{y,y^*\in L}
and \math{L} is a subspace. We have
\mand{\norm{CR y}_1=\norm{CR y^*+CR(y-y^*)}_1
\ge\norm{CR y^*}_1-\norm{CR(y-y^*)}_1
\ge 2\norm{y^*}_1-\kappa\norm{y-y^*}_1,
}
where  we used the upper bound in 
the last inequality  
\math{\kappa'=\frac1\delta\cdot O(d\log (r_1 d))}.
By choosing \math{\gamma=1/\kappa'} and
recalling that \math{\norm{y^*}_1=1}, we have that
\math{\norm{CR y}_1\ge 1}, with probability at least
\math{1-\delta-\expon(-r_1/12+d\log (2d\kappa'))}. 
Recall that \math{\kappa'=O(\frac{d}{\delta}\log (r_1d))}, so,
for \math{c} large enough, 
by picking \math{r_1=c\cdot d\log \frac{d}{\delta}},
we satisfy \math{\frac{r_1}{12}\ge \log\frac1\delta+d\log(2\frac{d^2}{\delta}\log(r_1 d))},
and so our bounds hold with probability 
at least \math{1-2\delta}.

%% file: appdx-pfSubspace.tex
\section{Proof of Lemma~\ref{thm:subspace}}
\label{sxn:app-pf-thm-subspace}

We will need some lemmas from prior work.
The first two lemmas are on properties of a \math{\gamma}-net, taken
directly from Lemma 4 of \cite{ahk06}. 
Let \math{U\in\R^{t\times d}} be a matrix whose columns are an orthonormal
basis for \math{L};
let \math{S} be the unit sphere in
\math{\R^d} 
and let \math{T} be the set of points in \math{S_L}, the intersection of $L$ and $S$, defined by
\mand{
T =\left\{w : w \in \frac{\gamma}{\sqrt{d}} \mathbb{Z}^d, \ \|w\|_2 \leq 1 
\right\},
}
where \math{\mathbb{Z}^d} is the \math{d}-dimensional integer lattice on 
(the orthonormal basis for) \math{L}. The set \math{T} is a \math{\gamma}-net on \math{S_L} because every
point in \math{S_L} is at most \math{\ell_2}-distance \math{\gamma} from some
point in \math{T}.
\begin{lemma}[Lemma 4 of \cite{ahk06}]\label{lem:netsize}
$|T|\le e^{cd}$ for $c = (\frac{1}{\gamma} + 2)$.
\end{lemma}
\begin{lemma}[Lemma 4 of \cite{ahk06}]\label{lem:netprod}
For any $d \times d$ matrix $M$, 
if for every $u, v \in T$ we have $|u^TMv| \leq \eps$, 
then for every unit vector $w$, we have
$|w^TMw| \leq \frac{\eps}{(1-\gamma)^2}$. 
\end{lemma}
Note that as $\gamma \rightarrow 0$, the inequality in Lemma \ref{lem:netprod} gets stronger,
but the bound on $|T|$ in Lemma \ref{lem:netsize} gets larger.

The next lemma demonstrates that a JLP distribution preserves matrix products.
\begin{lemma}[Theorem 19 of \cite{KN12}]\label{lem:conjl}
For \math{\eps\in(0,\frac12]},
let $G$ be an $s\times t$ matrix be drawn from an MJLP distribution 
as given in Definition~\ref{def:JLP}. Then for $A, B$ any 
real matrices with $t$ rows and
$\|A\|_F = \|B\|_F = 1$, 
$$\Prob_{G}[\|A^TG^TGB - A^TB\|_F > 3\eps/2] < c_1e^{-c_2s\eps^2}.$$
\end{lemma}

We now prove the first part of Lemma~\ref{thm:subspace}.
Let \math{M} be the $d \times d$ matrix
$M = U^TG^TGU - I$, and let \math{T} be the \math{\gamma}-net with
\math{\gamma=\frac12}. By Lemma~\ref{lem:netsize},
\math{|T|\le e^{4d}}.
Let $u, v \in T$ be any two points in \math{T}, and set
$A = Uu$, $B = Uv$ to be two 
matrices (actually vectors) with \math{t} rows.
Since \math{U} has orthonormal columns, \math{\norm{A}_F=\norm{B}_F=1}.
By Lemma~\ref{lem:conjl}, after relabeling \math{3\eps/2\rightarrow\eps},
$$\Prob_G[|u^T U^TG^TGUv - u^Tv| > \eps] \leq c_1e^{-4c_2s\eps^2/9}.$$
So, applying the union bound,
for every pair \math{x,y\in T}, 
\mand{
|x^TU^TGG^TUy-x^Ty|=|x^TMy|\le \eps
}
holds with probability at least \math{1-c_1|T|^2e^{-4c_2s\eps^2/9}}.
Let \math{G} be the \math{s\times t} MJLP 
matrix constructed as per Lemma~\ref{thm:al}. 
We will now derive a bound 
on~\math{s} for the first result (2-norm) to hold. 
For every unit-norm \math{x} in \math{L}, \math{x=Uw} for unit norm
\math{w\in\R^d}.
By Lemma~\ref{lem:netprod} (with 
\math{\gamma=\frac12}), 
for every unit vector \math{w\in S_L},
$$|w^TU^TG^TGUw - \|w\|_2^2| \leq 4\eps.$$
Since \math{w^TU^TG^TGUw=\norm{Gx}_2^2} and 
\math{\norm{w}_2^2=\norm{x}_2^2}, after rescaling 
\math{4\eps\rightarrow\eps},
 we have proved that with probability
at least \math{1 - c_1e^{8d}e^{-c_2 s 4 \eps^2/(9 \cdot 16)} = 1-c_1 e^{8d}e^{-c_2s\eps^2/36}},
\mand{
\sqrt{1-\eps}\norm{x}_2\le\norm{Gx}_2\le \sqrt{1+\eps}\norm{x}_2.
}

We now derive the second result (Manhattan norm), conditioning on the
high probability event that the result holds for the 2-norm as proved above.
Since \math{G} is an MJLP, we also have that with probability
at least \math{1-c_1|T|e^{-c_2s\eps^2}}, for every \math{w\in T} with
\math{x=Uw},
\mld{
c_3\sqrt{s}(1-\eps)\norm{x}_2\le\norm{Gx}_1\le c_3\sqrt{s}
(1+\eps)\norm{x}_2.\label{eq:manhattanAPP}
}
Now consider any unit 2-norm \math{x\in L};
\math{x=U(w+\Delta)}, where \math{w\in T} has 2-norm at most $1$, 
\math{w+\Delta} has unit 2-norm,
 and \math{\norm{\Delta}_2\le\gamma}
because \math{T} is a \math{\gamma}-net on \math{S}. Then,
\mand{
\norm{Gx}_1=
\norm{GUw+GU\Delta}_1=\norm{GUw}_1+\Delta',
}
where \math{\abs{\Delta'}\le \norm{GU\Delta}_1}. We can bound the first term
on the RHS using \r{eq:manhattanAPP}. To bound the second term, use the
2-norm bound as follows:
\mand{
\norm{GU\Delta}_1
\le\sqrt{s}\norm{GU\Delta}_2
\le \sqrt{s(1+\eps)}\norm{U\Delta}_2
=
\sqrt{s(1+\eps)}\norm{\Delta}_2
\le
\sqrt{2s}\gamma,
}
(the last inequality is because \math{\eps\le 1}).
Thus, for every unit norm \math{x\in L},
\mand{
c_3\sqrt{s}(1-\eps)-2\gamma\sqrt{s}\le\norm{Gx}_1
\le c_3\sqrt{s}(1+\eps)+2\gamma\sqrt{s}.
}
Choosing \math{\gamma=c_3\eps/2}, \math{|T|=\expon
\left(2d(1+\frac{1}{c_3\eps\sqrt{2}})\right)}.
Since \math{c_3<(1+\eps)/(1-\eps)} (as otherwise by the two properties of an MJLP, $\|Gx\|_1 > \sqrt{s}\|Gx\|_2$ 
for some $x$, a contradiction) and \math{\eps\le\frac13},
with probability at least \math{1-c_1 e^{4d/c_3\eps}e^{-c_2s\eps^2}},
\mand{
c_3\sqrt{s}(1-2\eps)\le\norm{Gx}_1
\le c_3\sqrt{s}(1+2\eps).
}
After rescaling \math{2\eps\rightarrow\eps}, the probability
becomes at least \math{1-c_1 e^{8d/c_3\eps}e^{-c_2s\eps^2/4}}.
Taking a union bound over the 2-norm result and the Manhattan norm result,
and using \math{8d\le 8d/c_3\eps}, given that $c_3 \leq (1+\eps)/(1-\eps)$ and $\eps \leq 1/3$, we finally have that for any
unit 2-norm \math{x}, both
the inequalities
\eqan{
&(1-\eps)\le\norm{Gx}_2\le (1+\eps)\\
&c_3\sqrt{s}(1-\eps)\le\norm{Gx}_1
\le c_3\sqrt{s}(1+\eps)&
}
hold with probability at least 
\math{1-2c_1 e^{8d/c_3\eps}e^{-c_2s\eps^2/36}= 1-e^{-k}},
where the last equality follows by setting 
\math{s=36(k+\frac{8d}{c_3\eps}+
\log (2c_1))/c_2\eps^2=O(\frac{k}{\eps^2}+\frac{d}{\eps^3})}.
Since the result holds for any unit norm \math{x}, it holds for any \math{x} by scaling by \math{\norm{x}_2}.